\documentclass[11pt,twoside,letterpaper]{article} 
\usepackage{times,fancyhdr}
\usepackage[dvips]{graphicx}
\usepackage{color}

\usepackage[english]{babel}

\usepackage[pdfborder={0 0 0}, colorlinks=true, linkcolor=blue, citecolor=blue, urlcolor=cyan]{hyperref}


\makeatletter
\def\cleardoublepage{\clearpage\if@twoside \ifodd\c@page\else%
    \hbox{}%
    \thispagestyle{empty}%
    \newpage%
    \if@twocolumn\hbox{}\newpage\fi\fi\fi}
\makeatother

\def\figurename{Figure}
\makeatletter
\renewcommand{\fnum@figure}[1]{\figurename~\thefigure.}
\makeatother

\def\tablename{Table}
\makeatletter
\renewcommand{\fnum@table}[1]{\tablename~\thetable.}
\makeatother

\usepackage{epsfig}
\usepackage{amsmath}
\usepackage{amstext}
\usepackage{amssymb}
\usepackage{amsfonts}
\usepackage[square,comma,numbers]{natbib}

\def\beq{\begin{equation}}
\def\eeq{\end{equation}}
\def\bea{\begin{eqnarray}}
\def\eea{\end{eqnarray}}

\def\ra{\rangle}
\def\la{\langle}

\def\q{\mathbf{q}}
\def\k{\mathbf{k}}

\def\p{\mathbf{p}}

\def\Q{\mathbf{Q}}

\def\dxy{$d_{xy}$\ }
\def\dxz{$d_{xz}$\ }
\def\dyz{$d_{yz}$\ }
\def\dzz{$d_{3 z^2-r^2}$\ }
\def\dxxyy{$d_{x^2-y^2}$\ }
\def\Jp{J^{\prime}}
\def\Up{U^{\prime}}
\def\imsigma{\Sigma^{\prime\prime}}

\newcommand{\eps}{\varepsilon}
\newcommand{\s}{\sigma}
\newcommand{\su}{\uparrow}
\newcommand{\sd}{\downarrow}
\newcommand{\sgn}[1] {\mathrm{sgn}\left({#1}\right)}

\newcommand{\abs}[1] {\left|{#1}\right|}
\newcommand{\ii}{{\mathrm{i}}}

\newcommand{\nn}{\nonumber}
\newcommand{\vect}[1] {\mathbf{#1}}
\newcommand{\half}{\frac{1}{2}}
\newcommand{\sixth}{\frac{1}{6}}

\newcommand{\ppm}{{+-}}
\newcommand{\xx}{{xx}}
\newcommand{\yy}{{yy}}
\newcommand{\zz}{{zz}}
\newcommand{\uu}{{\uparrow\uparrow}}
\newcommand{\dd}{{\downarrow\downarrow}}
\newcommand{\ud}{{\uparrow\downarrow}}
\newcommand{\du}{{\downarrow\uparrow}}
\newcommand{\av}[1]{\left<{#1}\right>}

\newcommand{\ket}[1]{\left|{#1}\right>}
\newcommand{\brro}[1]{\left({#1}\right)}
\newcommand{\brsq}[1]{\left[{#1}\right]}
\newcommand{\Gammappm}{{\hat\Gamma}}

\newcommand{\GG}{\mathcal{G}}
\newcommand{\FF}{\mathcal{F}}

\newcommand{\GDLL}[1]{\textcolor{red}{\underline{#1}}}
\newcommand{\DLS}[1]{\textbf{#1}}
\newcommand{\DLL}[1]{\textit{#1}}

\begin{document}
\title{Itinerant spin fluctuations in iron-based superconductors
{\begin{flushleft}
\vskip 0.45in
{\normalsize\bfseries\textit{Chapter~1}}
\end{flushleft}
\vskip 0.45in
\bfseries\scshape Itinerant spin fluctuations in iron-based superconductors}}
\author{\bfseries\itshape Maxim M. Korshunov\thanks{E-mail address: mkor@iph.krasn.ru}\\
Kirensky Institute of Physics, Federal Research Center KSC SB RAS, Krasnoyarsk, Russia}

\date{}
\maketitle
\thispagestyle{empty}
\setcounter{page}{1}
\thispagestyle{fancy}
\fancyhead{}
\fancyhead[L]{In: Book Title \\
Editor: Editor Name, pp. {\thepage-\pageref{lastpage-01}}} 
\fancyhead[R]{ISBN 0000000000  \\
\copyright~2007 Nova Science Publishers, Inc.}
\fancyfoot{}
\renewcommand{\headrulewidth}{0pt}

\vspace{2in}

\noindent \textbf{PACS} 74.20.Rp, 74.25.-q, 74.62.Dh
\vspace{.08in} \noindent \textbf{Keywords:} Fe-based superconductors, unconventional superconductivity, spin fluctuations, spin resonance peak


\pagestyle{fancy}
\fancyhead{}
\fancyhead[EC]{Maxim M. Korshunov}
\fancyhead[EL,OR]{\thepage}
\fancyhead[OC]{Itinerant spin fluctuations in iron-based superconductors}
\fancyfoot{}
\renewcommand\headrulewidth{0.5pt}
\begin{abstract}
Multiband systems, which possess a wide parameter space, allow to explore a variety of competing ground states.
Bright examples are the Fe-based pnictides and chalcogenides, which demonstrate metallic, superconducting, and various magnetic phases. Here I discuss only one of the many interesting topics, namely, spin fluctuations in metallic multiband systems. I show how to calculate the effect of itinerant spin excitations on the electronic properties and formulate a theory of spin fluctuation-induced superconductivity. The superconducting state is unconventional and thus the system demonstrates unusual spin response with the spin resonance feature. I discuss its origin, consequences, and relation to experimental observations. Role of the spin-orbit coupling is specifically emphasized.
\end{abstract}


\section{Introduction}
\label{sec:intro}

The presence of several electronic orbitals in bands near the Fermi level of a metallic system provides both a rich set of properties and complications in revealing the underlying physics. One of the most widely discussed examples of such systems is iron-based materials~\cite{y_kamihara_08,zren08a}. Discovered in 2008, they become a new player in the field of high temperature superconductivity. In general, these materials can be divided into two subclasses, pnictides and chalcogenides, with the square lattice of Fe as the basic element, though with orthorhombic distortions in lightly doped materials below temperatures comparable with the transition temperature to the antiferromagnetic (AFM) spin-density wave state $T_{SDW}$. Iron is surrounded by As or P situated in the tetrahedral positions within the first subclass and by Se, Te, or S within the second subclass. Pnictides are represented by 1111 systems (LaFeAsO, LaFePO, Sr$_2$VO$_3$FeAs, etc.) and 111 systems (LiFeAs, LiFeP, and others) with the single iron layer per unit cell, and 122 systems containing two FeAs layers per unit cell (BaFe$_2$As$_2$, KFe$_2$As$_2$, and so on). Chalcogenides can be of 11 type (Fe$_{1-\delta}$Se, Fe$_{1+y}$Te$_{1-x}$Se$_x$, monolayers of FeSe) and of 122 type (KFe$_2$Se$_2$). The structure and physical properties of iron-based materials have been discussed in detail in many reviews (see, e.g. Refs.~\cite{SadovskiiReview2008,IvanovskiiReview2008,IzyumovReview2008,IshidaReview,JohnstonReview,PaglioneReview,MazinReview,LumsdenReview,WenReview,BasovReview,StewartReview,HirschfeldKorshunov2011,Hirschfeld2016,Inosov2016,Sadovskii2016,FernandesReview2017}).

Apart from the exact solution that is impossible for the complicated multiband models, there are two distinct approaches to the description of spin excitations in the interacting system. One of them starts with the localized spin limit. It has some success in describing magnetic excitations in the AFM phase of slightly doped iron pnictides and chalcogenides~\cite{Fang2008,Xu2008,SiAbrahams,Han2008}. Those studies usually rely on Heisenberg model with exchange up to the third nearest neighbors, $J_1-J_2-J_3$, and sometimes include quartic exchange $K$. It is quite convenient to describe spin wave dispersion revealed by neutron scattering within the Heisenberg-type models, though necessity for long-range and quartic interactions signify the drawbacks of purely localized models, see discussion in Ref.~\cite{MazinPRB2008}.

Another approach is to consider the system as itinerant. Surely, it works quite well for the nonmagnetic part of the phase diagram. It is also capable of describing the appearance of the SDW state~\cite{Dong2008,KorshunovEreminEPL2008,Cvetkovic2009,Eremin2010,Knolle2010,Eremin2014}, as well as the emergence of the nematic phase without preliminary orbital ordering~\cite{Fernandes_10,Fernandes2012,Fernandes2014,Glasbrenner2015}. Problems with the itinerant approach arise in the minor details, e.g., the measured bands are narrower than the ones obtained within the LDA (local density approximation). In most cases this can be cured by introducing correlations like it is done in LDA+DMFT (dynamical mean field theory), where it is possible to reproduce the experimental bands and corresponding spectral weights~\cite{Nekrasov2017}.

The reality is somewhere in-between the pure localized and pure itinerant approaches, so Fe-based superconductors (FeBS) represent a challenge for a theory as being a correlated system. However, since they tend to be moderately correlated with the pronounced metallic behaviour, I believe that the most reasonable way to describe FeBS lies within the correlated itinerant approach. With this in mind, I discuss below how spin excitations arise in the multiband metallic system, how they affect electrons in the normal state, how they produce superconductivity, and how they are changed in the superconducting state.

\subsection{General properties of iron-based systems}

Iron under normal conditions is ferromagnetic. Under the pressure, however, once the Fe atoms form an hcp lattice, iron becomes nonmagnetic and even superconducting at $T<2$К~\cite{Shimizu2001} most probably due to the electron-phonon interaction~\cite{Bose2003}. On the other hand, iron-based superconductors are the quasi-two-dimensional materials with the conducting lattice of Fe ions. The Fe-based superconductors, with $T_c$ up to 58~K in bulk materials (SmFeAsO$_{1-x}$F$_{x}$,~\cite{SmFeAsOTc}) and up to 110~K in monolayer FeSe at the SrTiO$_3$ substrate~\cite{FeSeTc,FeSeARPES,HeFeSeAnneal,TanFeSeARPES,GeFeSe100K},
stand in second place after high-$T_c$ cuprates. Latter are known for their high critical temperature, unconventional superconducting state, and unusual normal state properties.

At the first glance, the phase diagrams of cuprates and Fe-based superconductors are similar. In both cases the undoped materials exhibit antiferromagnetism that vanishes with doping; superconductivity occurs at some nonzero doping and then disappears, such that $T_c$ forms a ``dome''. While in hole-doped cuprates the long range ordered N\'eel phase vanishes before superconductivity occurs, in iron-based materials the competition between these orders can take several forms. In LaFeAsO, for example, there appears to be a transition between the magnetic and superconducting states at a critical doping value, whereas in the 122 systems the superconducting phase coexists with magnetism over a finite range and then persists to higher doping. It is tempting to conclude that the two classes of superconducting materials show generally very similar behavior, but there are profound differences as well. The first striking difference is that the undoped cuprates are Mott insulators, but iron-based materials are metals. This suggests that the Mott-Hubbard physics of a half-filled Hubbard model is not a good starting point for pnictides. It does not of course exclude effects of correlations in iron-based materials, but they may be moderate or small. Another characteristic feature of iron-based materials is a qualitative or sometimes even quantitative agreement of their Fermi surface measured by the angle-resolved photoemission spectroscopy (ARPES) and by the quantum oscillations with the Fermi surface calculated via density functional theory-based approaches. Therefore, the natural starting point for their description is the model of itinerant electrons.

The second important difference is related to normal state properties. Underdoped cuprates reveal the pseudogap behavior in both one-particle and two-particle charge and/or spin excitations, while the similar robust behavior is absent in iron-based materials. Generally speaking, the term ``pseudogap'' imply the dip in the density of states near the Fermi level. There are, however, a wide variety of unusual features in the pseudogap state of cuprates. For example, a strange metal phase near optimal doping in hole-doped cuprates is characterized by linear-$T$ resistivity over a wide range of temperatures. In iron-based materials, different temperature power laws for the resistivity, including linear $T$-dependence of the resistivity for some materials, have been observed near optimal doping and interpreted as being due to multiband physics and interband scattering~\cite{Golubov_10eng}. On the other hand, there are indications of a pseudogap formation in densities of states in some pnictides, see, e.g., Refs.~\cite{KordyukPseudogapReview,Kuchinskii2008eng}.

\subsection{Bird's eye view of the band structure}

In iron-based materials, the Fermi level is occupied by the $3d^6$ states of Fe$^{2+}$. This was established in the early density functional theory (DFT) calculations~\cite{s_lebegue_07,d_singh_08,Mazin2008}, which are in a quite good agreement with the results of quantum oscillations and ARPES. All five orbitals, $d_{x^2-y^2}$, $d_{3z^2-r^2}$, $d_{xy}$, $d_{xz}$, and $d_{yz}$, are near or at the Fermi level. This results in the significantly multiorbital and multiband low-energy electronic structure, which could not be described within the single-band model. Since the conductivity is provided by the iron layer, the discussion of physics in terms of quasi two-dimensional system in most cases gives reasonable results~\cite{HirschfeldKorshunov2011}. At the same time, the presence of a few pockets and the multiorbital band character significantly affect the superconducting pairing.

Excluding the cases of extreme hole and electron doping, Fermi surface consists of two hole sheets around the $\Gamma=(0,0)$ point and two electron sheets around the $(\pi,0)$ and $(0,\pi)$ points in the two-dimensional Brillouin zone (BZ) corresponding to one Fe per unit cell (the so-called 1-Fe BZ)~\cite{HirschfeldKorshunov2011}. In the 2-Fe BZ, electron pockets are centered at the $M=(\pi,\pi)$ point. Such $\k$-space geometry results in the possibility of the spin-density wave (SDW) instability and the enhanced antiferromagnetic fluctuations due to the nesting between hole and electron Fermi surface sheets at the wave vector $\Q$ equal to $(\pi,0)$ or $(0,\pi)$ in the 1-Fe BZ or to $(\pi,\pi)$ in the 2-Fe BZ. Upon doping $x$, the long-range SDW order is destroyed. If electrons are doped, then for the large $x$, hole pockets disappear leaving only electron Fermi surface sheets that are observed in K$_x$Fe$_{2-y}$Se$_2$ and in FeSe monolayers~\cite{FeSeARPES}. Upon increase of the hole doping, first, a new hole pocket appears around $(\pi,\pi)$ point, and then electron sheets vanish. KFe$_2$As$_2$ corresponds to the latter case. ARPES confirms that the maximal contribution to the bands at the Fermi level comes from the $d_{xz,yz}$ and $d_{xy}$ orbitals~\cite{Kordyuk,Brouet}.

\subsection{Basic features of superconductivity in FeBS}

In the cuprates, the $d_{x^2-y^2}$ symmetry of the gap with $\cos k_x-\cos k_y$ structure was empirically established by penetration depth, ARPES, NMR and phase sensitive Josephson tunneling experiments. No similar consensus on any universal gap structure has been reached after several years of intensive research on the high-quality monocrystals of iron-based superconductors. There are strong evidences that small differences in electronic structure can lead to a considerable diversity in superconducting gap structures, including nodal states and states with a full gap at the Fermi surface. The actual symmetry class of most of the materials may be of generalized $A_{1g}$ (extended $s$-wave symmetry) type, probably involving a sign change of the order parameter between Fermi surface sheets or its parts~\cite{HirschfeldKorshunov2011}. Moreover, nothing forbids different iron-based materials from having different order parameter symmetries. Indeed, there are several theories claiming that, while most iron-based systems have the $s$-wave gap symmetry, those with unusual Fermi surfaces with either electron or hole pockets can have the $d$-wave symmetry of the order parameter~\cite{Maier2011,Wang2011,Das2011,MaitiKorshunovPRL2011,Mazin2011,Korshunov2014eng}. Nevertheless, it seems quite possible that the ultimate source of the pairing interaction in both cuprates and FeBS is fundamentally similar, although essential details such as pairing symmetry and the gap structure in FeBS depend on the Fermi surface geometry, orbital character, and degree of correlations~\cite{HirschfeldKorshunov2011,Hirschfeld2016}.

\begin{figure}
 \centering
 \includegraphics[width=\textwidth]{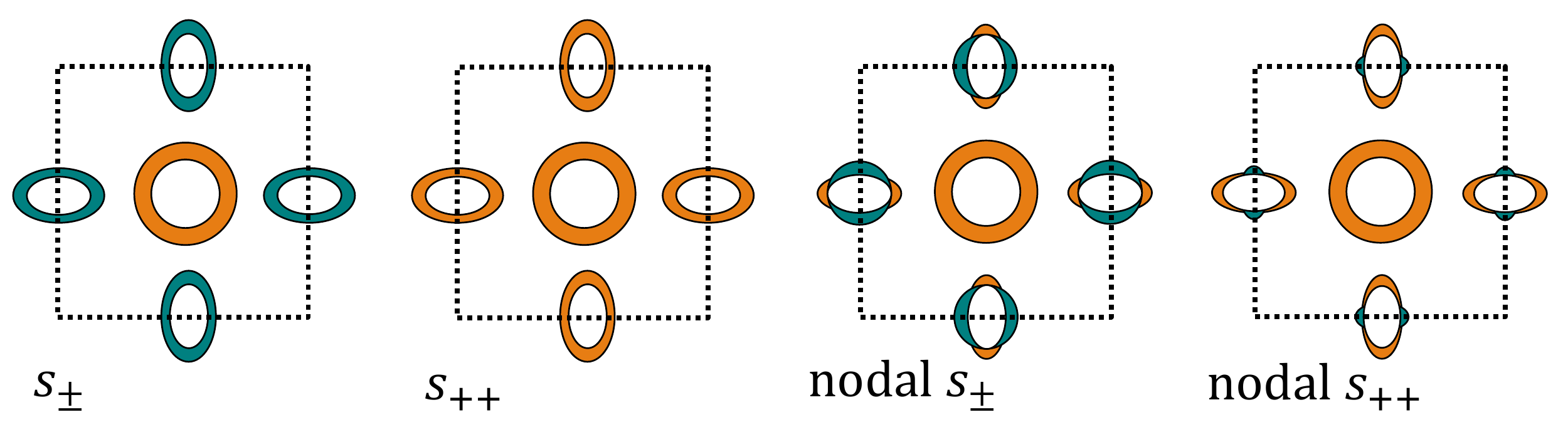} 
 \caption{Cartoon of order parameter \textit{structures} having the $s$-wave symmetry in the two-dimensional Brillouin zone (dashed square) corresponding to one iron per unit cell. Different colors stands for different signs of the gap.}
 \label{fig:gaps}
\end{figure}

Note that the term \textit{symmetry} should be distinguished from the term \textit{structure} of the gap. Latter we use to designate the $\k$-dependent variation of an order parameter within a given symmetry class. Gaps with the same symmetry may have very different structures. Let us illustrate this for the $s$-wave ($A_{1g}$) symmetry, see Fig.~\ref{fig:gaps}. Fully gapped $s$-states without nodes at the Fermi surface differ only by a relative gap sign between the hole and electron pockets that is positive in the $s_{++}$ state and negative in the $s_\pm$ state. On the other hand, in the nodal $s$-states, the gap vanishes at certain points on the electron pockets. These states are called ``nodal $s_\pm$'' (``nodal $s_{++}$'') and are characterized by the opposite (same) averaged signs of the order parameter on the hole and electron pockets. Nodes of this type are sometimes described as ``accidental'', since their existence is not dictated by symmetry in contrast to the symmetry nodes of the $d$-wave gap. Therefore, they can be removed continuously, resulting in either an $s_\pm$ or an $s_{++}$ state~\cite{v_mishra_09,Mizukami2014}. In general, understanding the symmetry character of the superconducting ground states as well as the detailed structure of the order parameter should provide clues to the microscopic pairing mechanism in the iron-based materials and thereby lead to a deeper understanding of the phenomenon of high-$T_c$ superconductivity.

Soon after the discovery of superconductivity in pnictides, estimates were made of the possibility of pairing due to the electron-phonon interaction. The coupling constant appears to be even smaller than that for aluminum~\cite{Boeri_08}, although $T_c$ in
some pnictides is significantly higher. This led to the conclusion that it is unlikely that the pairing caused by electron-phonon interaction could play a leading role in the emergence of superconductivity, although a more thorough analysis is probably required to take into account some specific features of the electronic band structure~\cite{Eschrig}. Such a situation immediately led to searching for alternative theories of superconducting pairing. The interactions that are analyzed in the theories vary from spin and orbital fluctuations to Hubbard repulsion and Hund's exchange. It is unrealistic to describe or even simply mention all these theories here; therefore, we focus on one of the most promising theories, namely, the spin-fluctuation theory of the superconducting pairing.

The spin-fluctuation theory of superconductivity is promising for a number of reasons: (1) this theory is based on the model of itinerant electrons that serves as a good starting point for the description of iron compounds; (2) the superconducting phase arises directly after the AFM phase or coexists with it; in this case, the character of the spin-lattice relaxation rate $1/T_1T$ gradually changes from the Curie-Weiss to Pauli behavior with an increase of doping~\cite{Ning2009} that indicates a decrease in the role of spin fluctuations; (3) the description of various experimentally observed properties of the pnictides and chalcogenides does not require the introduction of additional parameters in the theory; rather, only some specific features of the band structure and the interactions in different classes of FeBS should be taken into account.

\subsection{Chapter structure}

The chapter is organized as follows. First, I discuss the tight-binding model for iron-based systems and Hubbard-type interactions in Section~\ref{sec:model}. Then I describe the details of dynamical spin susceptibility calculation in Section~\ref{sec:susceptibility}. Role of scattering by spin excitations in the normal state properties is discussed in Section~\ref{sec:selfenergy}. How the superconductivity arises from spin fluctuations is described in Section~\ref{sec:sc}. It is divided in two parts: first one, subsection~\ref{subsec:sconeband}, deals with the simple single-band picture, and the second one, subsection~\ref{subsec:scmultiband}, is devoted to multiband approach to spin fluctuation mediated pairing. Spin resonance peak as a signature of $s_\pm$ state is discussed in Section~\ref{sec:spinres}. Conclusion is given in the last Section~\ref{sec:conclusion}.

\section{Multiorbital tight-binding model and interactions}
\label{sec:model}

We write the model in the general form separating the `bare' and the interacting parts of the Hamiltonian,
\beq
H = H_{0} + H_{int}.
\label{eq:H}
\eeq

\subsection{`Bare' part}
\label{subsec:H0}

\begin{figure}
 \centering
 \includegraphics[width=1.0\columnwidth]{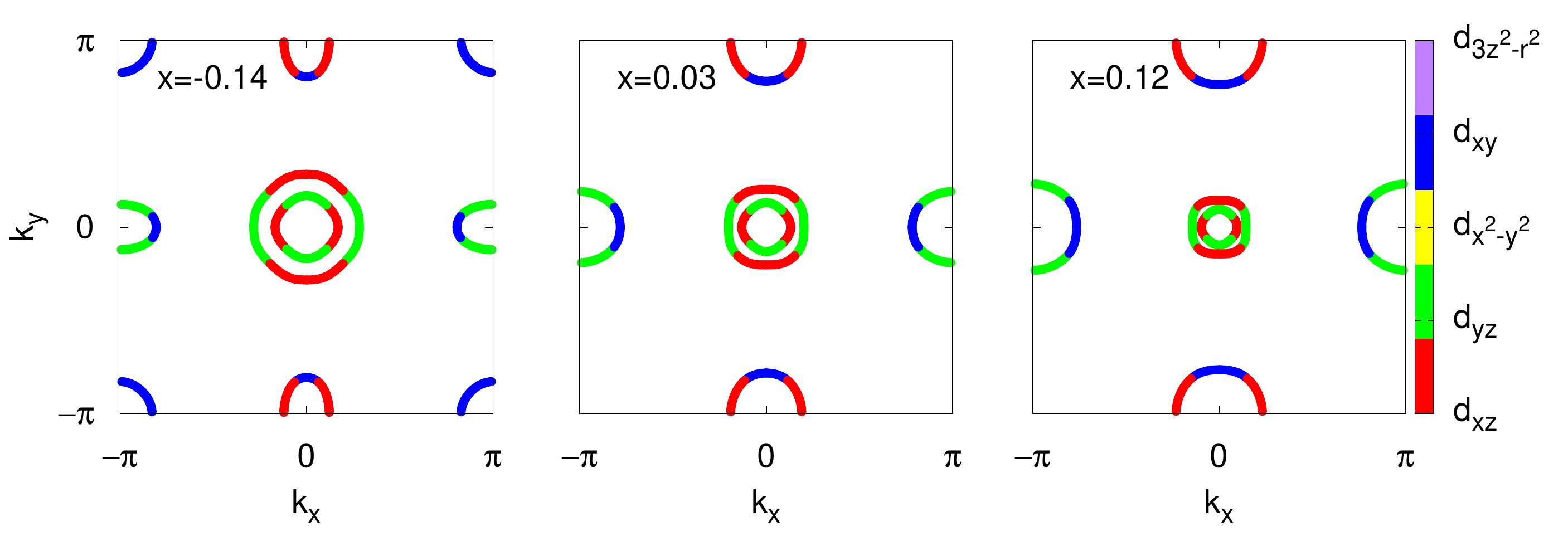}
 \caption{Fermi surfaces for electron doped (dopings $x=0.03$ and $0.12$) and hole doped (doping $x=-0.14$) systems calculated within the five-orbital model~\cite{Graser2009}. Different colors indicate the major orbital contribution to the particular point at the Fermi surface.}
 \label{fig:FS}
\end{figure}

As the kinetic energy $H_0$, we use the five-orbital tight-binding model of Graser \textit{et al.}~\cite{Graser2009} that is based on the DFT band structure calculations \cite{c_cao_08} within LDA for the prototypical iron pnictide, LaFeAsO. $H_0$ is described by a tight-binding model spanned by five Fe $d$-orbitals (\dxz, \dyz, \dxxyy, \dxy, and \dzz). Total number of electrons is given by $n = n_0 \pm x$, where electron filling $n_0 = 6$ corresponds to the fully occupied $d^6$-orbital and $x$ is the doping concentration. The \dxz, \dyz, and \dxy bands dominate near the Fermi level, as seen in Fig.~\ref{fig:FS}, where we show the Fermi surface arising from $H_0$ in the 1-Fe BZ. For the electron-doped and undoped systems, the Fermi surface consists of two small hole pockets, $\alpha_1$ and $\alpha_2$, around the $\Gamma=(0,0)$ point, and two small electron pockets, $\beta_1$ and $\beta_2$, around the $X=(\pi,0)$ and $Y=(0,\pi)$ points, respectively. Upon hole doping a new hole Fermi surface pocket, $\gamma$, emerges around the $(\pi,\pi)$ point.

Explicit form of $H_0$ is
\begin{equation}
 H_0 = \sum_{\k \sigma} \sum_{l l'} \left[ \left( \epsilon_{l} - \mu_0 \right) \delta_{l l'} + t_{l l'}(\k) \right] d_{\k l \sigma}^\dag d_{\k l' \sigma},
 \label{eq:H0}
\end{equation}
where $d_{\k l \sigma}$ is the annihilation operator of the electron with momentum $\k$, spin $\sigma$, and orbital index $l$, $t_{l l'}(\k)$ is the hopping matrix element, $\epsilon_{l}$ is the one-electron energy, and $\mu_0$ is the chemical potential. Later we use numerical values of hopping matrix elements $t_{l l'}(\k)$ and one-electron energies $\epsilon_{l}$ from Ref.~\cite{Graser2009}. Similar model for iron pnictides was proposed in Ref.~\cite{Kuroki2008}.

There is an important consequence of the multiorbital nature of the system. The single-particle noninteracting Green's function is diagonal in the band space but not in the orbital space. It makes sense to transform to the band basis that is constructed using operators of electron's creation and annihilation, $b_{\k\mu\sigma}^\dag$ and $b_{\k\mu\sigma}$, with the band index $\mu$. Green's function for $H_0$ is then diagonal in the band basis,
\beq
 G_{\mu\sigma}(\k,\ii\omega_n) = 1 / \left( \ii\omega_n - \eps_{\k\mu\sigma} \right),
 \label{eq:G0}
\eeq
where $\omega_n$ is the Matsubara frequency and $\eps_{\k\mu\sigma}$ is the energy.

The transition from the orbital to the band basis is implemented with the aid of the orbital matrix elements $\varphi^{\mu}_{\k l}$,
\beq
 \ket{\sigma l \k} = \sum\limits_{\mu} \varphi^{\mu}_{\k l} \ket{\sigma \mu \k}.
 \label{eq:varphi}
\eeq
In this case, $d_{\k l \sigma} = \sum\limits_{\mu} \varphi^{\mu}_{\k l} b_{\k \mu \sigma}$.

The noninteracting part of the Hamiltonian, $H_0$, is a complex matrix, which in general has complex eigenvectors $\varphi_{\k l}^{\mu}$, although the eigenvalues $\eps_{\k\mu\sigma}$ are real. In order to utilize a simple form of the Green's function spectral representation, in some cases it is useful to choose a gauge in which the Hamiltonian is real by performing a unitary transformation
\beq
 \tilde{H_0} = \hat{\psi}^{-1} \hat{H}_0 \hat{\psi},
 \label{eq:Hpsi}
\eeq
where $\hat{\psi}$ is the diagonal matrix, $\hat{\psi} = \mathrm{diag}\left(\ii, \ii, 1, 1, 1\right)$. The interaction part of the Hamiltonian must then also be rotated by $\hat{\psi}$.

\subsection{Interaction part}
\label{subsec:Hint}

As the interaction part of the model, we take the on-site Coulomb (Hubbard) electron-electron repulsion written for the the multiorbital systems~\cite{Graser2009,Kuroki2008,Castallani1978,Oles1983,Kemper2010},
\bea
 H_{int} &=& U \sum_{f, l} n_{f l \su} n_{f l \sd} + U' \sum_{f, l < l'} n_{f l} n_{f l'} \nn \\
 &+& J \sum_{f, l < l'} \sum_{\sigma,\sigma'} d_{f l \sigma}^\dag d_{f l' \sigma'}^\dag d_{f l \sigma'} d_{f l' \sigma}
 + J' \sum_{f, l \neq l'} d_{f l \su}^\dag d_{f l \sd}^\dag d_{f l' \sd} d_{f l' \su},
\label{eq:Hint}
\eea
where $n_{f l} = n_{f l \su} + n_{f l \sd}$, $d_{f l \sigma}$ is the electron annihilation operator, $n_{f l \sigma} = d_{f l \sigma}^\dag d_{f l \sigma}$ is the number of particles operator, $f$ is the site index, $l$ and $l'$ are orbital indices, $U$ and $U'$ are intra- and interorbital Hubbard repulsions, $J$ is the Hund's exchange, and $J'$ is the pair-hopping. Usually, parameters obey the spin-rotational invariance (SRI) that leads to relations $U' = U - 2J$ and $J' = J$ thus reducing the number of free parameters in the theory. Sometimes $V = U' - J/4$ is referred to as the interorbital Hubbard repulsion; the SRI relation is then written as $V = U - 5J/4$.

The following way to write $H_{int}$ is useful for extracting explicit expressions for interaction lines in diagrams for spin susceptibility,
\bea
 H_{int} &=& \sum_{f} \left[ U \sum_{l} d_{f l \su}^\dag d_{f l \su} d_{f l \sd}^\dag d_{f l \sd} - J' \sum_{l \neq l'} d_{f l \su}^\dag d_{f l' \sd} d_{f l \sd}^\dag d_{f l' \su} \right. \nn\\
 &+& (U' - J)/2 \sum_{l \neq l'} \left( d_{f l \su}^\dag d_{f l \su} d_{f l' \su}^\dag d_{f l' \su} + d_{f l \sd}^\dag d_{f l \sd} d_{f l' \sd}^\dag d_{f l' \sd} \right) \nn\\
 &+& U'/2 \sum_{l \neq l'} \left( d_{f l \su}^\dag d_{f l \su} d_{f l' \sd}^\dag d_{f l' \sd} + d_{f l \sd}^\dag d_{f l \sd} d_{f l' \su}^\dag d_{f l' \su} \right) \nn\\
 &-& \left. J/2 \sum_{l \neq l'} \left( d_{f l \su}^\dag d_{f l \sd} d_{f l' \sd}^\dag d_{f l' \su} + d_{f l \sd}^\dag d_{f l \su} d_{f l' \su}^\dag d_{f l' \sd} \right) \right].
\label{eq:Hintexpl}
\eea

The interactions in Hamiltonian~(\ref{eq:Hint}) have a complicated orbital structure. To compactify the expressions we define the local matrix interaction in orbital space,
\beq
 U_{nn'}^{ll'} d_{f l \sigma_1}^\dag d_{f n' \sigma_2}^\dag d_{f l' \sigma_3} d_{f n \sigma_4},
 \label{eq:Ueff}
\eeq
which accounts for all the quartic terms. It will be used later for calculations of susceptibility and self-energy diagrams.

\subsection{Spin-orbit coupling}
\label{subsec:socmodel}

One of the interesting features of FeBS, which was initially observed in Ba(Fe$_{1-x}$Ni$_x$)$_2$As$_2$, is the anisotropy in the spin space~\cite{Lipscombe}. It was found that $\mathrm{Im}\chi_{+-}$ and $2\mathrm{Im}\chi_{zz}$ are different. More generally, $xx$, $yy$, and $zz$ components of the spin susceptibility differ by some amount. This definitely contradicts the spin-rotational invariance, $\la S_x S_x \ra = \la S_y S_y \ra = \la S_z S_z \ra$ and $\la S_+ S_- \ra> = 2 \la S_z S_z \ra$, which have to be obeyed in the disordered paramagnetic system. The relation $\mathrm{Im}\chi_{+-} > 2\mathrm{Im}\chi_{zz}$ was also confirmed in measurements of the NMR spin-lattice relaxation rate~\cite{MatanoBKFA,LiBKFA}. One of the ways to solve the puzzle is to assume the presence of the spin-orbit (SO) interaction that can break the SRI like it does in Sr$_2$RuO$_4$~\cite{Eremin2002}. In pnictides, it is indeed provides splitting between $\mathrm{Im}\chi_{+-}$ and $2\mathrm{Im}\chi_{zz}$ components of the spin susceptibility~\cite{KorshunovTogushovaSO2013}. Here I discuss how to incorporate the effect of the SO coupling in the susceptibility calculation.

First step it to include SO coupling in the model. Since it is a local interaction that can be exactly diagonalized due to its $\phi^2$ structure, it is natural to include the SO coupling in the `bare' part of the Hamiltonian thus renormalizing it,
\beq
 H_{0\lambda} = H_0 + H_{\lambda}.
 \label{eq:H0SO}
\eeq
Here the spin-orbit Hamiltonian $H_{\lambda}$ is given by
\beq
 H_{\lambda} = \frac{\lambda}{2} \sum\limits_{f} \vect{L}_f\cdot\vect{S}_f,
 \label{eq:HSO}
\eeq
where $\lambda$ is the SO coupling strength, $f$ is the site number, $\vect{L}_f$ and $\vect{S}_f$ are the on-site orbital and the spin operators, respectively.

In this case it is easier to work with a matrix form of the Hamiltonian, so we define a vector in the orbital-spin space,
\beq
 \hat{d}_\k^\dag = \left( d_{\k 1 \su}^\dag, d_{\k 2 \su}^\dag, d_{\k 3 \su}^\dag, d_{\k 4 \su}^\dag, d_{\k 5 \su}^\dag, d_{\k 1 \sd}^\dag, d_{\k 2 \sd}^\dag, d_{\k 3 \sd}^\dag, d_{\k 4 \sd}^\dag, d_{\k 5 \sd}^\dag\right).
 \label{eq:dvect}
\eeq
The order of $d$-orbitals that we are going to use is the following: $\left\{d_{xz}, d_{yz}, d_{x^2-y^2}, d_{xy}, d_{3z^2-r^2} \right\}$. Then we have
\beq
 H_{0\lambda} = \sum\limits_{\k} \hat{d}_\k^\dag \hat{H}_{0\lambda} \hat{d}_\k,
 \label{eq:H0SOmatk}
\eeq
where the spin-resolved Hamiltonian matrix has the form
\beq
 \hat{H}_{0\lambda} = \left(
      \begin{array}{cc}
      \hat{H}_{0\lambda}^{\su\su} & \hat{H}_{0\lambda}^{\su\sd} \\
      \hat{H}_{0\lambda}^{\sd\su} & \hat{H}_{0\lambda}^{\sd\sd} \\
      \end{array}
     \right).
 \label{eq:H0SOmat}
\eeq

Omitting the site index since SO interaction is always local, dot product in Eq.~(\ref{eq:H0SO}) can be conveniently expressed in terms of the $\ppm$ components of spin and orbital vectors,
\beq
 \hat{H}_{\lambda f} = \frac{\lambda}{2} \vect{L}\cdot\vect{S} = \frac{\lambda}{2} \left[ \hat{S}_z \otimes \hat{L}_z + \half \left( \hat{S}_+ \otimes \hat{L}_- + \hat{S}_- \otimes \hat{L}_+ \right) \right],
\eeq
Here $\hat{S}_{z,+,-}$ are the spin matrices equal to Pauli matrices, $\hat{S}_z = \hat{\sigma}^z$, $\hat{S}_\pm = \hat{\sigma}^\pm$, where $\hat{\sigma}^\pm = \hat{\sigma}^x \pm \ii \hat{\sigma}^y$,
\bea
\label{eq:sigmaM}
&&\hat{\sigma}^x=\left(
                 \begin{array}{cc}
                   0 & 1 \\
                   1 & 0 \\
                 \end{array}
               \right),
\hat{\sigma}^y=\ii \left(
                 \begin{array}{cc}
                   0 & -1 \\
                   1 & 0 \\
                 \end{array}
               \right),
\hat{\sigma}^z=\left(
                 \begin{array}{cc}
                   1 & 0 \\
                   0 & -1 \\
                 \end{array}
               \right), \nn\\
&&\hat{\sigma}^+ =\left(
                 \begin{array}{cc}
                   0 & 2 \\
                   0 & 0 \\
                 \end{array}
               \right),
\hat{\sigma}^- =\left(
                 \begin{array}{cc}
                   0 & 0 \\
                   2 & 0 \\
                 \end{array}
               \right).
\eea
$\hat{L}_{z,+,-}$ are the $5 \times 5$ matrices describing $d$-orbitals in some basis. It is easy to express them in the atomic basis defined by spherical harmonic functions of degree $l$ and order $m$, $Y_l^m(\theta,\varphi)$. In general, matrix elements of $\hat{L}$ are $\left(\hat{L}_{z,+,-}\right)_{m',m} = L_{z,+,-}(j,m',m)$, where
$L_+(j,m',m) = \delta_{m' - 1, m} \sqrt{j(j + 1) - m(m + 1)}$, $L_-(j,m,m') = \delta_{m, m' - 1} \sqrt{j(j + 1) - m(m + 1)}$, and $L_z(j,m',m) = m \delta_{m', m}$.
Values $j=2$ and $l=2$ correspond to $d$-orbitals, thus $m = l \ldots -l$ takes the values $2, 1, 0, -1, -2$. We denote the explicit form of $\hat{L}$ in this basis by $\hat{L}'$ that is equal to
\bea
&\hat{L}'_+ = \left(
            \begin{array}{ccccc}
             0 & 2 & 0 & 0 & 0 \\
             0 & 0 & \sqrt{6} & 0 & 0 \\
             0 & 0 & 0 & \sqrt{6} & 0 \\
             0 & 0 & 0 & 0 & 2 \\
             0 & 0 & 0 & 0 & 0
            \end{array}
            \right), \
\hat{L}'_- = \left(
            \begin{array}{ccccc}
             0 & 0 & 0 & 0 & 0 \\
             2 & 0 & 0 & 0 & 0 \\
             0 & \sqrt{6} & 0 & 0 & 0 \\
             0 & 0 & \sqrt{6} & 0 & 0 \\
             0 & 0 & 0 & 2 & 0
            \end{array}
            \right),& \nn\\
&\hat{L}'_z = \mathrm{diag}\left(2, 1, 0, -1, -2\right).&
\eea

Since the tight-binding Hamiltonian $H_0$ is written in the $d$-orbital basis, we need to rewrite the matrix $\hat{L}'$ in that basis. The general structure of the SO Hamiltonian matrix is the following,
\beq
 \hat{H}_{\lambda f} = \frac{\lambda}{2} \left(
                    \begin{array}{c|c}
                     \hat{SO}_z & \hat{SO}_x \\
                     \hline
                     \hat{SO}_y & - \hat{SO}_z \\
                    \end{array}
                   \right).
\label{eq:HSOmatrix}
\eeq
The spin-resolved Hamiltonian of the system with the SO coupling becomes
\beq
 \hat{H}_{0\lambda} = \left(
      \begin{array}{cc}
      \hat{H}_0 + \frac{\lambda}{2} \hat{SO}_z & \frac{\lambda}{2} \hat{SO}_x \\
      \frac{\lambda}{2} \hat{SO}_y & \hat{H}_0^* - \frac{\lambda}{2} \hat{SO}_z \\
      \end{array}
     \right).
 \label{eq:H0SOgen}
\eeq
Here $\hat{H}_{0\lambda}^{\sd\sd}$ contains complex conjugated term $\hat{H}_{0}^*$. It is done to preserve the Kramers degeneracy, $\eps_{\k l \su} = \eps_{\k l \sd}$, that should not be broken by the SO coupling of the form~(\ref{eq:HSO}).

There are, however, several definitions of $d$-orbitals through the $Y_l^m$'s that give different forms of the SO matrices $\hat{H}_{\lambda}$. Let us discuss few of them.

First popular definition is the following: $d_{xz} = \frac{1}{\sqrt{2}} (Y_2^{-1} - Y_2^{1})$, $d_{yz} = \frac{\ii}{\sqrt{2}} (Y_2^{-1} + Y_2^{1})$, $d_{x^2-y^2} = \frac{1}{\sqrt{2}} (Y_2^{-2} + Y_2^{2})$, $d_{xy} = \frac{\ii}{\sqrt{2}} (Y_2^{-2} - Y_2^{2})$, and $d_{3z^2-r^2} = Y_2^{0}$. Then the transformation matrix from the $Y_l^m$ basis to the $d$-orbital basis is
\beq
\hat{\Psi} = \left(
            \begin{array}{ccccc}
             0 & -\frac{1}{\sqrt{2}} & 0 & \frac{1}{\sqrt{2}} & 0 \\
             0 & \frac{i}{\sqrt{2}} & 0 & \frac{i}{\sqrt{2}} & 0 \\
             \frac{1}{\sqrt{2}} & 0 & 0 & 0 & \frac{1}{\sqrt{2}} \\
             -\frac{i}{\sqrt{2}} & 0 & 0 & 0 & \frac{i}{\sqrt{2}} \\
             0 & 0 & 1 & 0 & 0
            \end{array}
          \right).
\eeq
Using this matrix, we transform the $\hat{L}'$ matrix in the $Y_l^m$ basis to the matrix $\hat{L}$ in the $d$-orbitals basis, $\hat{L} = \hat{\Psi} \hat{L}' \hat{\Psi}^\dag$. Multiplying the latter by the Pauli matrices, we find
\beq
 \hat{H}_{\lambda f} = \frac{\lambda}{2} \left(
                    \begin{array}{ccccc|ccccc}
                     0 & i & 0 & 0 & 0 & 0 & 0 & -1 & -i & \sqrt{3} \\
                     -i & 0 & 0 & 0 & 0 & 0 & 0 & i & -1 & i \sqrt{3} \\
                     0 & 0 & 0 & 2 i & 0 & 1 & -i & 0 & 0 & 0 \\
                     0 & 0 & -2 i & 0 & 0 & i & 1 & 0 & 0 & 0 \\
                     0 & 0 & 0 & 0 & 0 & -\sqrt{3} & -i \sqrt{3} & 0 & 0 & 0 \\
                     \hline
                     0 & 0 & 1 & -i & -\sqrt{3} & 0 & -i & 0 & 0 & 0 \\
                     0 & 0 & i & 1 & i \sqrt{3} & i & 0 & 0 & 0 & 0 \\
                     -1 & -i & 0 & 0 & 0 & 0 & 0 & 0 & -2 i & 0 \\
                     i & -1 & 0 & 0 & 0 & 0 & 0 & 2 i & 0 & 0 \\
                     \sqrt{3} & -i \sqrt{3} & 0 & 0 & 0 & 0 & 0 & 0 & 0 & 0
                    \end{array}
                   \right).
 \label{eq:HSOwien2k}
\eeq

Second definition
states that $d_{xz} = \frac{1}{\sqrt{2}} (Y_2^{1} - Y_2^{-1})$, $d_{yz} = -\frac{\ii}{\sqrt{2}} (Y_2^{1} + Y_2^{-1})$, $d_{x^2-y^2} = \frac{1}{\sqrt{2}} (Y_2^{2} + Y_2^{-2})$, $d_{xy} = -\frac{\ii}{\sqrt{2}} (Y_2^{2} - Y_2^{-2})$, and $d_{3z^2-r^2} = Y_2^{0}$. The transformation matrix is
\beq
\hat{\Psi} = \left(
            \begin{array}{ccccc}
             0 & \frac{1}{\sqrt{2}} & 0 & -\frac{1}{\sqrt{2}} & 0 \\
             0 & -\frac{i}{\sqrt{2}} & 0 & -\frac{i}{\sqrt{2}} & 0 \\
             -\frac{i}{\sqrt{2}} & 0 & 0 & 0 & \frac{i}{\sqrt{2}} \\
             \frac{1}{\sqrt{2}} & 0 & 0 & 0 & \frac{1}{\sqrt{2}} \\
             0 & 0 & 1 & 0 & 0
            \end{array}
          \right)
\eeq
that leads to the following SO coupling Hamiltonian:
\beq
 \hat{H}_{\lambda f} = \frac{\lambda}{2} \left(
                    \begin{array}{ccccc|ccccc}
                     0 & i & 0 & 0 & 0 & 0 & 0 & i & 1 & -\sqrt{3} \\
                     -i & 0 & 0 & 0 & 0 & 0 & 0 & 1 & -i & -i \sqrt{3} \\
                     0 & 0 & 0 & -2 i & 0 & -i & -1 & 0 & 0 & 0 \\
                     0 & 0 & 2 i & 0 & 0 & -1 & i & 0 & 0 & 0 \\
                     0 & 0 & 0 & 0 & 0 & \sqrt{3} & i \sqrt{3} & 0 & 0 & 0 \\
                     \hline
                     0 & 0 & i & -1 & \sqrt{3} & 0 & -i & 0 & 0 & 0 \\
                     0 & 0 & -1 & -i & -i \sqrt{3} & i & 0 & 0 & 0 & 0 \\
                     -i & 1 & 0 & 0 & 0 & 0 & 0 & 0 & 2 i & 0 \\
                     1 & i & 0 & 0 & 0 & 0 & 0 & -2 i & 0 & 0 \\
                     -\sqrt{3} & i \sqrt{3} & 0 & 0 & 0 & 0 & 0 & 0 & 0 & 0
                    \end{array}
                    \right).
 \label{eq:HSOIlya}
\eeq

The third definition,
$d_{xz} = \frac{1}{\sqrt{2}} (Y_2^{1} + Y_2^{-1})$, $d_{yz} = -\frac{\ii}{\sqrt{2}} (Y_2^{1} - Y_2^{-1})$, $d_{x^2-y^2} = \frac{1}{\sqrt{2}} (Y_2^{2} + Y_2^{-2})$, $d_{xy} = -\frac{\ii}{\sqrt{2}} (Y_2^{2} - Y_2^{-2})$, and $d_{3z^2-r^2} = Y_2^{0}$, results in the following transformation matrix:
\beq
\hat{\Psi} = \left(
            \begin{array}{ccccc}
             0 & \frac{1}{\sqrt{2}} & 0 & \frac{1}{\sqrt{2}} & 0 \\
             0 & -\frac{i}{\sqrt{2}} & 0 & \frac{i}{\sqrt{2}} & 0 \\
             \frac{1}{\sqrt{2}} & 0 & 0 & 0 & \frac{1}{\sqrt{2}} \\
             -\frac{i}{\sqrt{2}} & 0 & 0 & 0 & \frac{i}{\sqrt{2}} \\
             0 & 0 & 1 & 0 & 0
            \end{array}
          \right).
\eeq
Using it, we find
\beq
 \hat{H}_{\lambda f} = \frac{\lambda}{2} \left(
                        \begin{array}{ccccc|ccccc}
                         0 & i & 0 & 0 & 0 & 0 & 0 & 1 & i & \sqrt{3} \\
                         -i & 0 & 0 & 0 & 0 & 0 & 0 & -i & 1 & i \sqrt{3} \\
                         0 & 0 & 0 & 2 i & 0 & 1 & -i & 0 & 0 & 0 \\
                         0 & 0 & -2 i & 0 & 0 & i & 1 & 0 & 0 & 0 \\
                         0 & 0 & 0 & 0 & 0 & \sqrt{3} & i \sqrt{3} & 0 & 0 & 0 \\
                         \hline
                         0 & 0 & 1 & -i & \sqrt{3} & 0 & -i & 0 & 0 & 0 \\
                         0 & 0 & i & 1 & -i \sqrt{3} & i & 0 & 0 & 0 & 0 \\
                         1 & i & 0 & 0 & 0 & 0 & 0 & 0 & -2 i & 0 \\
                         -i & 1 & 0 & 0 & 0 & 0 & 0 & 2 i & 0 & 0 \\
                         \sqrt{3} & -i \sqrt{3} & 0 & 0 & 0 & 0 & 0 & 0 & 0 & 0
                        \end{array}
                        \right).
 \label{eq:HSOwikirus}
\eeq
Note that in this particular case, $\hat{H}_{0\lambda}^{\sd\sd}$ in Eq.~(\ref{eq:H0SOgen}) should contains the non-conjugated term $\hat{H}_{0}$ to preserve the Kramers degeneracy. It results in the band structure without additional splitting.

All the mentioned definitions of $d$-orbitals' wave functions differ by the phase. Altering the phase should not change the observable properties of the system. Therefore, different forms of the SO matrices $\hat{H}_{\lambda}$ should lead to the same set of eigenvalues and in general it doesn't matter which form to use.

The addition of the SO coupling leads to the important consequence: the Hamiltonian and the single-particle noninteracting Green's function are diagonal in the band-pseudospin basis, but not in the orbital-spin space. Therefore, Eqs.~(\ref{eq:G0}) and~(\ref{eq:varphi}) relying on the spin $\sigma$ are not valid anymore and we have to define a new band-pseidospin basis.
Let's choose operators $b_{\k \mu s}$ and $b_{\k \mu s}^\dag$ that correspond to the Green's function diagonal in band ($\mu$) and pseudospin ($s$) basis,
\beq
 G_{\mu s}(\k,\ii\omega_n) = 1 / \left( \ii\omega_n - \eps_{\k \mu s} \right),
 \label{eq:G0SO}
\eeq
where $\eps_{\k \mu s}$ is the energy calculated as the eigenvalue of matrix $\hat{H}_{0\lambda}$~(\ref{eq:H0SOmat}).

Transformation from the orbital-spin basis to the band-pseudospin basis is done via the matrix elements $\varphi^{\mu s}_{\k l \sigma}$,
\beq
 \ket{\sigma l \k} = \sum\limits_{\mu,s} \varphi^{\mu s}_{\k l \sigma} \ket{s \mu \k}.
 \label{eq:varphiSO}
\eeq
The relation between $d$ and $b$ operators is the following:
\beq
    d_{\k l \sigma} = \sum\limits_{\mu,s} \varphi^{\mu s}_{\k l \sigma} b_{\k \mu s}.
\eeq

\subsection{Spin-orbit coupling in a ten-orbital model}
\label{subsec:socmodel10orb}

There are several subtle points regarding the correct treatment of the SO coupling effects in FeBS. One of them is that in the pseudospin basis electron spin states become mixed and thus spin-singlet and spin-triplet representations can mix as well. Such a mixed state involving singlet $A_{1g}$ and triplet $A_{2g}$ pairing components was discussed in Ref.~\cite{Cvetkovic2013}; application of these ideas to an enhancement of the $A_{1g}$ pairing channel in KFe$_2$As$_2$ was studied in Ref.~\cite{Vafek2017}.

Another subtle point involves real crystallographic unit cell inherent to FeBS. It corresponds to 2-Fe BZ that can be obtained from 1-Fe BZ by the folding procedure. Later involves rotation of the BZ by $\pi/2$~\cite{HirschfeldKorshunov2011} in 1111 systems. Since electron pockets at $(\pi,0)$ and $(0,\pi)$ are elongated, there are four points where they overlap after the folding. Degeneracy at points of overlap can be lifted by the SO interaction. Thus the topology of the Fermi surface changes and that can results in the dominance of the so-called bonding-antibonding $s_\pm$ state~\cite{HirschfeldKorshunov2011,Mazin2011,Khodas2012}. While actual calculations within the DFT-based model put $d$-wave state to be the leading instability, the bonding-antibonding $s_\pm$ state becomes a runner-up~\cite{Kreisel2013}. It is hard to catch such effects in the five-orbital model without additional manipulations since the folding in it is essentially ignored being an external procedure applied after the ground state calculation. Therefore to demonstrate the topological changes of the Fermi surface, we have to consider a ten-orbital model composed of five $d$-orbitals from each iron in the unit cell.

\begin{figure}
 \centering
 \includegraphics[width=0.9\textwidth]{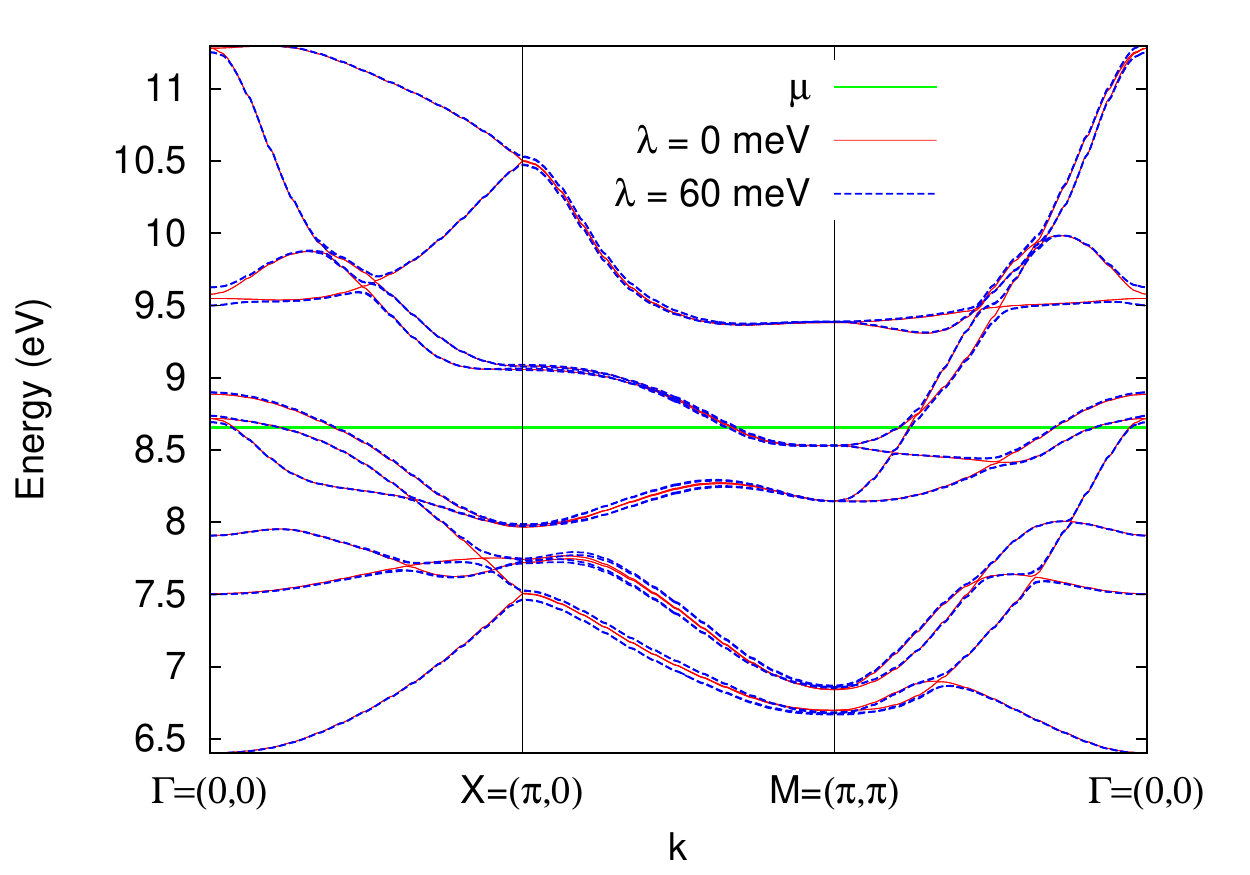}
 \caption{Comparison of energy dispersions for zero (red solid curves) and finite (blue dashed curves) SO coupling constant $\lambda$. Note the splitting of bands at high-symmetry points in the latter case.}
 \label{fig:10orbEk}
\end{figure}
\begin{figure}
 \centering
 \includegraphics[width=0.9\textwidth]{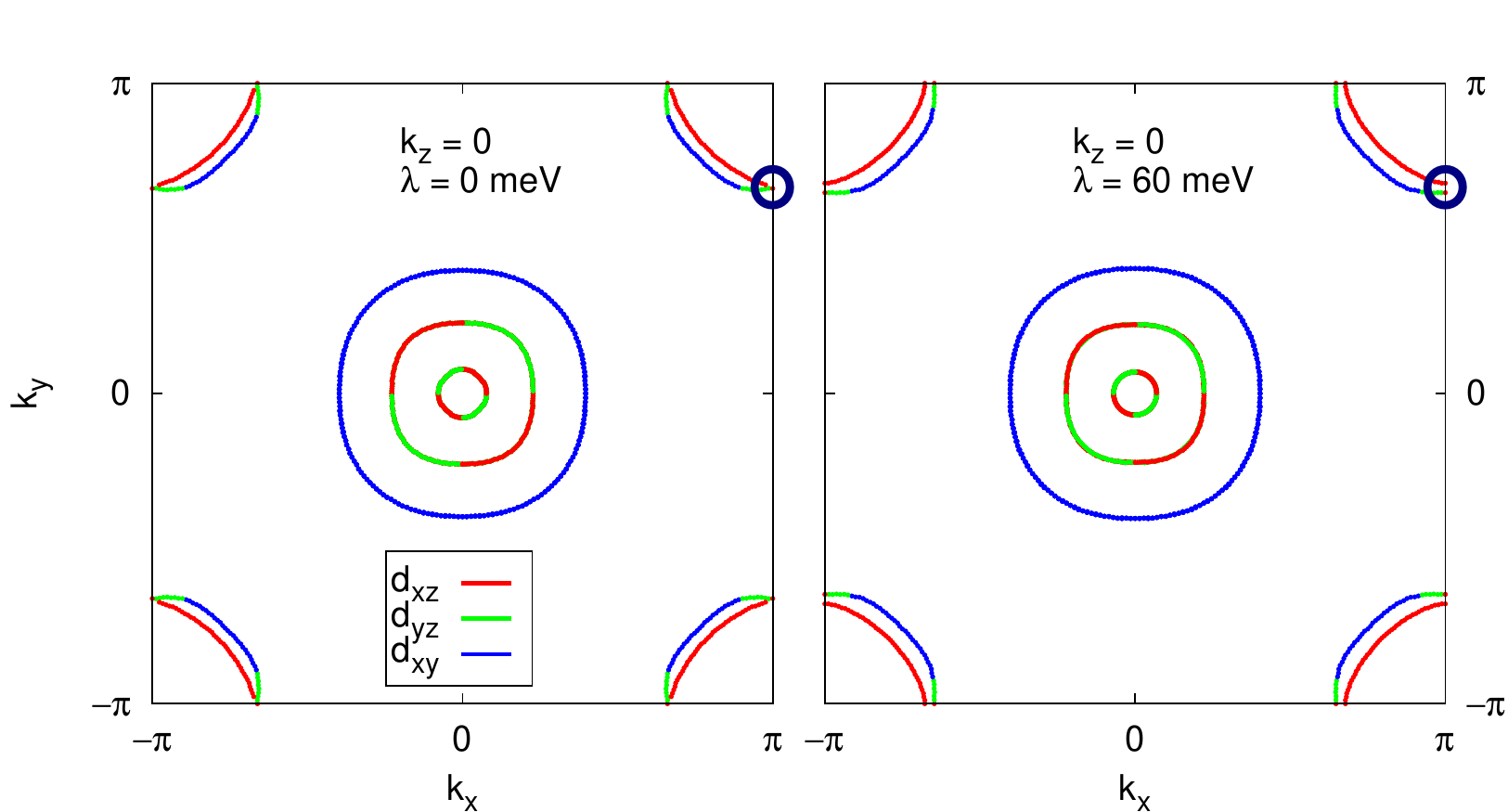}
 \caption{Fermi surface cuts at $k_z=0$ for zero (left) and finite (right) SO coupling constant $\lambda$. Different colors mark the dominant orbital contribution to the Fermi surface sheets. The main difference is the lifting of the degeneracy at electron pockets at the edges of the Brillouin zone for $\lambda > 0$. One such point is marked by a circle.}
 \label{fig:10orbFS}
\end{figure}

For the calculations below I used the LiFeAs model~\cite{Wang2013} provided by A.~Kreisel. The generalization of the SO matrix for the ten-orbital case is trivial; all one has to do is to duplicate matrix~(\ref{eq:H0SOmat}) for the second iron in the unit cell. In Figs.~\ref{fig:10orbEk} and~\ref{fig:10orbFS}, I compare band structures and Fermi surfaces with $\lambda = 0$ and with finite SO coupling constant chosen to be 60meV. The main difference in band structures is the splitting of bands due to the SO interaction. It is especially noticeable near the $\Gamma$ point where bands forming hole Fermi surface sheets become split. The other important contribution of finite $\lambda$ is the topological change of electron Fermi surface sheets seen in the right panel of Fig.~\ref{fig:10orbFS}. In particular, four crossing points of two `Fermi ellipses' at the Brillouin zone boundary are eliminated revealing a two concentric electron pockets instead. This opens up a set of alternative pairing scenarios including $s$-wave, $d$-wave, and $d + \ii s$-wave states, discussed in details in Ref.~\cite{Khodas2012}. Whether any of these states are going to be realized depends on many details of a system and only calculations in each particular case could give us the answer.

\section{Dynamical spin susceptibility}
\label{sec:susceptibility}

Cornerstone of the itinerant approach is the dynamical spin susceptibility. For the noninteracting system the `bare' susceptibility can be calculated straightforwardly though in the multiorbital case the procedure is a little more complicated and computationally more intensive. As for the interacting system, for the on-site Hubbard-like interaction of the from~(\ref{eq:Hint}), the most common approach is to use the random phase approximation (RPA). Diagrammatically, it is the exact summation of the infinite series of `bubbles' and `ladders'. While in most cases there are no formal grounds for keeping only such kind of diagrams (no small parameter), practically RPA appears quite useful. For example, results of the parquet renormalization group (pRG) approach~\cite{Xing2017,Classen2017} generally agrees with the RPA conclusions obtained previously for FeBS~\cite{HirschfeldKorshunov2011}. Keeping this in mind, I stick with the RPA in the following discussion.

\subsection{`Bare' susceptibility}

In the one-band noninteracting system the `bare' susceptibility is de-facto given by the simple electron-hole bubble,
\beq
 \chi_0(\q,\ii\omega_n) = \sum_\p \frac{f(\eps_{\p+\q}) - f(\eps_{\p})}{\ii\omega_n - \eps_{\p+\q} + \eps_{\p}},
\label{eq:chi0}
\eeq
where $f(\eps_{\p})$ is the Fermi distribution function for the electron dispersion $\eps_{\p}$ and $\omega_n$ is the Matsubara frequency.

In the multiorbital case, the dynamical spin susceptibility is a tensor with respect to the orbital indices $l$, $l'$, $m$, and $m'$,
\begin{equation} \label{eq.chi_def}
 \chi^{ll',mm'}_{ss'}(\q,\ii\Omega) = \int\limits_0^\beta d\tau e^{\ii\Omega\tau} \av{T_\tau S^{s}_{ll'}(\q,\tau) S^{s'}_{m'm}(-\q,0)}.
\end{equation}
Here, $\Omega$ is the Matsubara frequency and $\tau$ is the Matsubara time, $\beta=1/T$ is the inverse temperature, $S^{s}_{ll'}(\q,\tau)$ is the $s$'th component of the spin operator vector,
\begin{equation} \label{eq.S_def}
 \vec{S}_{ll'}(\q,\tau) = \frac{1}{2} \sum_{\p,\gamma,\delta} d^\dag_{\p l \gamma}(\tau) \vec{\hat\sigma}_{\gamma\delta} d_{\p+\q l' \delta}(\tau),
\end{equation}
where $\vec{\hat\sigma}$ is a vector composed of Pauli matrices $\hat\sigma$, $\gamma$ and $\delta$ are spin indices.

Substitution of Eq.~(\ref{eq.S_def}) into~(\ref{eq.chi_def}) results in the following ensemble average:
\beq \label{eq.chi_av}
 \sum_{\p,\p',\gamma,\delta,\gamma',\delta'} \av{T_\tau d^\dag_{\p l \gamma}(\tau) d_{\p+\q l' \delta}(\tau) d^\dag_{\p' m' \gamma'}(0) d_{\p'-\q m \delta'}(0)} \hat\sigma^{s}_{\gamma\delta} \hat\sigma^{s'}_{\gamma'\delta'}.
\eeq
%


Average here and in~(\ref{eq.chi_def}) is taken over the grand canonical ensemble with the full Hamiltonian $H$ from Eq.~(\ref{eq:H}). To obtain a zero's order approximation, we replace it with the ensemble averaging with the noninteracting Hamiltonian $H_0$ and then decouple it via Wick's theorem,
%
\begin{eqnarray}\label{eq.Wick}
 & \av{T_\tau d^\dag_{\p l \gamma}(\tau) d_{\p+\q l' \delta}(\tau) d^\dag_{\p' m' \gamma'}(0) d_{\p'-\q m \delta'}(0)} & \nn\\
 & = -\av{T_\tau d_{\p'-\q m \delta'}(0) d^\dag_{\p l \gamma}(\tau)} \av{T_\tau d_{\p+\q l' \delta}(\tau) d^\dag_{\p' m' \gamma'}(0)} & \nn\\
 & -\av{T_\tau d^\dag_{\p l \gamma}(\tau) d^\dag_{\p' m' \gamma'}(0)} \av{T_\tau d_{\p+\q l' \delta}(\tau) d_{\p'-\q m \delta'}(0)} & \nn\\
 & = -\delta_{\p',\p+\q} G_{m l \delta'\gamma}(\p,0-\tau) G_{l' m' \delta\gamma'}(\p+\q,\tau-0) & \nn\\
 & -\delta_{\p',-\p} F^\dag_{l m' \gamma\gamma'}(\p,\tau-0) F_{l' m \delta\delta'}(\p+\q,\tau-0),&
\end{eqnarray}
where we have introduced spin-resolved normal and anomalous (Gor'kov) Green's functions,
\bea \label{eq:G}
 G_{m l \sigma\sigma'}(\k,\tau) = - \av{T_\tau d_{\k m \sigma}(\tau) d^\dag_{\k l \sigma'}(0)}, \\
 F^\dag_{m l \sigma\sigma'}(\k,\tau) = \av{T_\tau d^\dag_{\k m \sigma}(\tau) d^\dag_{-\k l \sigma'}(0)}, \\
 F_{m l \sigma\sigma'}(\k,\tau) = \av{T_\tau d_{\k m \sigma}(\tau) d_{-\k l \sigma'}(0)}.
 \label{eq:F}
\eea
From now on, for simplicity by $\av{...}$ we assume the ensemble average with the noninteracting Hamiltonian.

Defining a Fourier transform for a Green's function $O$ as $O_{m l \sigma}(\k,\tau) = T \sum\limits_{\omega_n} O_{m l \sigma}(\k,\ii\omega_n) e^{-\ii\omega_n\tau}$ and applying it to the combination of~(\ref{eq.chi_av}) and~(\ref{eq.Wick}), we find
\bea \label{eq.chi1}
 \chi^{ll',mm'}_{0ss'}(\q,\ii\Omega) 
 &=& -\frac{T}{4} \sum_{\omega_n,\p,\gamma,\delta,\gamma',\delta'} \left[ G_{m l \delta'\gamma}(\p,\ii\omega_n) G_{l' m' \delta\gamma'}(\p+\q,\ii\Omega+\ii\omega_n) \right.\nn\\
 &&\left. +F^\dag_{l m' \gamma\gamma'}(\p,\ii\omega_n) F_{l' m \delta\delta'}(\p+\q,\ii\Omega-\ii\omega_n) \right] \hat\sigma^{s}_{\gamma\delta} \hat\sigma^{s'}_{\gamma'\delta'} .
\eea
The last term can be rewritten as $F^\dag(\p,-\ii\omega_n) F(\p+\q,\ii\Omega+\ii\omega_n)$ or even as $F^\dag(\p,\ii\omega_n) F(\p+\q,\ii\Omega+\ii\omega_n)$ only if there is a symmetry $F^\dag(\p,-\ii\omega_n)=F^\dag(\p,\ii\omega_n)$.

From the explicit form of Pauli matrices~(\ref{eq:sigmaM}), we immediately derive the following expressions for the combinations of two $\sigma$-matrices:
\bea
 &&\hat\sigma^x_{\gamma\delta} \hat\sigma^x_{\gamma'\delta'} =
                \left\{
                 \begin{split}
                   &1, \gamma=-\delta \;\mathrm{and}\; \gamma'=-\delta' \\
                   &0, \mathrm{otherwise}
                 \end{split}
                \right., 
 \hat\sigma^y_{\gamma\delta} \hat\sigma^y_{\gamma'\delta'} =
                \left\{
                 \begin{split}
                   &1, \gamma=-\delta=-\gamma'=\delta' \\
                   &-1, \gamma=-\delta=\gamma'=-\delta' \\
                   &0, \mathrm{otherwise}
                 \end{split}
                \right., \nn\\
 &&\hat\sigma^z_{\gamma\delta} \hat\sigma^z_{\gamma'\delta'} =
                \left\{
                 \begin{split}
                   &1, \gamma=\delta=\gamma'=\delta' \\
                   &-1, \gamma=\delta=-\gamma'=-\delta' \\
                   &0, \mathrm{otherwise}
                 \end{split}
                \right.,
 \hat\sigma^+_{\gamma\delta} \hat\sigma^-_{\gamma'\delta'} =
                \left\{
                 \begin{split}
                   &4, \gamma=-\delta=-\gamma'=\delta' \\
                   &0, \mathrm{otherwise}
                 \end{split}
                \right.. 
\eea

Then the transverse ($+-$) component of Eq.~(\ref{eq.chi1}) becomes~\cite{Korshunov2014eng}
\begin{eqnarray}\label{eq.chipm}
 \chi^{ll',mm'}_{0+-}(\q,\ii\Omega) = -T \sum_{\omega_n,\p} &&\left[ G_{m l \su\su}(\p,\ii\omega_n) G_{l' m' \sd\sd}(\p+\q,\ii\Omega+\ii\omega_n) \right.\nn\\
 &&\left. +F^\dag_{l m' \su\sd}(\p,\ii\omega_n) F_{l' m \sd\su}(\p+\q,\ii\Omega-\ii\omega_n) \right].
\end{eqnarray}

Here are expressions for $xx$ and $yy$ components,
\begin{eqnarray}\label{eq.chixx}
 \chi^{ll',mm'}_{0xx}(\q,\ii\Omega) = -\frac{T}{4} \sum_{\omega_n,\p,\sigma} &&\left[ G_{m l \sigma\sigma}(\p,\ii\omega_n) G_{l' m' \bar\sigma\bar\sigma}(\p+\q,\ii\Omega+\ii\omega_n)
 \right.\nn\\
 &&\left. +F^\dag_{l m' \sigma\bar\sigma}(\p,\ii\omega_n) F_{l' m \bar\sigma\sigma}(\p+\q,\ii\Omega-\ii\omega_n)
 \right.\nn\\
 &&\left. +G_{m l \bar\sigma\sigma}(\p,\ii\omega_n) G_{l' m' \bar\sigma\sigma}(\p+\q,\ii\Omega+\ii\omega_n)
 \right.\nn\\
 &&\left. +F^\dag_{l m' \sigma\sigma}(\p,\ii\omega_n) F_{l' m \bar\sigma\bar\sigma}(\p+\q,\ii\Omega-\ii\omega_n)
 \right],
\end{eqnarray}
\begin{eqnarray}\label{eq.chiyy}
 \chi^{ll',mm'}_{0yy}(\q,\ii\Omega) = -\frac{T}{4} \sum_{\omega_n,\p,\sigma} &&\left[ G_{m l \sigma\sigma}(\p,\ii\omega_n) G_{l' m' \bar\sigma\bar\sigma}(\p+\q,\ii\Omega+\ii\omega_n)
 \right.\nn\\
 &&\left. +F^\dag_{l m' \sigma\bar\sigma}(\p,\ii\omega_n) F_{l' m \bar\sigma\sigma}(\p+\q,\ii\Omega-\ii\omega_n)
 \right.\nn\\
 &&\left. -G_{m l \bar\sigma\sigma}(\p,\ii\omega_n) G_{l' m' \bar\sigma\sigma}(\p+\q,\ii\Omega+\ii\omega_n)
 \right.\nn\\
 &&\left. -F^\dag_{l m' \sigma\sigma}(\p,\ii\omega_n) F_{l' m \bar\sigma\bar\sigma}(\p+\q,\ii\Omega-\ii\omega_n)
 \right].
\end{eqnarray}
Here $\bar\sigma \equiv -\sigma$.
Obviously, since $(\hat{\sigma}^+ \hat{\sigma}^- + \hat{\sigma}^- \hat{\sigma}^+)/2 = \hat{\sigma}^x \hat{\sigma}^x + \hat{\sigma}^y \hat{\sigma}^y$, we have $\chi^{ll',mm'}_{0xx}(\q,\ii\Omega) + \chi^{ll',mm'}_{0yy}(\q,\ii\Omega) = \chi^{ll',mm'}_{0+-}(\q,\ii\Omega)$ if we consider that $G_{\sd\sd} = G_{\su\su}$ and $F_{\sd\su} = F_{\su\sd}$.

Longitudinal ($zz$) component are
\begin{eqnarray}\label{eq.chizz}
 \chi^{ll',mm'}_{0zz}(\q,\ii\Omega) = -\frac{T}{4} \sum_{\omega_n,\p,\sigma} &&\left[ G_{m l \sigma\sigma}(\p,\ii\omega_n) G_{l' m' \sigma\sigma}(\p+\q,\ii\Omega+\ii\omega_n) \right.\nn\\
 &&\left. -F^\dag_{l m' \sigma\bar\sigma}(\p,\ii\omega_n) F_{l' m \sigma\bar\sigma}(\p+\q,\ii\Omega-\ii\omega_n)  \right.\nn\\
 &&\left. -G_{m l \bar\sigma\sigma}(\p,\ii\omega_n) G_{l' m' \sigma\bar\sigma}(\p+\q,\ii\Omega+\ii\omega_n) \right.\nn\\
 &&\left. +F^\dag_{l m' \sigma\sigma}(\p,\ii\omega_n) F_{l' m \sigma\sigma}(\p+\q,\ii\Omega-\ii\omega_n)
 \right].
\end{eqnarray}

Note that the last two terms in $xx$, $yy$, and $zz$ components are non-zero only when spin non-diagonal Green's functions, $G_{\su\sd}$ and $F_{\su\su}$, are present. In the paramagnetic state with the spin-singlet gap symmetry considered here this can be true
only for non-zero spin-orbit coupling.

Physical (observable) susceptibility corresponds to the case of coinciding orbital indices of two Green's functions entering the vertex, i.e., $l'=l$ and $m'=m$. Thus,
\beq \label{eq.chiphys}
 \chi_{ss'}(\q,\ii\Omega) = \half \sum_{l,m} \chi^{ll,mm}_{ss'}(\q,\ii\Omega).
\eeq

\subsection{From orbitals and spins to bands and pseudospins}

Now we introduce Green's functions that are diagonal in the band-pseudospin basis and consider the pseudospin-singlet pairing only,
\bea \label{eq:Gmu}
 G_{\mu s}(\k,\ii\Omega) = - \int\limits_0^\beta d\tau e^{\ii\Omega\tau} \av{T_\tau b_{\k \mu s}(\tau) b^\dag_{\k \mu s}(0)}, \\
 F^\dag_{\mu s}(\k,\ii\Omega) = \int\limits_0^\beta d\tau e^{\ii\Omega\tau} \av{T_\tau b^\dag_{\k \mu s}(\tau) b^\dag_{-\k \mu \bar{s}}(0)}, \\
 F_{\mu s}(\k,\ii\Omega) = \int\limits_0^\beta d\tau e^{\ii\Omega\tau} \av{T_\tau b_{\k \mu s}(\tau) b_{-\k \mu \bar{s}}(0)}.
 \label{eq:Fmu}
\eea

The transformation between band-pseudospin Green's functions~(\ref{eq:Gmu})-(\ref{eq:Fmu}) and orbital-spin Green's functions~(\ref{eq:G})-(\ref{eq:F}) is done via the matrix elements~(\ref{eq:varphiSO}),
\bea
 G_{m l \sigma\sigma'}(\k,\ii\Omega) = \sum\limits_{\mu,s} \varphi^{\mu s}_{\k m \sigma} {\varphi^*}^{\mu s}_{\k l \sigma'} G_{\mu s}(\k,\ii\Omega), \\
 F^\dag_{m l \sigma\sigma'}(\k,\ii\Omega) = \sum\limits_{\mu,s} {\varphi^*}^{\mu s}_{\k m \sigma} {\varphi^*}^{\mu \bar{s}}_{-\k l \sigma'} F^\dag_{\mu s}(\k,\ii\Omega), \\
 F_{m l \sigma\sigma'}(\k,\ii\Omega) = \sum\limits_{\mu,s} \varphi^{\mu s}_{\k m \sigma} \varphi^{\mu \bar{s}}_{-\k l \sigma'} F_{\mu s}(\k,\ii\Omega).
\eea

Now we can express the spin susceptibility through the band-pseudospin Green's functions. To shorten the notations (also helpful in the actual calculations), it is useful to introduce the following Matsubara sums of the Green's functions product,
\bea \label{eq:GG}
 \GG_{\mu s}^{\nu s'}(\p,\q,\Omega) = -T \sum_{\omega_n} G_{\mu s}(\p,\ii\omega_n) G_{\nu s'}(\p+\q,\ii\Omega+\ii\omega_n), \\
 \FF_{\mu s}^{\nu s'}(\p,\q,\Omega) = -T \sum_{\omega_n} F^\dag_{\mu s}(\p,\ii\omega_n) F_{\nu s'}(\p+\q,\ii\Omega-\ii\omega_n).
 \label{eq:FF}
\eea
To emphasize that some of the components are finite only when spin-orbit coupling are present, we also introduce the factor $(1-\delta_{\lambda,0})$, where $\lambda$ is the SO coupling strength~(\ref{eq:HSO}). Using these notations, from the expression~(\ref{eq.chipm}) we have~\cite{Korshunov2014eng,KorshunovPRB2016}
\bea \label{eq.chipmmu}
 \chi^{ll',mm'}_{0+-}(\q,\ii\Omega) &=& \sum_{\p, \mu,\nu,s,s'} \left[ \varphi^{\mu s}_{\p m \su} {\varphi^*}^{\mu s}_{\p l \su}
 \GG_{\mu s}^{\nu s}(\p,\q,\Omega)
 \varphi^{\nu s'}_{\p+\q l' \sd} {\varphi^*}^{\nu s'}_{\p+\q m' \sd}
 \right.\nn\\
 &+&\left. {\varphi^*}^{\mu s}_{\p l \su} {\varphi^*}^{\mu \bar{s}}_{-\p m' \sd}
 \FF_{\mu s}^{\nu s'}(\p,\q,\Omega)
 \varphi^{\nu s'}_{\p+\q l' \sd} \varphi^{\nu \bar{s}'}_{-\p-\q m \su}
 \right].
\eea
From~(\ref{eq.chixx}),~(\ref{eq.chiyy}), and~(\ref{eq.chizz}) we find
\begin{eqnarray}\label{eq.chixxmu}
 \chi^{ll',mm'}_{0xx}(\q,\ii\Omega) &=& \frac{1}{4} \sum_{\p,\sigma, \mu,\nu,s,s'} \left[ \varphi^{\mu s}_{\p m \sigma} {\varphi^*}^{\mu s}_{\p l \sigma}
 \GG_{\mu s}^{\nu s'}(\p,\q,\Omega)
 \varphi^{\nu s'}_{\p+\q l' \bar\sigma} {\varphi^*}^{\nu s'}_{\p+\q m' \bar\sigma}
 \right.\nn\\
 &+&\left. {\varphi^*}^{\mu s}_{\p l \sigma} {\varphi^*}^{\mu \bar{s}}_{-\p m' \bar\sigma}
 \FF_{\mu s}^{\nu s'}(\p,\q,\Omega)
 \varphi^{\nu s'}_{\p+\q l' \bar\sigma} \varphi^{\nu \bar{s}'}_{-\p-\q m \sigma}
 \right.\nn\\
 &+&\left. (1-\delta_{\lambda,0}) \varphi^{\mu s}_{\p m \bar\sigma} {\varphi^*}^{\mu s}_{\p l \sigma}
 \GG_{\mu s}^{\nu s'}(\p,\q,\Omega)
 \varphi^{\nu s'}_{\p+\q l' \bar\sigma} {\varphi^*}^{\nu s'}_{\p+\q m' \sigma}
 \right.\nn\\
 &+&\left. (1-\delta_{\lambda,0}) {\varphi^*}^{\mu s}_{\p l \sigma} {\varphi^*}^{\mu \bar{s}}_{-\p m' \sigma}
 \FF_{\mu s}^{\nu s'}(\p,\q,\Omega)
 \varphi^{\nu s'}_{\p+\q l' \bar\sigma} \varphi^{\nu \bar{s}'}_{-\p-\q m \bar\sigma}
 \right],
\end{eqnarray}
\begin{eqnarray}\label{eq.chiyymu}
 \chi^{ll',mm'}_{0yy}(\q,\ii\Omega) &=& \frac{1}{4} \sum_{\p,\sigma, \mu,\nu,s,s'} \left[ \varphi^{\mu s}_{\p m \sigma} {\varphi^*}^{\mu s}_{\p l \sigma}
 \GG_{\mu s}^{\nu s'}(\p,\q,\Omega)
 \varphi^{\nu s'}_{\p+\q l' \bar\sigma} {\varphi^*}^{\nu s'}_{\p+\q m' \bar\sigma}
 \right.\nn\\
 &+&\left. {\varphi^*}^{\mu s}_{\p l \sigma} {\varphi^*}^{\mu \bar{s}}_{-\p m' \bar\sigma}
 \FF_{\mu s}^{\nu s'}(\p,\q,\Omega)
 \varphi^{\nu s'}_{\p+\q l' \bar\sigma} \varphi^{\nu \bar{s}'}_{-\p-\q m \sigma}
 \right.\nn\\
 &-&\left. (1-\delta_{\lambda,0}) \varphi^{\mu s}_{\p m \bar\sigma} {\varphi^*}^{\mu s}_{\p l \sigma}
 \GG_{\mu s}^{\nu s'}(\p,\q,\Omega)
 \varphi^{\nu s'}_{\p+\q l' \bar\sigma} {\varphi^*}^{\nu s'}_{\p+\q m' \sigma}
 \right.\nn\\
 &-&\left. (1-\delta_{\lambda,0}) {\varphi^*}^{\mu s}_{\p l \sigma} {\varphi^*}^{\mu \bar{s}}_{-\p m' \sigma}
 \FF_{\mu s}^{\nu s'}(\p,\q,\Omega)
 \varphi^{\nu s'}_{\p+\q l' \bar\sigma} \varphi^{\nu \bar{s}'}_{-\p-\q m \bar\sigma}
 \right],
\end{eqnarray}
\begin{eqnarray}\label{eq.chizzmu}
 \chi^{ll',mm'}_{0zz}(\q,\ii\Omega) &=& \frac{1}{4} \sum_{\p,\sigma, \mu,\nu,s,s'} \left[ \varphi^{\mu s}_{\p m \sigma} {\varphi^*}^{\mu s}_{\p l \sigma}
 \GG_{\mu s}^{\nu s'}(\p,\q,\Omega)
 \varphi^{\nu s'}_{\p+\q l' \sigma} {\varphi^*}^{\nu s'}_{\p+\q m' \sigma}
 \right.\nn\\
 &-&\left. {\varphi^*}^{\mu s}_{\p l \sigma} {\varphi^*}^{\mu \bar{s}}_{-\p m' \bar\sigma}
 \FF_{\mu s}^{\nu s'}(\p,\q,\Omega)
 \varphi^{\nu s'}_{\p+\q l' \sigma} \varphi^{\nu \bar{s}'}_{-\p-\q m \bar\sigma} \right.\nn\\
 &-&\left. (1-\delta_{\lambda,0}) \varphi^{\mu s}_{\p m \bar\sigma} {\varphi^*}^{\mu s}_{\p l \sigma}
 \GG_{\mu s}^{\nu s'}(\p,\q,\Omega)
 \varphi^{\nu s'}_{\p+\q l' \sigma} {\varphi^*}^{\nu s'}_{\p+\q m' \bar\sigma}
 \right.\nn\\
 &+&\left. (1-\delta_{\lambda,0}) {\varphi^*}^{\mu s}_{\p l \sigma} {\varphi^*}^{\mu \bar{s}}_{-\p m' \sigma}
 \FF_{\mu s}^{\nu s'}(\p,\q,\Omega)
 \varphi^{\nu s'}_{\p+\q l' \sigma} \varphi^{\nu \bar{s}'}_{-\p-\q m \sigma}
 \right].
\end{eqnarray}
Again, the last two terms in $xx$, $yy$, and $zz$ components are non-zero only for finite SO coupling.

\subsection{Explicit form of Green's functions product}

For the noninteracting system's Green's functions $G_{\mu s}(\p,\ii\omega_n)$ and $F^\dag_{\mu s}(\p,\ii\omega_n)$, one can evaluate Matsubara sums in Eqs.~(\ref{eq:GG}) and~(\ref{eq:FF}) exactly. After solving Gor'kov equations in the pseudospin-singlet superconducting state, instead of~(\ref{eq:G0SO}) we have~\cite{allen,MineevSamokhin1998eng}
\bea
 G_{\mu s}(\p,\ii\omega_n) = \frac{\ii\omega_n + \eps_{\p\mu s}} {\brro{\ii\omega_n - E_{\p\mu s}} \brro{\ii\omega_n + E_{\p\mu s}}}, \\
 F_{\mu s}^\dag(\p,\ii\omega_n) = \frac{\Delta_{\p\mu s}^\dag} {\brro{\ii\omega_n - E_{\p\mu s}} \brro{\ii\omega_n + E_{\p\mu s}}},
\eea
where $E_{\p\mu s} = \sqrt{\eps_{\p\mu s}^2 + \abs{\Delta_{\p\mu s}}^2}$ is the energy spectrum, $\Delta_{\p\mu s}^\dag = \Delta_{\p\mu}^* (\ii\hat\sigma^y)_{\bar{s}s}$ and $\Delta_{\p\mu s} = \Delta_{\p\mu} (\ii\hat\sigma^y)_{s\bar{s}}$ are gap functions.


Then from Eqs.~(\ref{eq:GG}) and~(\ref{eq:FF}) we obtain
\bea
 \GG_{\mu s}^{\nu s'}(\p,\q,\Omega) &=& \frac{1}{4} \brsq{1 + \frac{\eps_{\p\mu s}\eps_{\p+\q \nu s'} + \eps_{\p\mu s}E_{\p+\q \nu s'} + E_{\p\mu s}\eps_{\p+\q \nu s'}}{E_{\p\mu s}E_{\p+\q \nu s'}} } f_1 \nn\\
 &-& \frac{1}{4} \brsq{1 + \frac{\eps_{\p\mu s}\eps_{\p+\q \nu s'} - \eps_{\p\mu s}E_{\p+\q \nu s'} - E_{\p\mu s}\eps_{\p+\q \nu s'}}{E_{\p\mu s}E_{\p+\q \nu s'}} } f_2 \nn\\
 &-& \frac{1}{4} \brsq{1 + \frac{-\eps_{\p\mu s}\eps_{\p+\q \nu s'} - \eps_{\p\mu s}E_{\p+\q \nu s'} + E_{\p\mu s}\eps_{\p+\q \nu s'}}{E_{\p\mu s}E_{\p+\q \nu s'}} } f_3 \nn\\
 &+& \frac{1}{4} \brsq{1 + \frac{-\eps_{\p\mu s}\eps_{\p+\q \nu s'} + \eps_{\p\mu s}E_{\p+\q \nu s'} - E_{\p\mu s}\eps_{\p+\q \nu s'}}{E_{\p\mu s}E_{\p+\q \nu s'}} } f_4, \\
 \FF_{\mu s}^{\nu s'}(\p,\q,\Omega) &=& \frac{1}{2} \frac{\Delta_{\p\mu s}^\dag \Delta_{\p+\q \nu s'}}{E_{\p\mu s}E_{\p+\q \nu s'}} \brsq{f_1 - f_2 + f_3 - f_4}.
\eea
where combination of Fermi functions and frequency $\Omega$ are included in functions $f_1 = \frac{f\brro{E_{\p+\q \nu s'}} - f\brro{E_{\p\mu s}}}{\ii\Omega - E_{\p+\q \nu s'} + E_{\p\mu s}}$,
$f_2 = \frac{f\brro{E_{\p+\q \nu s'}} - f\brro{E_{\p\mu s}}}{\ii\Omega + E_{\p+\q \nu s'} - E_{\p\mu s}}$,
$f_3 = \frac{1 - f\brro{E_{\p+\q \nu s'}} - f\brro{E_{\p\mu s}}}{\ii\Omega - E_{\p+\q \nu s'} - E_{\p\mu s}}$, and
$f_4 = \frac{1 - f\brro{E_{\p+\q \nu s'}} - f\brro{E_{\p\mu s}}}{\ii\Omega +- E_{\p+\q \nu s'} + E_{\p\mu s}}$.

\subsection{Random phase approximation}

\begin{figure}
 \centering
 \includegraphics[width=0.55\textwidth]{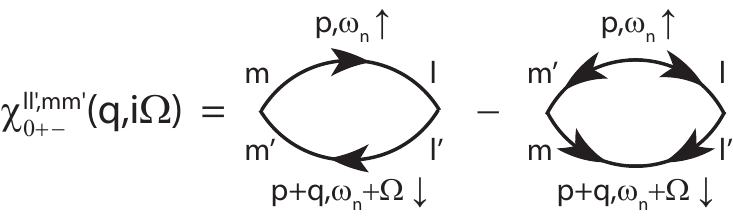}
 \includegraphics[width=0.55\textwidth]{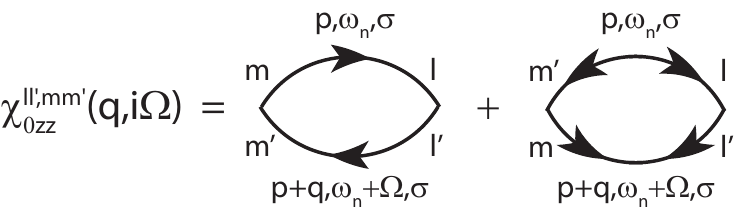}
 \caption{Diagrams for `bare' susceptibilities defined by Eqs.~(\ref{eq.chipm}) and~(\ref{eq.chizz}).}
 \label{fig:chibubble}
\end{figure}

\begin{figure}[t]
 \centering
 \includegraphics[width=0.8\textwidth]{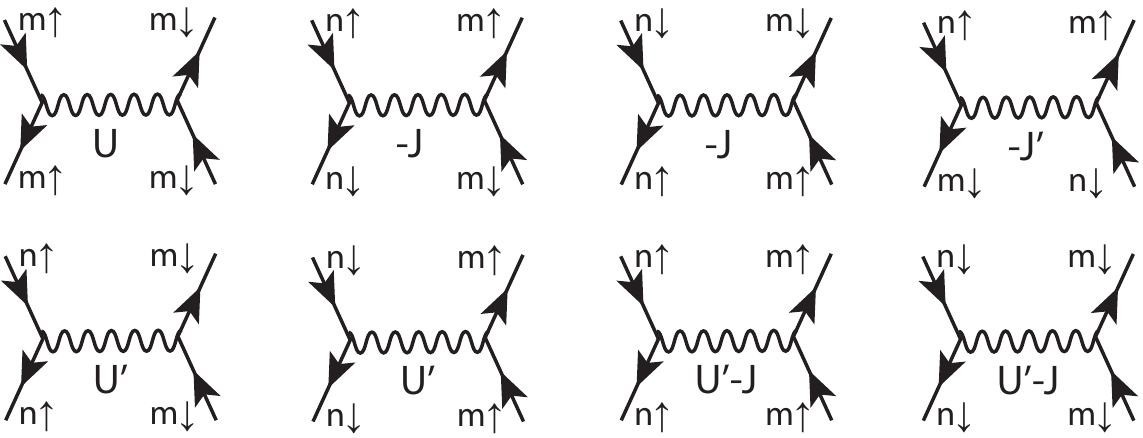}
 \caption{Interaction lines corresponding to the following expressions in Hamiltonian~(\ref{eq:Hintexpl}):
 $U d_{f m \su}^\dag d_{f m \su} d_{f m \sd}^\dag d_{f m \sd}$,
 $- J d_{f n \su}^\dag d_{f n \sd} d_{f m \sd}^\dag d_{f m \su}$,
 $- J d_{f n \sd}^\dag d_{f n \su} d_{f m \su}^\dag d_{f m \sd}$,
 $J' d_{f n \su}^\dag d_{f m \sd} d_{f n \sd}^\dag d_{f m \su}$,
 $U' d_{f n \su}^\dag d_{f n \su} d_{f m \sd}^\dag d_{f m \sd}$,
 $U' d_{f n \sd}^\dag d_{f n \sd} d_{f m \su}^\dag d_{f m \su}$,
 $(U' - J) d_{f n \su}^\dag d_{f n \su} d_{f m \su}^\dag d_{f m \su}$,
 and $(U' - J) d_{f n \sd}^\dag d_{f n \sd} d_{f m \sd}^\dag d_{f m \sd}$.
 Here $m$ and $n$ are orbital indices.
 }
 \label{fig:diagr_Hint}
\end{figure}

To describe spin response in normal and superconducting states of FeBS, we use RPA with local Coulomb interactions. Also, for convenience, we consider the system without spin-orbit coupling. In this case, diagrams for $\chi_\ppm$ and $\chi_\zz$ are shown in Fig.~\ref{fig:chibubble}. According to Hamiltonian~(\ref{eq:Hintexpl}), we define interaction lines as shown in Fig.~\ref{fig:diagr_Hint}.


It is convenient to utilize a matrix form of the RPA equations so later we denote matrices in the orbital space as $\hat\chi$, that is, $\hat\chi_\xx$, $\hat\chi_\yy$, $\hat\chi_\zz$, and $\hat\chi_\ppm$. To make matrices out of tensors with four indices, we introduce the following correspondence between matrix indices ($\imath$, $\jmath$) and orbital indices ($l$, $l'$, $m$, $m'$), $\imath = l + l' n_O$, $\jmath = m + m' n_O$, where $n_O$ is the number of orbitals.
In the following expressions for susceptibilities we omit momenta and frequency arguments since they are conserved in RPA for local Coulomb interactions.

As the next step, we introduce RPA series as vertex corrections and derive and solve equations for transverse and longitudinal components of spin susceptibility separately.
Since the right-hand side vertex of susceptibility `bubbles' composed of both normal Green's and Gor'kov functions are the same (see Fig.~\ref{fig:chibubble}), later we draw diagrams only for normal state `bubbles' assuming that the same conclusions apply for the full susceptibilities.

\subsubsection{RPA for $\hat\chi_{+-}$}

\begin{figure}
 \centering
 \includegraphics[width=0.99\textwidth]{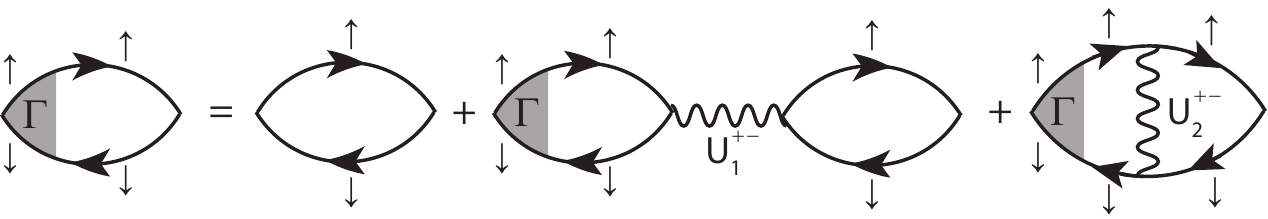}
 \caption{`Ladder' and `bubble' diagrams constituting RPA equations for the $\ppm$ susceptibility vertex $\Gammappm$~(\ref{eq.rpa.chippm}). Only spin indices are shown for clarity.}
 \label{fig:chipmRPA}
\end{figure}

RPA vertex $\Gammappm$ for the transverse spin susceptibility is introduced in the following way:
\begin{equation}\label{eq.rpa.chippm}
 \hat\chi_\ppm = \Gammappm \hat\chi_{0\ppm}.
\end{equation}
Graphically, the infinite RPA series can be represented as equation shown in Fig.~\ref{fig:chipmRPA}. Analytical expression for the vertex is
%
 $\Gammappm = \Gammappm_0 + \Gammappm \hat\chi_{0\ppm} \hat{U}_1^\ppm + \Gammappm \hat\chi_{0\ppm} \hat{U}_2^\ppm$.
%
One can simplify the expressions by combining two types of interaction lines,
\begin{equation}\label{eq.rpa.uppm}
  \hat{U}^\ppm = -\hat{U}_1^\ppm + \hat{U}_2^\ppm,
\end{equation}
where ``$-$'' sign is due to the additional bubble in the diagram with the horizontal interaction line. Since $\Gammappm_0 = \hat{1}$, we have
%
 $\Gammappm = \left[ \hat{1} - \hat\chi_{0\ppm} \hat{U}^\ppm \right]^{-1}$,
%
and, finally, the solution for the transverse RPA spin susceptibility is the following:
\begin{equation}\label{eq.rpa.chippmsol}
 \hat\chi_\ppm = \left[\hat{1} - \hat\chi_{0\ppm} \hat{U}^\ppm \right]^{-1} \hat\chi_{0\ppm}.
\end{equation}
Here $\hat{U}^\ppm$ is the interaction matrix in the $\ppm$ channel.

\subsubsection{RPA for $\hat\chi_{zz}$}

\begin{figure}[t]
 \centering
 \includegraphics[width=0.99\textwidth]{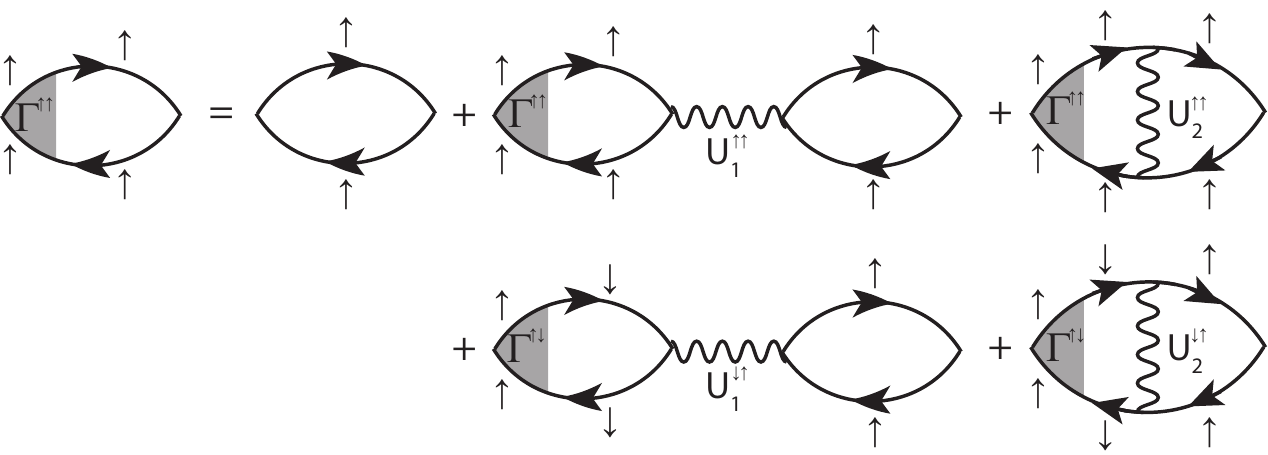}
 \includegraphics[width=0.85\textwidth]{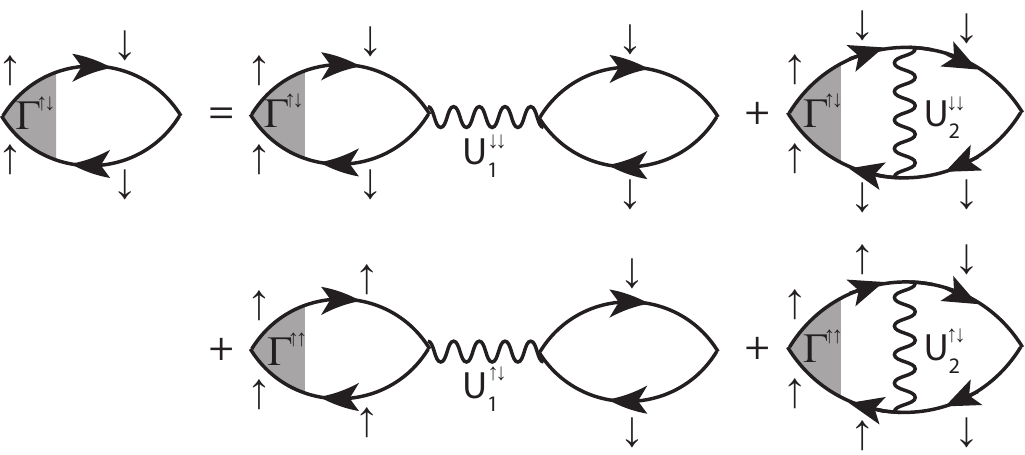}
 \caption{Diagrammatic representation of the coupled RPA equations for the $\zz$ susceptibility vertices $\hat\Gamma^\uu$ and $\hat\Gamma^\ud$~(\ref{eq.rpa.chizzgamma}). Only spin indices are shown for clarity.}
 \label{fig:chizzRPA}
\end{figure}

Equations for the longitudinal spin susceptibility is slightly more complicated than for the $\ppm$ component because former includes four terms having different spin structures,
\begin{equation}
 \hat\chi_\zz = \frac{1}{4} \left( \hat\chi_\uu + \hat\chi_\dd - \hat\chi_\ud - \hat\chi_\du \right).
 \label{eq.rpa.chizz}
\end{equation}
%
These terms can be expressed through vertices in the following way:
\beq
 \hat\chi_\uu = \hat\Gamma^\uu \hat\chi_{0\uu}, \,
 \hat\chi_\ud = \hat\Gamma^\ud \hat\chi_{0\dd}, \,
 \hat\chi_\dd = \hat\Gamma^\dd \hat\chi_{0\dd}, \,
 \hat\chi_\du = \hat\Gamma^\du \hat\chi_{0\uu}.
 \label{eq.rpa.chizzgamma}
\eeq
Equations for vertices follows from the RPA expansion shown in Fig.~\ref{fig:chizzRPA}. Analytical expressions for them are
\bea
 \hat\Gamma^\uu &=& \hat{1} + \hat\Gamma^\uu \hat\chi_{0\uu} \hat{U}^\uu + \hat\Gamma^\ud \hat\chi_{0\dd} \hat{U}^\du, \label{eq.rpa.gammazzuu} \\
 \hat\Gamma^\ud &=& \hat\Gamma^\ud \hat\chi_{0\dd} \hat{U}^\dd + \hat\Gamma^\uu \hat\chi_{0\uu} \hat{U}^\ud.
 \label{eq.rpa.gammazzud}
\eea
Similar to the case of the $\ppm$ susceptibility component, here we also combined interaction lines considering ``$-$'' sign due to the additional bubble in the diagram with the horizontal wavy line,
\beq
 \hat{U}^{\sigma\sigma'} = -\hat{U}_1^{\sigma\sigma'} + \hat{U}_2^{\sigma\sigma'}.
 \label{eq.rpa.uzz}
\eeq
%


Expressing $\hat\Gamma^\ud$ through $\hat\Gamma^\uu$ via Eq.~(\ref{eq.rpa.gammazzud}), $\hat\Gamma^\ud = \hat\Gamma^\uu \hat\chi_{0\uu} \hat{U}^\ud \left( \hat{1} - \hat\chi_{0\dd} \hat{U}^\dd \right)^{-1}$, and substituting it to Eq.~(\ref{eq.rpa.gammazzuu}), we find
\beq
 \hat\Gamma^\uu = \left[ \hat{1} - \hat\chi_{0\uu} \hat{U}^\uu - \hat\chi_{0\uu} \hat{U}^\ud \left( \hat{1} - \hat\chi_{0\dd} \hat{U}^\dd \right)^{-1} \hat\chi_{0\dd} U^\du \right]^{-1}.
 \label{eq.rpa.gammasol_uu}
\eeq
Using this vertex, we obtain the combination entering the susceptibility~(\ref{eq.rpa.chizz}),
\begin{equation}\label{eq.rpa.chisoluu}
 \hat\chi_\uu - \hat\chi_\ud = \hat\Gamma^\uu \hat\chi_{0\uu} - \hat\Gamma^\ud \hat\chi_{0\dd} = \hat\Gamma^\uu \hat\chi_{0\uu} \left[ \hat{1} - \hat{U}^\ud \left( \hat{1} - \hat\chi_{0\dd} \hat{U}^\dd \right)^{-1} \hat\chi_{0\dd} \right].
\end{equation}
For the opposite spin combination we similarly have
\begin{eqnarray}\label{eq.rpa.chisoldd}
 \hat\Gamma^\dd &=& \left[ \hat{1} - \hat\chi_{0\dd} \hat{U}^\dd - \hat\chi_{0\dd} \hat{U}^\du \left( \hat{1} - \hat\chi_{0\uu} \hat{U}^\uu \right)^{-1} \hat\chi_{0\uu} \hat{U}^\ud \right]^{-1}, \\
 \hat\chi_\dd - \hat\chi_\du &=& \hat\Gamma^\dd \hat\chi_{0\dd} \left[ \hat{1} - \hat{U}^\du \left( \hat{1} - \hat\chi_{0\uu} \hat{U}^\uu \right)^{-1} \hat\chi_{0\uu} \right].
\end{eqnarray}

In the simplified case of $\hat{U}^\uu = \hat{U}^\dd = 0$ we have $\hat\Gamma^{\s\s} = \left[ \hat{1} - \hat\chi_{0\s\s} \hat{U}^{\s\bar\s} \hat\chi_{0\bar\s\bar\s} \hat{U}^{\bar\s\s} \right]^{-1}$, $\hat\chi_{\s\s} - \hat\chi_{\s\bar\s} = \hat\Gamma^{\s\s} \hat\chi_{0\s\s} \left[ \hat{1} - \hat{U}^{\s\bar\s} \hat\chi_{0\bar\s\bar\s} \right]$, and $\hat\chi_{zz} = \frac{1}{4} \sum\limits_{\s} \left( \hat\chi_{\s\s} - \hat\chi_{\s\bar\s} \right)$.

To check the result in the single-band case we set $\hat{U}^{\s\bar\s} = -U$, $\hat{U}^{\s\s} = 0$, $\hat\chi_{0\s\s} = \chi_0$, and obtain
$\chi_{zz} = \frac{1}{2} \frac{\chi_0 \left( 1 + U \chi_0\right)}{1 - \left( U \chi_0 \right)^2} = \frac{1}{2} \frac{\chi_0}{1 - U \chi_0}$.
This is exactly the single-band RPA solution for $\chi_\zz$.

\subsubsection{Interaction lines}

To define interaction lines entering expressions for the susceptibility, we have to determine their orbital structure because spin structure is already defined via diagrams in Figs.~\ref{fig:chipmRPA} and~\ref{fig:chizzRPA}. Firstly, we introduce interaction matrix element $U^{l l'}_{n n'}$ as the factor in front of the combination of four operators, see Eq.~(\ref{eq:Ueff}).
Secondly, we fill this tensor with values defined in Fig.~\ref{fig:diagr_Hint} and match them with the interaction matrix elements entering spin susceptibilities thus establishing full spin and orbital structure of interaction $\hat{U}$.

\begin{figure}
 \centering
 \includegraphics[width=0.65\textwidth]{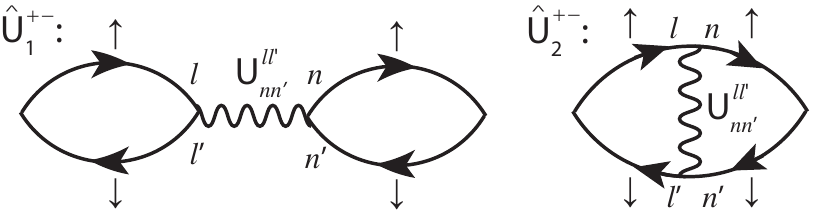}
 \includegraphics[width=0.65\textwidth]{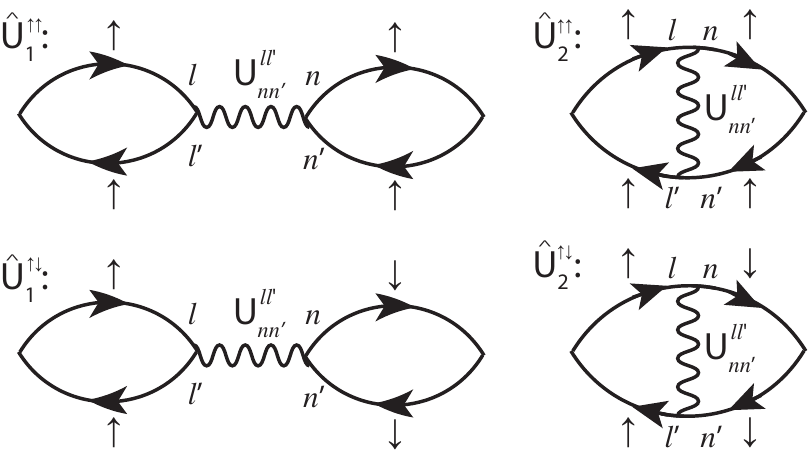}
 \caption{Combination of orbital and spin structure of interaction lines entering RPA equations for $\ppm$ and $\zz$ vertices.}
 \label{fig:RPAint}
\end{figure}

Fig.~\ref{fig:RPAint} represents a comparison between tensor $U^{l l'}_{n n'}$ and matrices $\hat{U}_{1,2}$. Thus we derive the following relations:
$\hat{U}_1^\ppm \to \left\{ U^{ll}_{nn}=-J, U^{ln}_{nl}=-J' \right\}$,
$\hat{U}_2^\ppm \to \left\{ U^{ll}_{ll}=U, U^{ln}_{ln}=U' \right\}$,
$\hat{U}_1^\uu \to \left\{ U^{ll}_{nn}=U'-J \right\}$,
$\hat{U}_2^\uu \to \left\{ U^{ln}_{ln}=U'-J \right\}$,
$\hat{U}_1^\ud \to \left\{ U^{ll}_{ll}=U, U^{ll}_{nn}=U' \right\}$,
$\hat{U}_2^\ud \to \left\{ U^{ln}_{ln}=-J, U^{ln}_{nl}=-J' \right\}$. Combining them into the matrices $\hat{U}$ defined by Eqs.~(\ref{eq.rpa.uppm}) and~(\ref{eq.rpa.uzz}), we finally obtain following nonvanishing matrix elements:
\bea
 \begin{array}{llll}
 (\hat{U}^\ppm)^{ll}_{ll}=U, & (\hat{U}^\ppm)^{ll}_{nn}=J, & (\hat{U}^\ppm)^{ln}_{nl}=J', & (\hat{U}^\ppm)^{ln}_{ln}=U', \\
 (\hat{U}^\ud)^{ll}_{ll}=-U, & (\hat{U}^\ud)^{ll}_{nn}=-U', & (\hat{U}^\ud)^{ln}_{nl}=-J', & (\hat{U}^\ud)^{ln}_{ln}=-J, \\
 (\hat{U}^\uu)^{ll}_{nn}=-(U'-J), & (\hat{U}^\uu)^{ln}_{ln}=U'-J. && \\
 \end{array}
 \label{eq.rpa.int}
\eea
Here $l \neq n$ and other matrix elements are zeros. These matrices are similar to the interaction matrix $U_s$ of Ref.~\cite{Graser2009} and spin-resolved interaction matrix elements of Ref.~\cite{KemperKorshunov2011}.

With Eq.~(\ref{eq.rpa.int}) we finish the definition of RPA for the multiorbital spin susceptibility.
%

\section{Self-energy effects and quasiparticle scattering}
\label{sec:selfenergy}

We discussed how electrons form spin excitations. Naturally, electrons then scatter by these excitations. Here we show the role of such inverse effect on the electron's self-energy and kinetic coefficients~\cite{KemperKorshunov2011}.

There are a simple reasoning and arguments from several group of experiments demonstrating importance of quasiparticle scattering in low-energy physics of FeBS. In particular, since the sizes of the hole and electron Fermi surface sheets are roughly identical in the undoped system, one might expect a vanishingly small Hall coefficient and a roughly electron-hole symmetric doping dependence. However, in the intensively studied 122 systems (Ba(Fe$_{1-x}$Co$_{x})_2$As$_2$, Ba(Fe$_{1-x}$Ni$_{x})_2$As$_2$) and 1111 systems (LaFeAsO$_{1-x}$F$_{x}$ and SmFeAsO$_{1-x}$F$_{x}$), Hall effect measurements find that transport is dominated by the electrons even for the parent compounds~\cite{f_rullier_albenque_09,l_fang_09,Kasahara_10,Liu_08,Riggs_09,Hess_09}. In the compensated case, this result can be explained only if the mobilities of holes and electrons are remarkably different that suggests an order of magnitude disparity in relaxation times, $\tau_e \gg \tau_h$~\cite{l_fang_09}. A similar large asymmetry of electronic and hole scattering rates has also been suggested in the analysis of the electronic Raman measurements that can selectively probe different parts of the Brillouin zone using various polarizations~\cite{b_muschler_09}. That is, the normal state Raman spectrum for the $A_{1g}$ polarization, which weights the hole pockets strongly, is markedly different from that of the $B_{2g}$ channel, which probes the electron Fermi surface, consistent with $\tau_e \gg \tau_h$~\cite{b_muschler_09}. Optical conductivity measured by THz spectrometry provides estimate $\tau_e \approx 4\tau_h$ \cite{Maksimov_10}. Theoretical analysis of the normal state resistivity $\rho$ in the two-band model for Ba$_{1-x}$K$_x$Fe$_2$As$_2$ shows that the experimental temperature dependence $\rho(T)$ can be reproduced only if one assumes order of magnitude larger scattering in the hole band~\cite{Golubov_10eng}. Finally, quantum oscillation experiments on P-doped systems indicate that the electron pockets have a longer mean free path~\cite{j_analytis_10,a_coldea_08,j_analytis_09a}.

There are two main sources for quasiparticle decay: i) electron-electron inelastic processes and ii) impurity scattering. We concentrate here on the first case and mention impurity scattering only briefly. Experimentally, the apparent disparity in mobilities for holes and electrons becomes smaller as one dopes away from the magnetically ordered
parent compounds~\cite{l_fang_09}. This suggests that the spin fluctuations, which also decrease upon doping, play an important role in the scattering rate asymmetry.



\subsection{Self-energy}

\begin{figure}
 \centering
 \includegraphics[width=0.99\columnwidth]{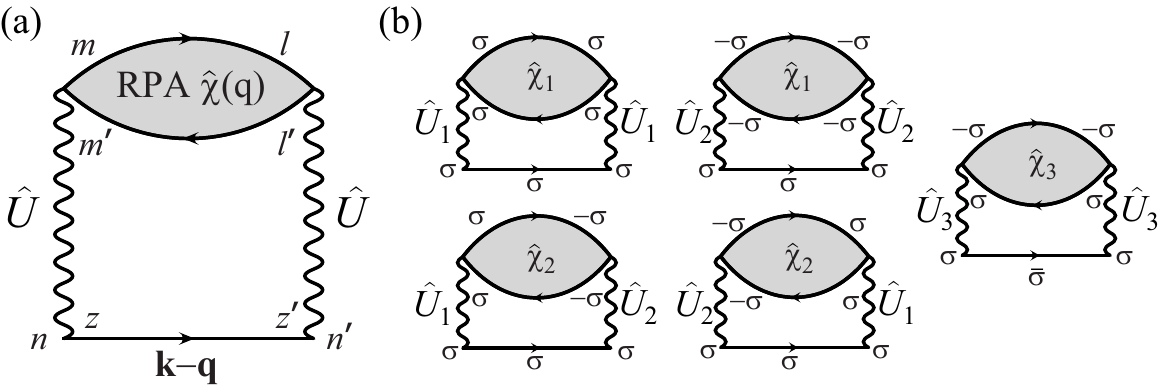}
 \caption{Orbital (a) and spin (b) structure of the second order diagram for the self-energy in the multiorbital system, $\Sigma_{\bar{n} \bar{n}'}(\k,\omega)$, where $\bar{n} = (n,\sigma_n)$. Interaction lines contain four orbital indices, $\hat{U}=U_{nz}^{mm'}$. Shaded bubble denote the RPA susceptibility, $\hat{\chi}(q)=\chi^{ll',mm'}(\q,\omega_q)$. Incoming and outgoing indices $\bar{n}$ and $\bar{n}'$ carry the same spin $\sigma$. $\hat{\chi}_1$, $\hat{\chi}_2$, and $\hat{\chi}_3$ are the different susceptibility channels, see Eq.~(\ref{eq:chi}).}
 \label{fig:bubble_diagram}
\end{figure}

The leading non-vanishing contribution to the quasiparticle scattering rate $1/\tau$ comes from the imaginary part of the second-order self-energy diagram ($\mathrm{Im}\Sigma$) with the polarization bubble (see Fig.~\ref{fig:bubble_diagram}). To take scattering from spin fluctuations into account we renormalize the bubble within the RPA. Note that second order diagrams with crossing interaction lines are not included in Fig.~\ref{fig:bubble_diagram}. This is done to preserve consistency with calculations of the spin fluctuation pairing vertex~\cite{Graser2009}. The bubble then represents the RPA susceptibility that in the multiorbital system is $\chi^{ll',mm'}(\q,\omega_q)$ with $l$, $l'$, $m$, $m'$ being the orbital indices, and $\q$ and $\omega_q$ are the momentum and frequency, respectively. The same susceptibility was calculated in Ref.~\cite{Graser2009} and was shown to produce superconductivity with the A$_{1g}$ order parameter symmetry, similar to other spin fluctuation calculations~\cite{Kuroki2008,Kemper2010,Kuroki2009,Graser2010}.

Since we focus on the lifetime effects, we consider only $\mathrm{Im}\Sigma$ neglecting
the real part of the self-energy, $\mathrm{Re}\Sigma$. The renormalization of the band structure due to the real part of the self-energy has been discussed in some detail in Refs.~\cite{l_ortenzi_09,Ikeda2010} and is not considered here. We note that present calculations are based on the LDA band structure that already contains important Hartree corrections and agrees fairly well with quantum oscillation experiments~\cite{j_analytis_10,a_coldea_08,j_analytis_09a}.

After the rotation~(\ref{eq:Hpsi}), the eigenvectors and interactions become real and calculation of the diagram in Fig.~\ref{fig:bubble_diagram} results in the multiband extension of the standard zero-temperature expressions for the self-energy,
\bea
\label{eq:imsigma}
 \mathrm{Im}\Sigma_{\bar{n}\bar{n}'}(\k,\omega) &=& \sum\limits_{\q,\mu} \sum\limits_{\bar{m},\bar{m}',\bar{l}',\bar{l},\bar{z},\bar{z}'}
  U_{\bar{n}\bar{z}}^{\bar{m}\bar{m}'} U_{\bar{z}'\bar{n}'}^{\bar{l}'\bar{l}} \varphi^{\mu}_{\k-\q z} \varphi^{\mu}_{\k-\q z'} \nn\\
  &\times& \mathrm{Im}\chi^{\bar{l}\bar{l}',\bar{m}\bar{m}'}(\q, \omega - \eps_{\k-\q \mu})
  \left[ \Theta\left(\eps_{\k-\q \mu}\right) - \Theta\left(\eps_{\k-\q \mu} - \omega\right) \right].
\eea
For simplicity, we have introduced the notation $\bar{n} = (n,\sigma_n)$, where $n$ and $\sigma_n$ are the orbital and spin index, respectively. The initial and final spins $\sigma_n$ and $\sigma_{n'}$ have been kept equal since we are considering the paramagnetic state.

The momentum dependence of the orbital matrix elements generates an effective momentum-dependent interaction from the bare local Coulomb interactions,
\begin{equation}
\label{eq:V}
 V_{\bar{n},\mu}^{\bar{m}\bar{m}'}\left(\k-\q \right) = \sum\limits_{\bar{z}} U_{\bar{n}\bar{z}}^{\bar{m}\bar{m}'} \varphi^{\mu}_{\k-\q z},
\end{equation}
in terms of which Eq.~(\ref{eq:imsigma}) may be written as
\bea
\label{eq:imsigmaVeff}
  \mathrm{Im}\Sigma_{\bar{n}\bar{n}'}(\k,\omega) &=& \sum\limits_{\q,\mu} \sum\limits_{\bar{m},\bar{m}',\bar{l},\bar{l}'}
  V_{\bar{n},\mu}^{\bar{m}\bar{m}'}(\q) V_{\bar{n}',\mu}^{\bar{l}\bar{l}'}(\q) \nn\\
  &\times& \mathrm{Im}\chi^{\bar{l}\bar{l}',\bar{m}\bar{m}'}(\k-\q, \omega - \eps_{\q\mu})
  \left[ \Theta\left(\eps_{\q\mu}\right) - \Theta\left(\eps_{\q\mu}-\omega\right) \right].
\eea
The effective interaction enhances the anisotropy of the scattering rate, as will be demonstrated below.

We now discuss briefly the spin structure of the diagram in Fig.~\ref{fig:bubble_diagram} which is important for the calculation of $\mathrm{Im}\Sigma$ using Eq.~(\ref{eq:imsigma}). The susceptibility can be divided into charge and spin channels, and subsequently into singlet and triplet parts,
\bea
\label{eq:chi}
 \chi^{\bar{l}\bar{l'},\bar{m}\bar{m}'} &=
   \half \chi_{c}^{ll',mm'} \delta_{\sigma_m\sigma_{m'}} \delta_{\sigma_l\sigma_{l'}} + \sixth \chi_{s}^{ll',mm'} \vec{\hat\sigma}_{\sigma_m\sigma_{m'}} \cdot \vec{\hat\sigma}_{\sigma_l\sigma_{l'}} \nn\\
 & = \left\{
   \begin{array}{cc}
    \hat\chi_{1,2} \equiv \half \chi_{c}^{ll',mm'} \pm \sixth \chi_{s}^{ll',mm'} \quad & \mathrm{triplet} \\
    \hat\chi_{3} \equiv \frac{1}{3} \chi_{s}^{ll',mm'} \quad & \mathrm{singlet}
   \end{array}
  \right.
\eea
where $\chi_c$ and $\chi_s$ are the charge and spin parts of the susceptibility, respectively, and $\vec{\hat\sigma}_{\sigma\sigma'}$ are vectors, composed of Pauli spin matrices.

For the purpose of the self-energy calculation, the interactions can be grouped into three channels. If we denote the incoming spins as $\sigma_1$ and $\sigma_3$, and the outgoing as $\sigma_2$ and $\sigma_4$, the channels are:
(1) $\sigma_1 = \sigma_2 = \sigma_3 = \sigma_4$,
(2) $\sigma_1 = \sigma_2 \ne \sigma_3 = \sigma_4$,
(3) $\sigma_1 \ne \sigma_2 = \sigma_3 \ne \sigma_4$.
Then the orbital part of interactions in each channel, $\hat{U}_1$, $\hat{U}_2$, and $\hat{U}_3$, are:
\bea
 \begin{array}{lll}
 (U_1)_{ll}^{ll} = 0 & (U_2)_{ll}^{ll} = U & (U_3)_{ll}^{ll} = -U \\
 (U_1)_{ll}^{nn} = \Up-J & (U_2)_{ll}^{nn} = \Up & (U_3)_{ll}^{nn} = -J \\
 (U_1)_{ln}^{ln} = 0 & (U_2)_{ln}^{ln} = \Jp &  (U_3)_{ln}^{ln} = -\Jp \\
 (U_1)_{ln}^{nl} = J-\Up & (U_2)_{ln}^{nl} = J &  (U_3)_{ln}^{nl} = -\Up
 \end{array} \nn
\eea
where orbital indices $l \ne n$.

To combine the interactions with the susceptibility, we first note that due to the spin structure of the diagram, the interaction channels (1)-(3) decouple. Second, we see by inspection that channels (1) and (2) couple to $\hat{\chi}_{1,2}$, and channel (3) couples to $\hat{\chi}_3$. Thus, the self-energy will contain the following matrix structure
\begin{equation}
 \hat{U} \hat{\chi} \hat{U} \propto \hat{U}_1 \hat{\chi}_1 \hat{U}_1 + \hat{U}_2 \hat{\chi}_1 \hat{U}_2 + \hat{U}_1 \hat{\chi}_2 \hat{U}_2 + \hat{U}_2 \hat{\chi}_2 \hat{U}_1 + \hat{U}_3 \hat{\chi}_3 \hat{U}_3.
\end{equation}
This expression by construction resolves the spin summation and only sums over orbital indices remain. Combining it with the calculation of $\chi^{ll',mm'}(\q,\omega_q)$ for a given doping $x=n_e-6$, we use Eq.~(\ref{eq:imsigma}) to obtain $\mathrm{Im}\Sigma_{nn'}$ straightforwardly. Then we convert it to a band representation,
\beq
 \mathrm{Im}\Sigma_{\mu\mu'}(\k,\omega) = \sum\limits_{n,n'} \varphi^{\mu}_{\k n} \mathrm{Im}\Sigma_{nn'}(\k,\omega) \varphi^{\mu'}_{\k n'}.
\eeq
For the energy range where there are no band crossings, there is a unique band $\mu$ corresponding to the momentum $\k$. The self-energy describes the scattering of the particle with $\k$ back to the same momentum $\k$, and thus back to the same band, $\mu' = \mu$. For the small energies around the Fermi level considered, there are no band crossings, so the major contribution to the scattering rate in the full Green's function in band space, $\hat{G} = (\hat{G}_0^{-1} - \hat{\Sigma})^{-1}$, comes from diagonal, $\mu = \mu'$, matrix elements of $\mathrm{Im}\hat{\Sigma}$. We denote them as
\beq
 \imsigma_\mu(\k,\omega) \equiv \mathrm{Im}\Sigma_{\mu\mu}(\k,\omega).
\eeq

\subsection{Results of calculations}

Because inter-band transitions are negligible in the range of energies considered here, the calculated scattering rate follows the Fermi liquid relation $\imsigma(\k,\omega) \propto \omega^2 + \pi^2 T^2$. Thus some finite frequency or temperature is needed for non-vanishing results. Here and below, the quantities we report will be calculated at $\omega = 20$meV that is equivalent to $T \approx 74$K at zero frequency. It was verified numerically that our results scale as $\omega^2$. The results below are qualitatively independent of the exact frequency chosen, since we are below the range of frequencies where inter-band scattering plays a large role.

\begin{figure}[ht]
 \centering
 \includegraphics[width=0.32\textwidth]{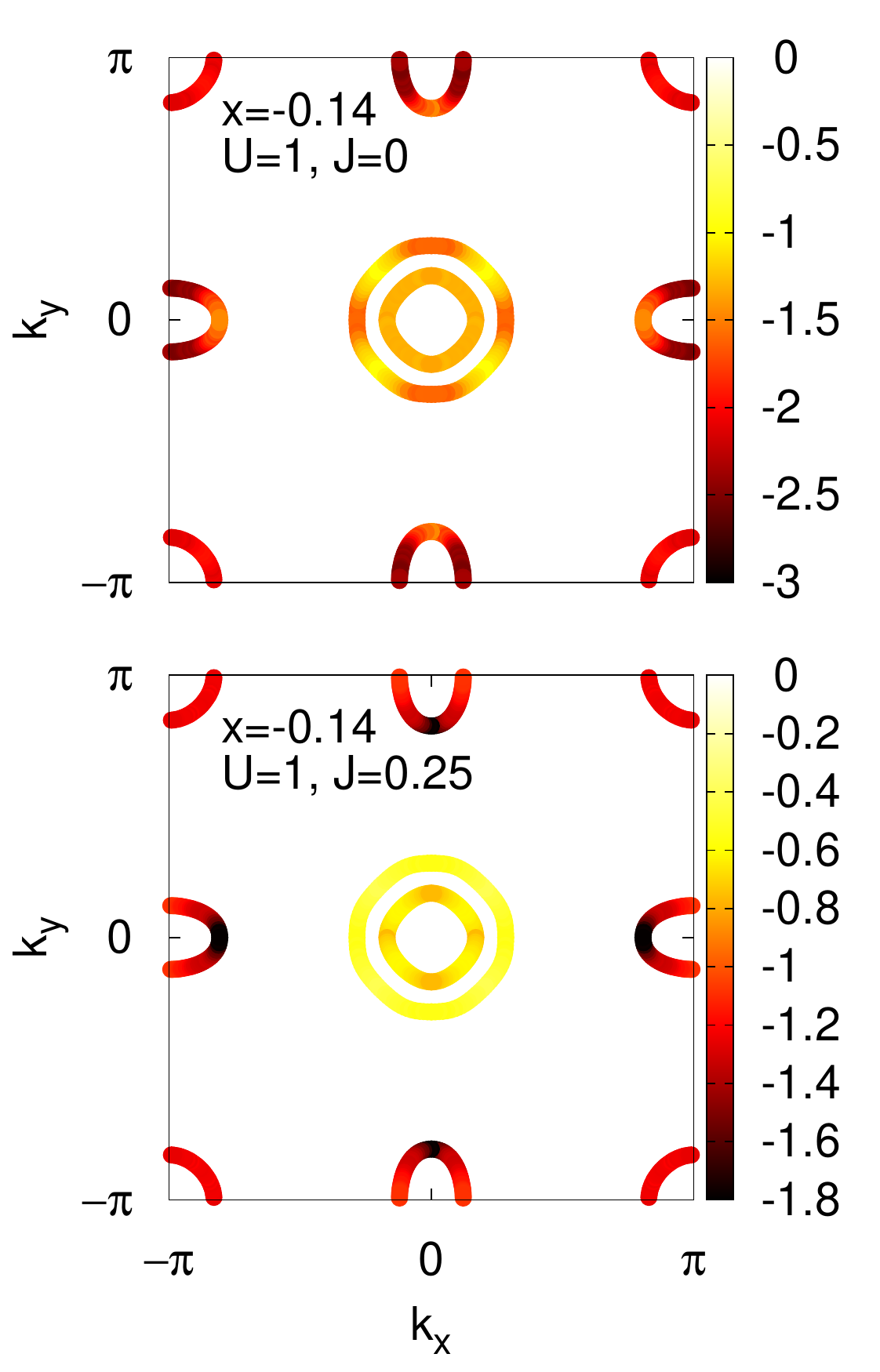}
 \includegraphics[width=0.32\textwidth]{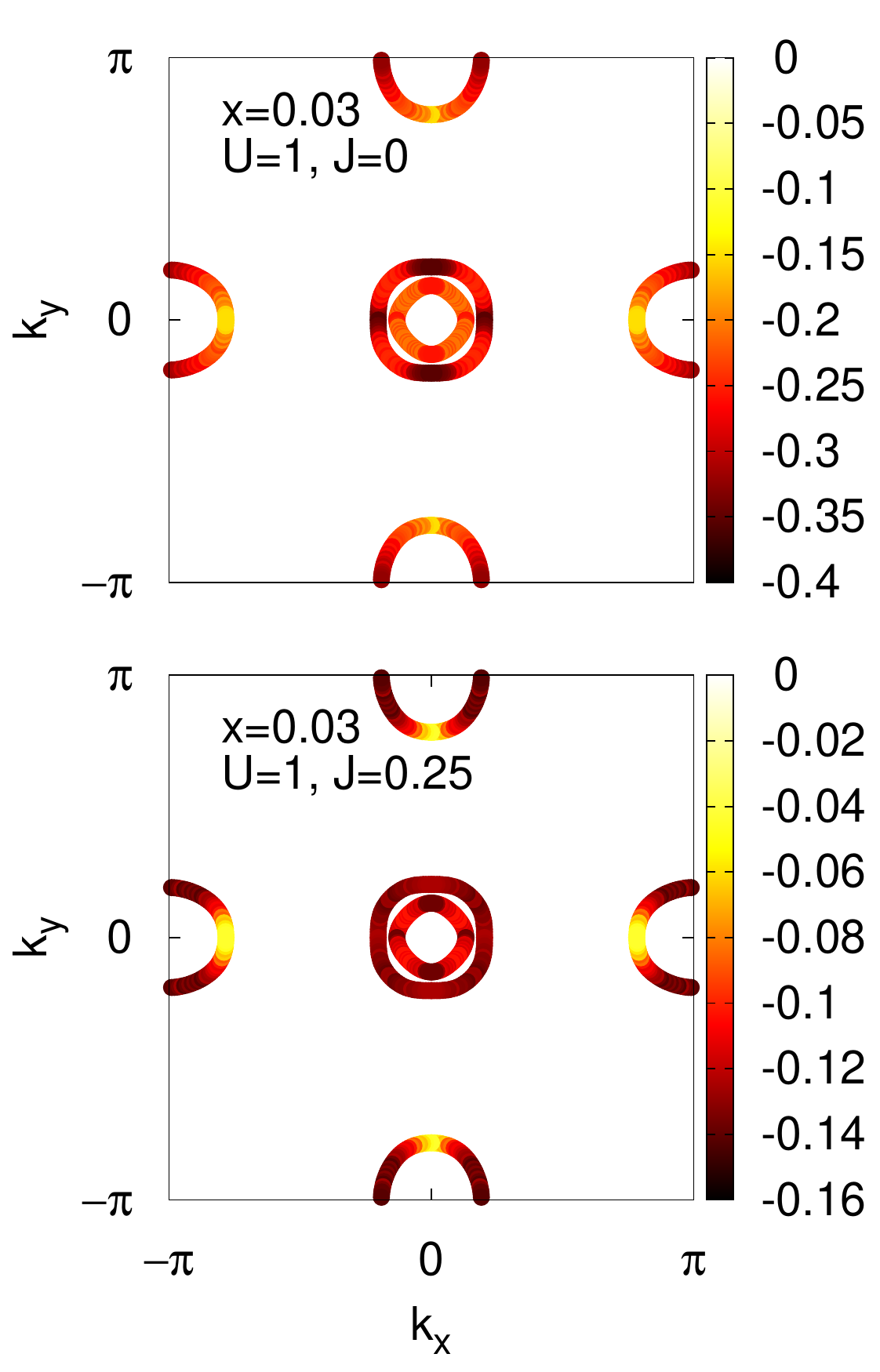}
 \includegraphics[width=0.32\textwidth]{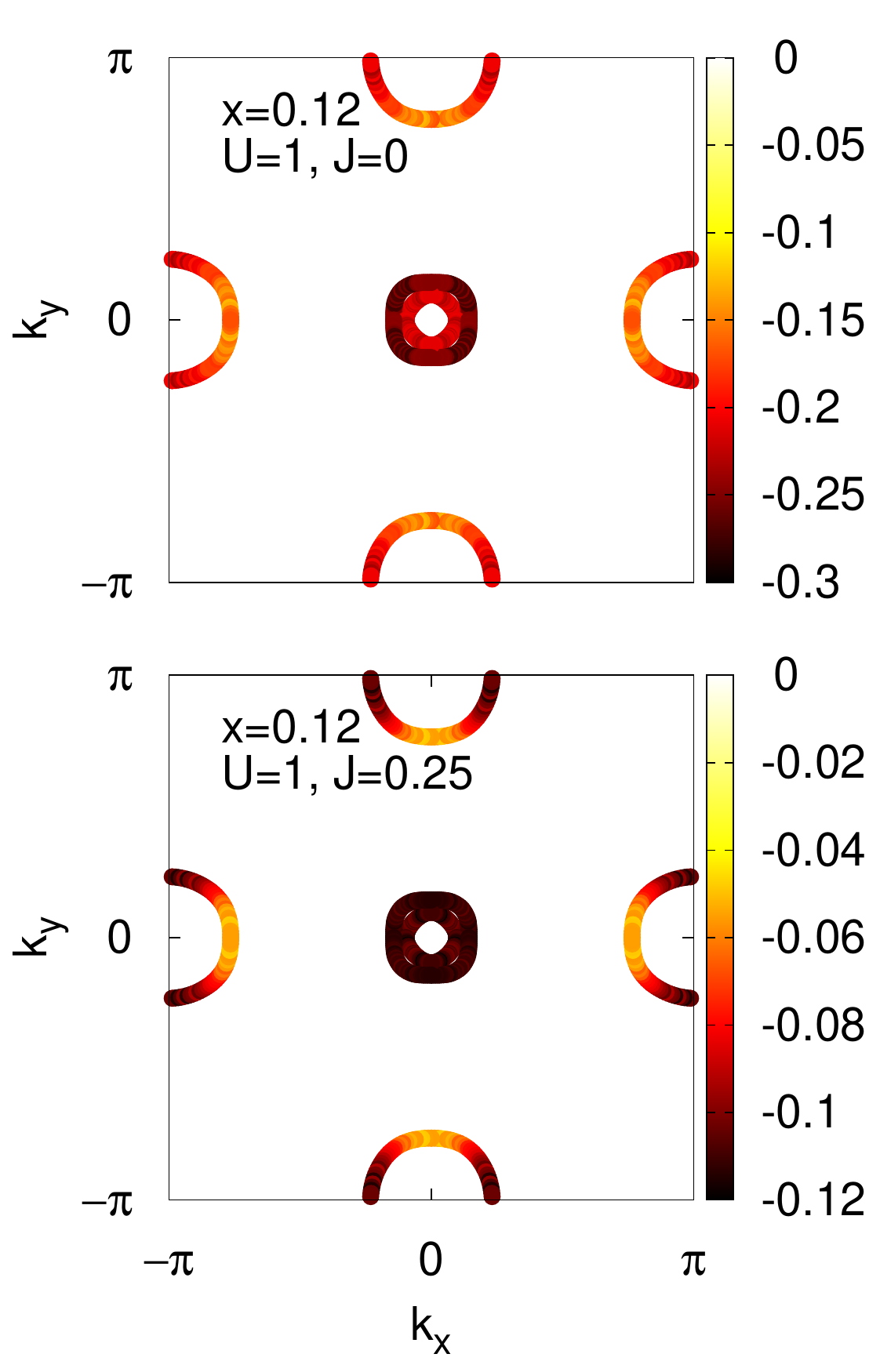}
 \caption{Imaginary part of the self-energy $\Sigma$ (in meV) at $\omega = 20$meV along the Fermi surface for various dopings ($x=-0.14$, $0.03$, and $0.12$ from left to right) and for two sets of interaction parameters (in eV). The color scale is different for each plot.}
 \label{fig:imsigma}
\end{figure}

For several dopings and few sets of interaction parameters, the calculated scattering rate along the Fermi surface is shown in Fig.~\ref{fig:imsigma}. Here, $U$ and $J$ are were chosen to be close to the SDW-instability in the spin susceptibility.

\begin{figure}[ht]
 \centering
 \includegraphics[width=0.7\textwidth]{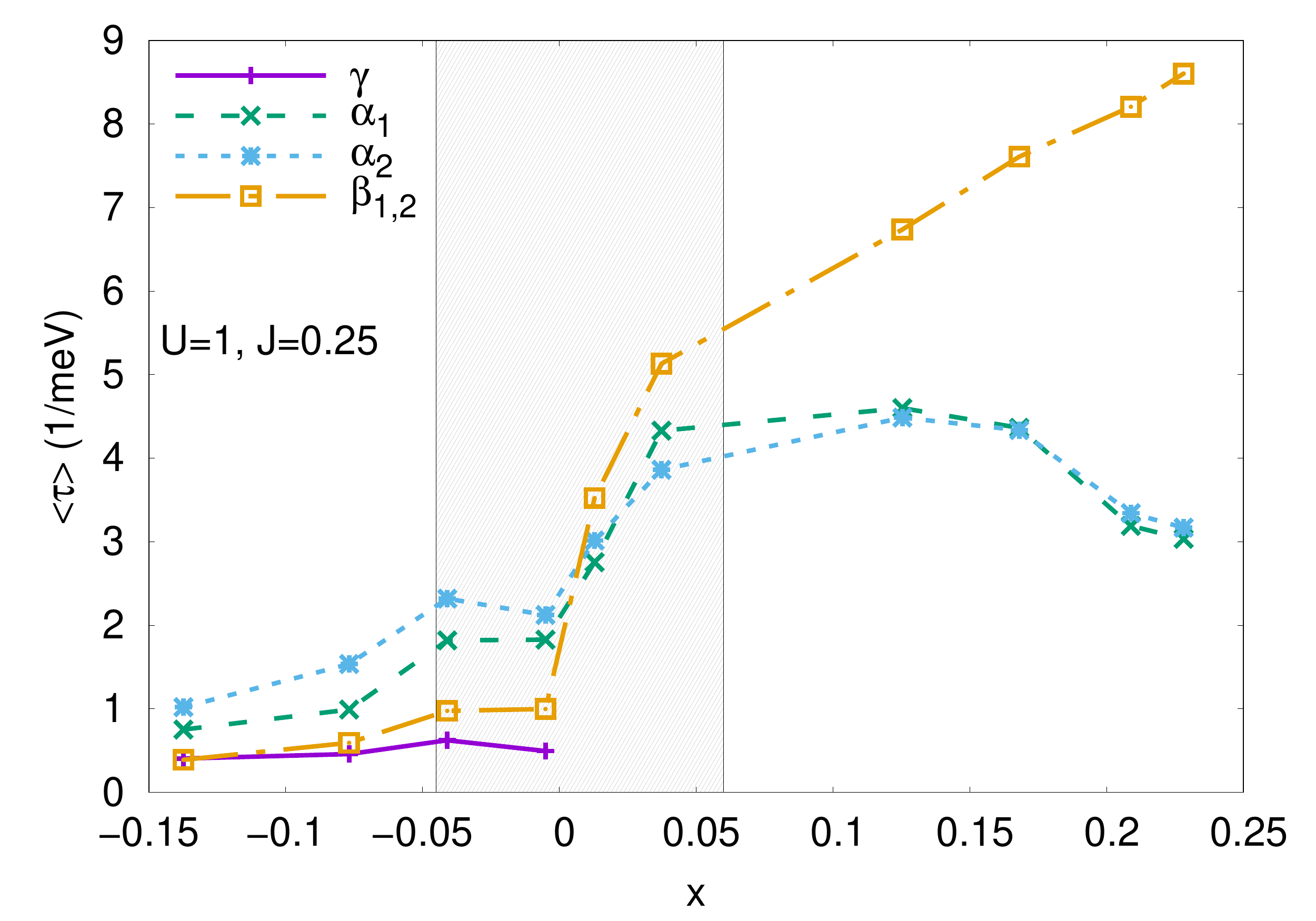}
 \caption{Average scattering rate $\la\tau\ra$ for holes ($\alpha_1$, $\alpha_2$, $\gamma$) and electrons ($\beta_1$, $\beta_2$) at $\omega = 20$ meV for $U = 1.0$eV and $J = 0.25$eV. The shaded region marks the rough experimental SDW region in 122 systems. Solid lines are guides to the eye.}
 \label{fig:tau_ave}
\end{figure}

We observe that the average scattering rate increases monotonically with doping. Fig.~\ref{fig:tau_ave} shows the average lifetime $\la\tau\ra$ with $\tau_k = - 1/2 \Sigma''(k,\omega)$ for holes and electrons on the Fermi surface. We see a clear increase in the quasiparticle lifetime on all Fermi surface sheets as the system is electron doped. On the electron-doped side, the average scattering rates are essentially controlled by the degree of nesting. As more electrons are doped into the system, the hole pockets shrink and the nesting between the $\alpha$ and $\beta$ sheets deteriorates. The hole-doped systems have a smaller lifetime due to the presence of the $\gamma$ pocket; in addition to $(\pi,0)$ scattering between $\alpha$ and $\beta$ sheets, new phase space for scattering opens up and the average rate increases. Thus, one expects the resistivity due to spin-fluctuations to increase with hole doping.

Aside from the overall change in scale, Fig.~\ref{fig:tau_ave} shows that the ratio of electron to hole scattering rate changes as one goes from hole to electron doping; electrons have a higher average scattering rate on the hole-doped side, and vice versa. Although there is already an anisotropy between the hole and electron pockets in terms of lifetimes, it is not enough to cause the experimentally observed anisotropy, as will be discussed below.

Next, we observe a clear anisotropy in the scattering rate going around the Fermi surfaces as shown in Fig.~\ref{fig:imsigma}. Focusing first on the undoped and electron-doped systems, the $\beta_1$ sheet exhibits strong anisotropy between the $\Gamma-X$ and $X-M$ directions. From Fig.~\ref{fig:FS}, we observe that this is where the Fermi surface orbital composition changes from \dxy to \dyz character. There is a strong minimum in the scattering rate in the \dxy portions of the $\beta$ sheets; this is due to the above-mentioned anisotropy of the effective interaction~(\ref{eq:V}). The orbital matrix elements tend to restrict scattering to be maximal for intra-orbital processes. For the \dxy electrons, there is very little phase space to scatter compared to other orbitals, see Fig.~\ref{fig:FS}, because the spin fluctuation scattering intensity $\chi(\q)$ is peaked at $\q=(\pi,0)$. Thus, they behave more like free electrons. When the system is sufficiently hole doped to create the (\dxy) $\gamma$ hole pocket, $(\pi,0)$ spin fluctuations couple them strongly to other \dxy states, causing the scattering rate there to increase. Throughout the doping range, \dxz and \dyz states on the $\alpha$ pockets scatter strongly with their counterparts on the $\beta$ pockets, and vice versa.

As for the interaction dependence in Fig.~\ref{fig:imsigma}, the top row of panels there shows a case where $J=0$, and the bottom is for finite $J=0.25$. As the Hund's rule coupling $J$ is turned on, we observe two effects. First, the overall scattering rate decreases (note that the color scale on each plot is different). This is due to the SRI relation $U'=U-2J$, so that $U'$ is decreased in the middle row of panels. Although new scattering channels open up through $J$ itself, this is more than compensated by the decrease in the inter-orbital scattering $U'$.

Let us consider the effect of $J$ on the $\beta$ sheet anisotropy for the hole-doped system. When $J=0$, the minimum scattering rate occurs near the \dxy sections of the Fermi surfaces for all dopings. Once $J$ is turned on, the anisotropy reverses, and instead a maximum scattering rate is found on the same sections. This reversal of anisotropy can be explained by the same argument as above. When $J=0$, the intra-orbital and inter-orbital scattering are the same since $U = U'$. Thus, there is a strong scattering from both the \dxz/\dyz portions as well as the \dxy portions of the $\beta$ sheets to the $\gamma$ pocket (of \dxy character). Since the \dxz and \dyz portions additionally scatter to the $\alpha$ sheets, a stronger scattering rate occurs there. When $J$ is finite, the effective inter-orbital scattering rate $U'$ decreases through the SRI relation. Thus, the scattering on the \dxz and \dyz portions is decreased while that on the \dxy sections remains the same. With sufficiently large $J$, the anisotropy on the $\beta$ sheets is reversed. Note, however, that this argument depends on the existence of the $\gamma$ pocket. When the pocket is not present, such as in the undoped and electron doped cases, no such reversal occurs, and thus the \dxy states have the longest lifetimes for the configurations investigated~\cite{KemperKorshunov2011}.

Similar momentum dependence of the lifetimes was found in Ref.~\cite{Onari}, where the scattering due to spin fluctuations was considered within the fluctuation-exchange approximation (FLEX).

\subsection{Conductivity}

Now we consider the effect of the calculated scattering rates on the electric conductivity. The total conductivity is the sum of the band conductivities, $\sigma(\omega) = \sum\limits_\mu \sigma_{x \mu}(\omega)$,
\begin{align}
 \sigma_{x \mu}(\omega) = \frac{e^2}{\pi h} \int\limits_{\k \in \k_{F\mu}} d\k N_\k v_{\k_x}^2 \tau_{\k}(\omega),
\label{eq:conductivity}
\end{align}
where $\tau_\k = -1/2\imsigma_\mu(\k,\omega)$, $\k_{F\mu}$ is the Fermi momentum for a particular band index $\mu$, we integrate over $\k_\parallel$ that is the component of momentum along the Fermi surface, $v_{\k}$ is the velocity, and $N_{\k_{F\mu}} = 1/|v_{\k_{F\mu}}|$ is the momentum- and band-dependent density of states (DOS) at the Fermi level. Note that we have approximated the transport lifetime with the one-electron lifetime $\tau_\k$, neglecting forward scattering corrections, as well as the distinction between normal and Umklapp processes. Such an approximation can only give the crude qualitative effect of the scattering from spin fluctuations on the conductivity.

\begin{figure}[ht]
 \centering
 \includegraphics[width=0.7\textwidth]{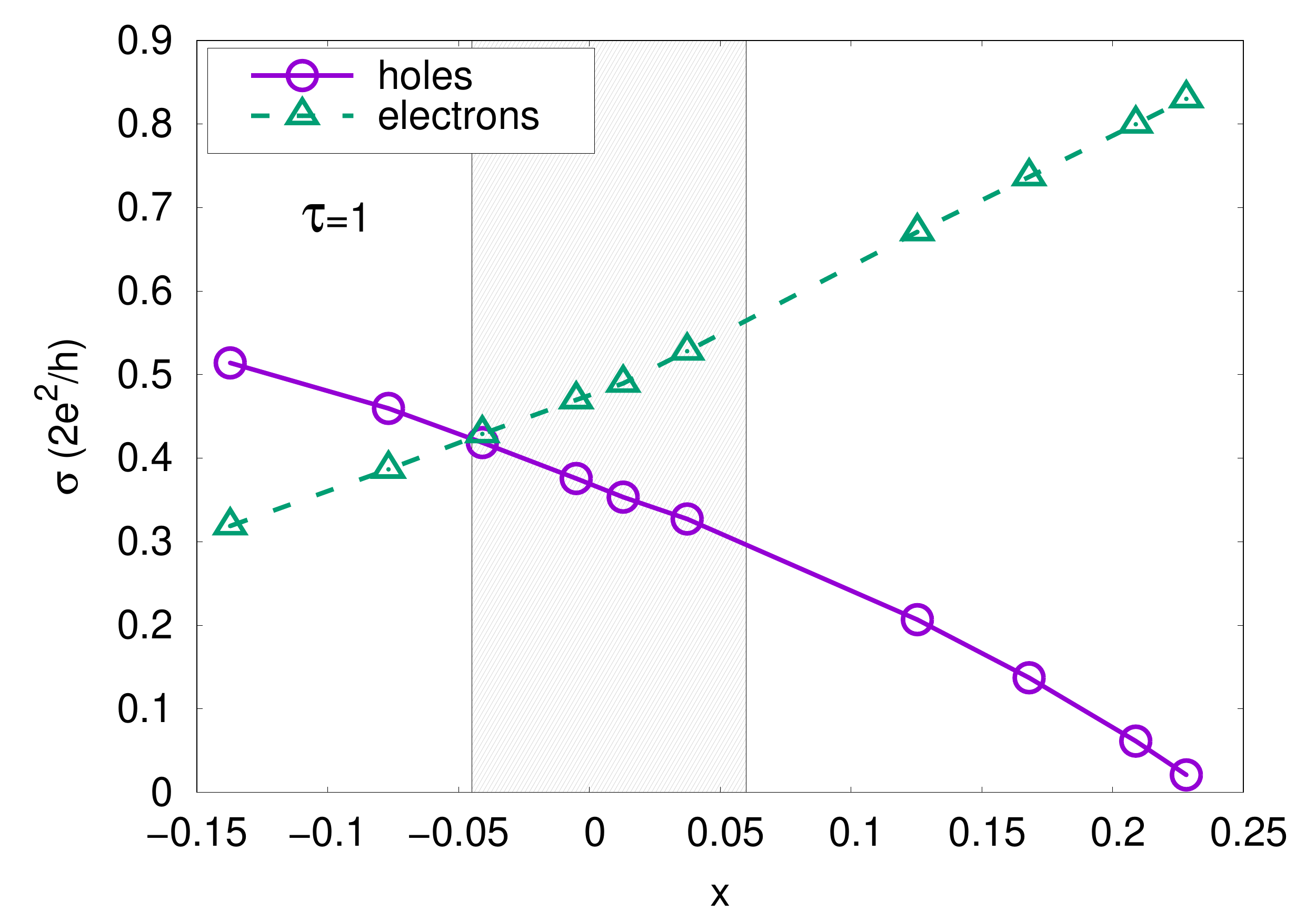}
 \caption{Conductivity for holes and electrons as a function of doping $x=n_e-6$ for constant relaxation rate $1/\tau=1$ eV. The shaded region marks the rough experimental SDW region in 122 systems. Solid lines are guides to the eye.}
 \label{fig:conductivity_tau1}
\end{figure}

To analyze the doping-dependence of the conductivity, we now keep the interactions constant at values that do not produce an RPA instability over the range of dopings considered. We evaluate the DC conductivities at finite temperature by replacing $1/\tau_\k(\omega)$ in Eq.~(\ref{eq:conductivity}) by $1/\tau_\k(\pi T)$. It is important to ask which aspects of the doping dependence of transport arise from purely kinematic effects such as carrier density and Fermi velocity, which evolve with doping, and which arise from interactions. To illustrate this, we first plot in Fig.~\ref{fig:conductivity_tau1} the separate contributions to the total conductivity from the electron and hole sheets, with an assumed constant relaxation time. Here the conductivities evolve more or less as expected with electron doping as the volumes of hole sheets shrink and electron sheets grow. On the other hand, it is important that the ``perfectly compensated'' situation of equal kinetic conductivity of electrons and holes does not occur for the undoped case, but rather for $x \simeq -0.05$ corresponding to a slight hole doping. We have indicated in the figure the range of doping over which the 122 systems display long range magnetic order, which is not included in the current theory, and thus where the results are not directly applicable.

\begin{figure}[ht]
 \centering
 \includegraphics[width=0.7\textwidth]{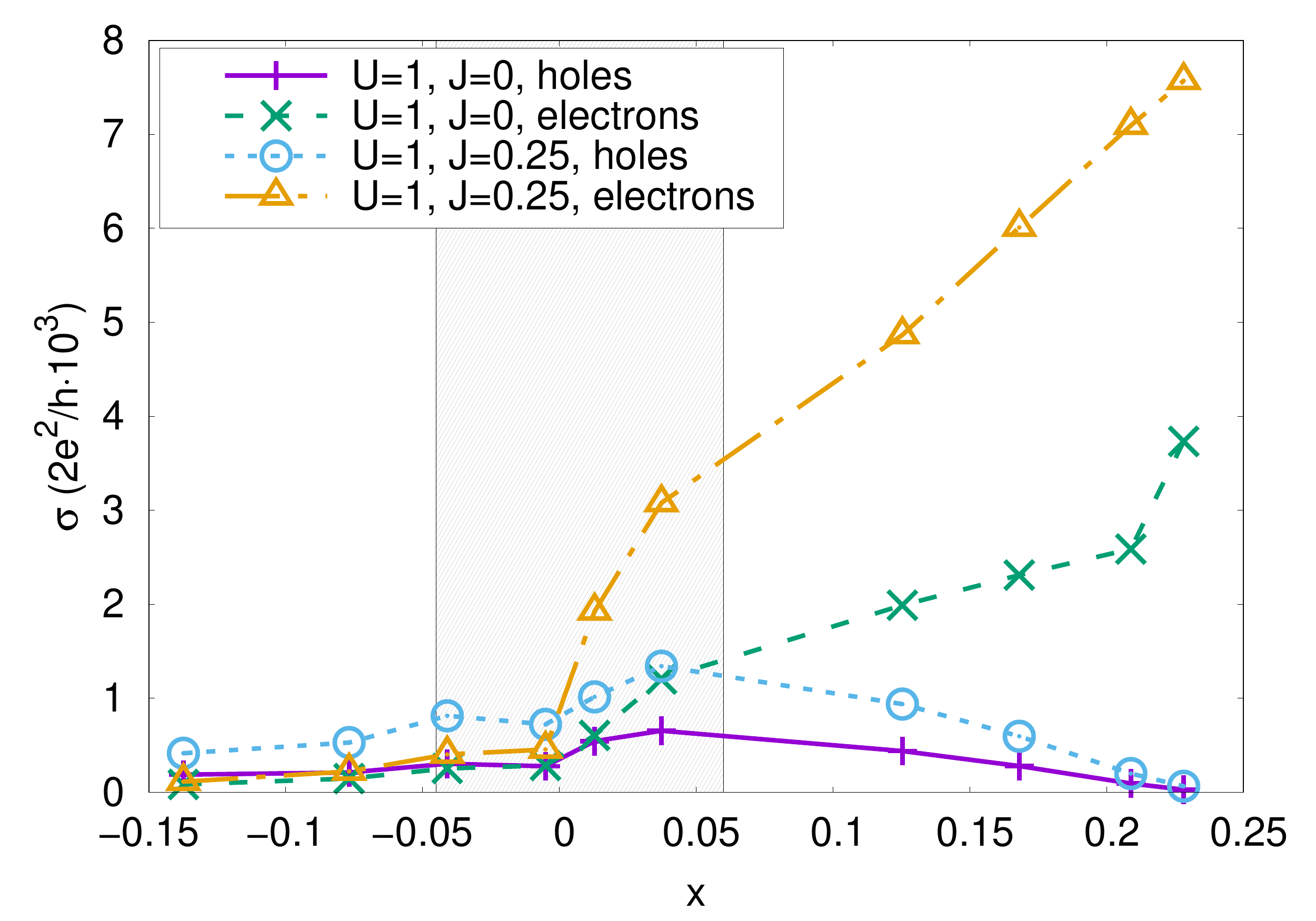}
 \caption{Conductivity for holes and electrons as a function of doping $x$ for the two sets of parameters (in eV): $U=1.0, J=0$ and $U=1.0, J=0.25$, at effective temperature $T=74K$. The shaded region marks the rough experimental SDW region in 122 systems. Solid lines are guides to the eye.}
 \label{fig:conductivity}
\end{figure}

By contrast, Fig.~\ref{fig:conductivity} shows the separate conductivities on the hole and electron Fermi surfaces as a function of doping. We immediately notice that conductivity for electrons grows quite strongly upon electron doping. Quite unlike the purely kinetic case in Fig.~\ref{fig:conductivity_tau1}, the hole conductivity varies only weakly compared to that of the electrons.
It is this asymmetry, due to a combination of the kinetic effects illustrated in Fig.~\ref{fig:conductivity_tau1} and lifetime effects calculated here, which lead to the rapid domination of the conductivity by electrons. This has led transport experiments for Co-doped Ba-122 being interpreted in terms of a one-band model with electrons only \cite{f_rullier_albenque_09,l_fang_09}. The feature that greatly affects the doping dependence is the fact that the maximum of the Fermi velocity is precisely where the lifetime is largest on the electron Fermi surface sheets, namely the \dxy sections of the $\beta$ sheets.

The calculated conductivity shown in Fig.~\ref{fig:conductivity} was obtained for interaction parameters chosen sufficiently small to show the effect of doping while avoiding the RPA instability. For these parameters, the absolute scale of $\sigma$ is much larger than in experiments on 1111 or 122 samples we have examined. Clearly increasing the overall scale of the interactions will increase the scattering rates and decrease the conductivity. However, to obtain the observed values of the conductivity requires approaching the RPA instability extremely closely. We have not attempted to fine tune the interaction strengths, but merely to illustrate the possible qualitative behavior. It seems more likely that a more complete theory will require a renormalization of the susceptibility akin to that seen in Quantum Monte Carlo (QMC) studies of the Hubbard model, which indicated that the RPA form of the dynamical magnetic response was qualitatively correct, but that the ``$U$'' driving the instability (through the RPA denominator) needed to be taken independent of the $U^2$ prefactor in the effective interaction~\cite{t_maier_07}. A similar effect should occur in multiorbital Hubbard models, such that the overall scales of scattering rates, and degree of proximity to the instability, should not be taken overly seriously.

\subsection{Hall coefficient}

Any disparity between the scattering rates of electrons and holes manifests itself in the Hall coefficient
\begin{equation}
 R_H = - \sigma_{H}(\omega)/\sigma^2(\omega),
\label{eq:RH}
\end{equation}
where $\sigma_{H}(\omega)$ is the Hall conductivity \cite{Schulz_92,Kim_98}. For a multiband system, $\sigma_{H}(\omega) = \sum\limits_{\mu} \sigma_{H \mu}(\omega)$ and the expression for the band Hall conductivity has the form
\begin{equation}
 \sigma_{H \mu}(\omega) = \frac{e^3}{\pi h} \int\limits_{\k \in \k_{F\mu}} d\k N_\k
 \mathbf{v}_{\k} \cdot \left[ \mathrm{Tr}(\mathbf{M}_{\k}^{-1}) - \mathbf{M}_{\k}^{-1} \right] \cdot \mathbf{v}_{\k}
 \tau_{\k}^2(\omega),
\label{eq:sigmaH}
\end{equation}
where $\left( \mathbf{M}_{\k}^{-1} \right)_{\alpha\beta} = \hbar^{-1} \partial v_{k_\alpha} / \partial k_\beta$ is the inverse mass tensor.

\begin{figure}[ht]
 \centering
 \includegraphics[width=0.7\textwidth]{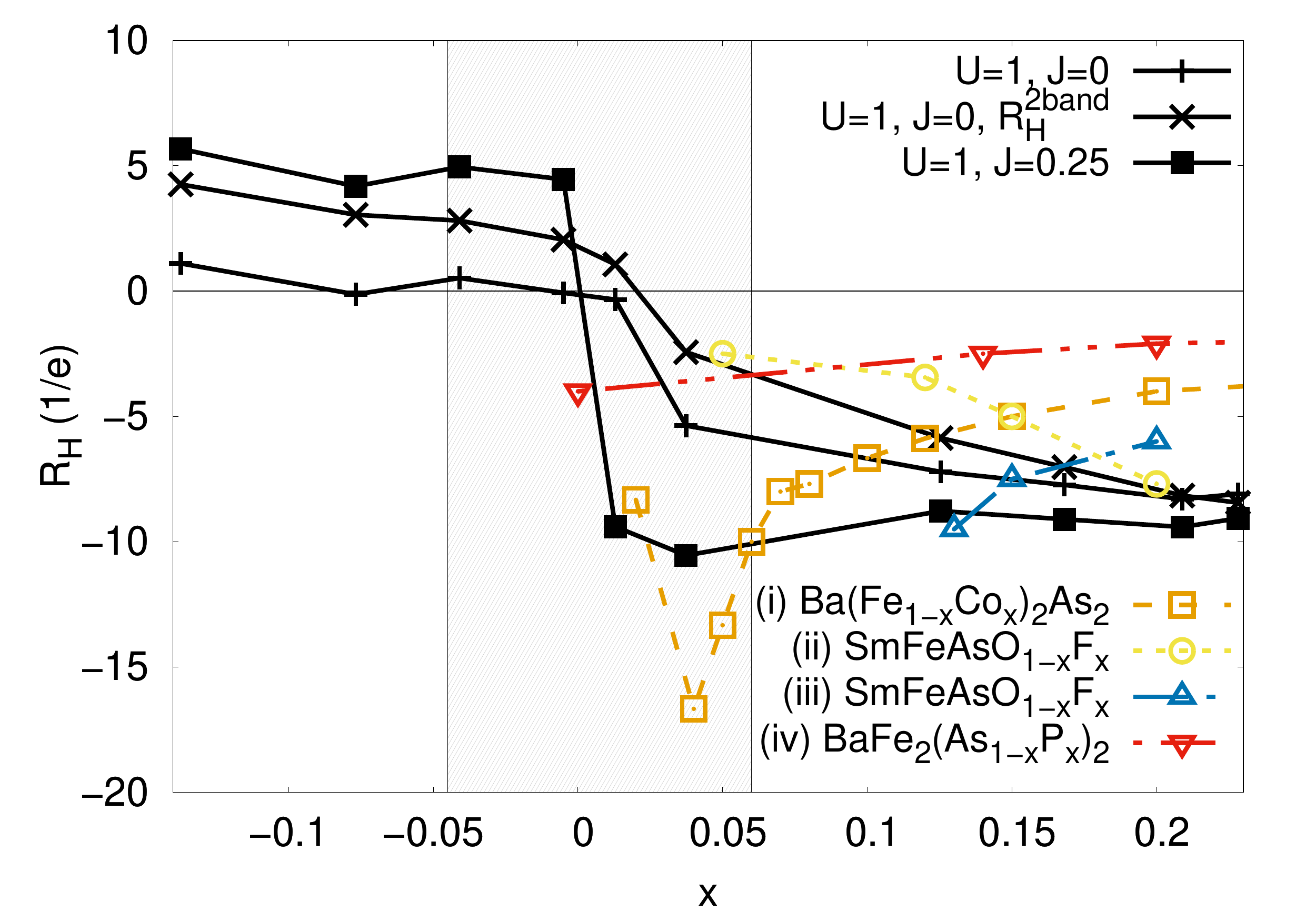}
 \caption{Doping dependence of the Hall coefficient. The theoretical calculations are for two sets of parameters (in eV): $U=1.0$, $J=0$ and $U=1.0$, $J=0.25$. For the first set we also show result of the multiband approximation for $R_H$ from Eq.~(\ref{eq:sigmaHapprox}). Experimental data points are from (i) Ref.~\cite{l_fang_09} for Ba(Fe$_{1-x}$Co$_x$)$_2$As$_2$ at 100K, (ii) Ref.~\cite{Riggs_09} and (iii) Ref.~\cite{Liu_08} for SmFeAsO$_{1-x}$F$_x$ at 125K, and (iv) Ref.~\cite{Kasahara_10} for BaFe$_2$(As$_{1-x}$P$_x$)$_2$ at 150K. The shaded region tentatively marks the experimental SDW region. Solid lines are guides for the eye.}
 \label{fig:Hall}
\end{figure}

\begin{figure}
 \centering
 \includegraphics[width=0.7\textwidth]{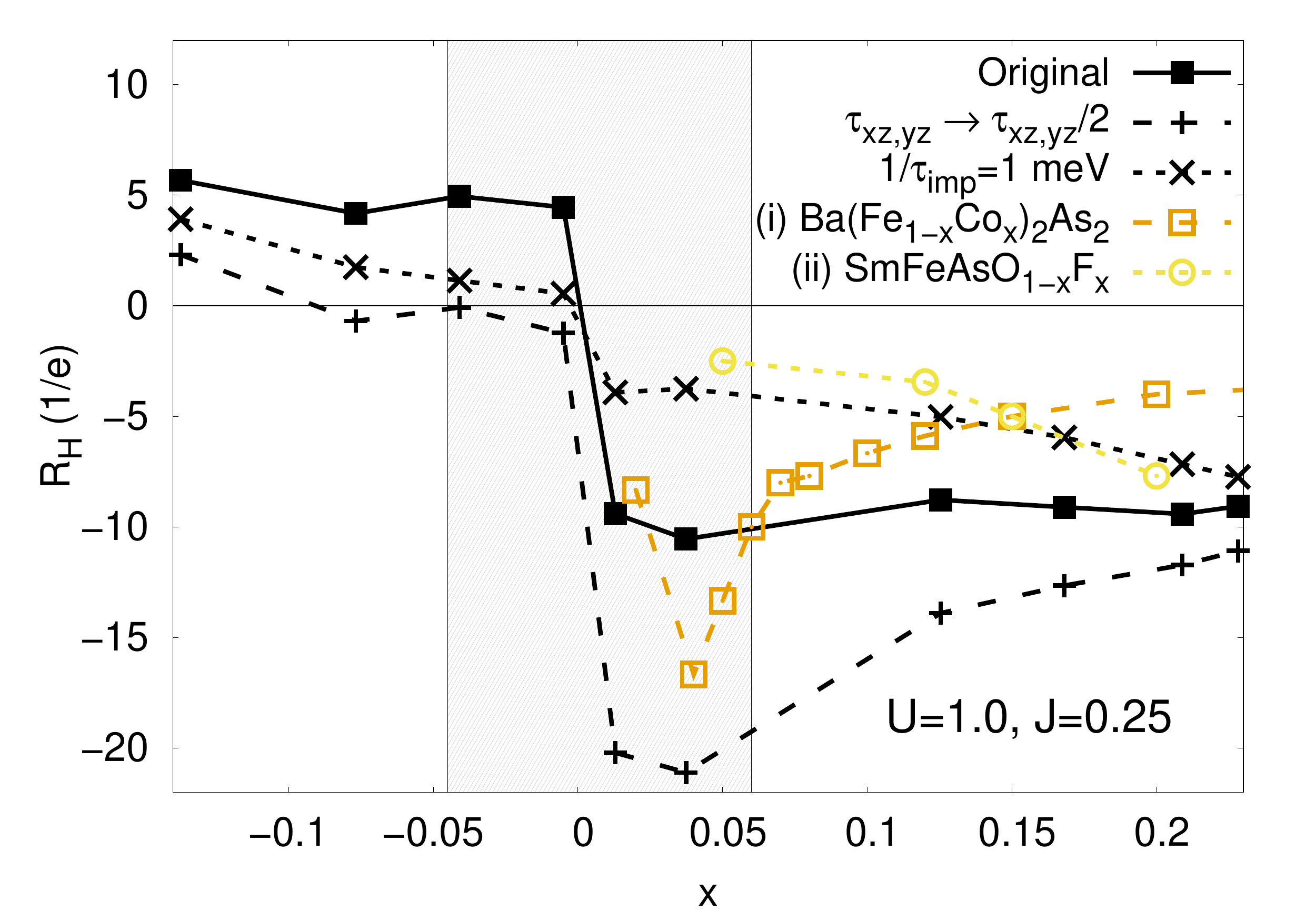}
 \caption{Doping dependence of the Hall coefficient for three distinct cases: (1) original calculated $R_H$ from Fig.~\ref{fig:Hall}, (2) the one with the artificially increased scattering rate for all orbitals except for \dxy, $\tau_{xz,yz} \to \tau_{xz,yz}/2$, and (3) $R_H$ with added constant impurity scattering $1/\tau_{imp}=1$meV. For all cases parameters are $U=1.0$eV, $J=0.25$eV. Experimental data points (i) are from Ref.~\cite{l_fang_09} for Ba(Fe$_{1-x}$Co$_x$)$_2$As$_2$ at 100K. The shaded region tentatively marks the experimental SDW region. Solid lines are guides for the eye.}
 \label{fig:Hall_artificial}
\end{figure}

Fig.~\ref{fig:Hall} shows calculated $R_H$ as a function of doping for $\omega = 20$meV (the corresponding effective temperature is 74K). One can qualitatively understand the doping dependence of $R_H$ by analyzing the approximate equation for the band Hall conductivity,
\begin{equation}
 \sigma_{H \mu}(\omega) \approx R_\mu \sigma_{\mu}^2(\omega).
\label{eq:sigmaHapprox}
\end{equation}
where $1/R_\mu = \pm e n_\mu$ is the Hall coefficient for an electron (hole) band $\mu$, and $n_\mu$ is the occupation of that band. For the simple case of two bands (hole and electron) we have
\begin{equation}
 R_H^{2band} = \frac{1}{e} \frac{\sigma_h^2/n_h-\sigma_e^2/n_e} {\left(\sigma_h+\sigma_e\right)^2}.
\label{eq:RH2band}
\end{equation}
Since conductivity for the hole band $\sigma_h \propto n_h \tau_h / m_h$ and for the electron band $\sigma_e \propto n_e \tau_e / m_e$ with $\tau_{h,e}$ and $m_{h,e}$ being the corresponding lifetimes and band masses, $R_H^{2band}$ is a \textit{decreasing} function of electron doping if $\tau_e \sim \tau_h$ and $m_e \sim m_h$. This is what we see in Fig.~\ref{fig:Hall} for the $U=1.0, J=0$ case. On the other hand, experimental data for 1111 and 122 compounds indicate that $R_H^{expt}$ is an \textit{increasing} function of electron doping (i.e., the magnitude $|R_H^{expt}|$ decreases with increasing $x$) away from the SDW state. According to the simple analysis of Eq.~(\ref{eq:RH2band}), this may be due to (i) $\tau_e \gg \tau_h$ and/or (ii) $m_h \gg m_e$. Note that use of Eq.~(\ref{eq:sigmaH}) gives a different result from Eq.~(\ref{eq:sigmaHapprox}) due to the mass anisotropy across the Fermi surface which contributes to factor (ii). Factor (i) starts to play a role when we consider finite $J$. For the case of $U=1.0$ and $J=0.25$, $R_H(x)$ becomes slightly increasing function of $x$ for $x>0$ (Fig.~\ref{fig:Hall}). However, it is not in quantitative agreement with experimental data. To see whether the present approach can provide the correct slope of $R_H(x)$, we artificially increased scattering rate on all orbitals except \dxy twice, so that the anisotropy between hole and electron sheets becomes more pronounced. The resulting doping dependence of the Hall coefficient is shown in Fig.~\ref{fig:Hall_artificial}. Now the slope of $R_H(x)$ is in good agreement with experimental data.


The fact that we underestimate the disparity between holes and electrons by a factor of two is not very discouraging. There are several factors not included in the present theory. In the interest of studying the doping dependence, we have kept the interactions fairly low to avoid the instability that occurs for relatively small interaction strengths on the hole-doped side. Furthermore, we have neglected impurity scattering. In multiband impurity models \cite{Golubov1997,Senga2008}, the ratio of intra- to interband scattering is taken as a parameter, and the scattering rate asymmetry between electrons and holes is weak. One might expect that an ``orbital impurity'' model, where an impurity introduces a local Coulomb potential for electrons in all $d$-orbitals, might produce a scattering rate anisotropy in $\k$-space due to the matrix elements $\varphi_{\k n}^{\mu}$, just as in the inelastic scattering case. By investigating simple models similar to those considered in Ref.~\cite{Onari2009}, we have similarly concluded that both average elastic scattering rate asymmetry, and elastic scattering rate anisotropy on any given Fermi surface sheet are small. To address the effect of isotropic impurities on the Hall coefficient, we introduced a constant impurity scattering with a strength comparable to the calculated spin-fluctuation scattering rate $1/\tau_\k$. Since concurring scattering processes add to the self-energy,
the scattering rate is $1/\tau_{\k}^\mathrm{total} = 1/\tau_\mathrm{imp} + 1/\tau_{\k}$. Substituting $\tau_{\k}^\mathrm{total}$ in Eqs.~(\ref{eq:conductivity}) and (\ref{eq:sigmaH}), we find $R_H(x)$ shown in Fig.~\ref{fig:Hall_artificial} for $1/\tau_\mathrm{imp}=0.1$meV. Clearly, increasing disorder leads to a monotonically decreasing Hall coefficient with doping similar to Eq.~(\ref{eq:RH2band}) with $\tau_e \simeq \tau_h$. Thus dirtier samples
will show a decrease of $R_H(x)$ with increasing electron doping.

\begin{figure}
 \centering
 \includegraphics[width=0.7\textwidth]{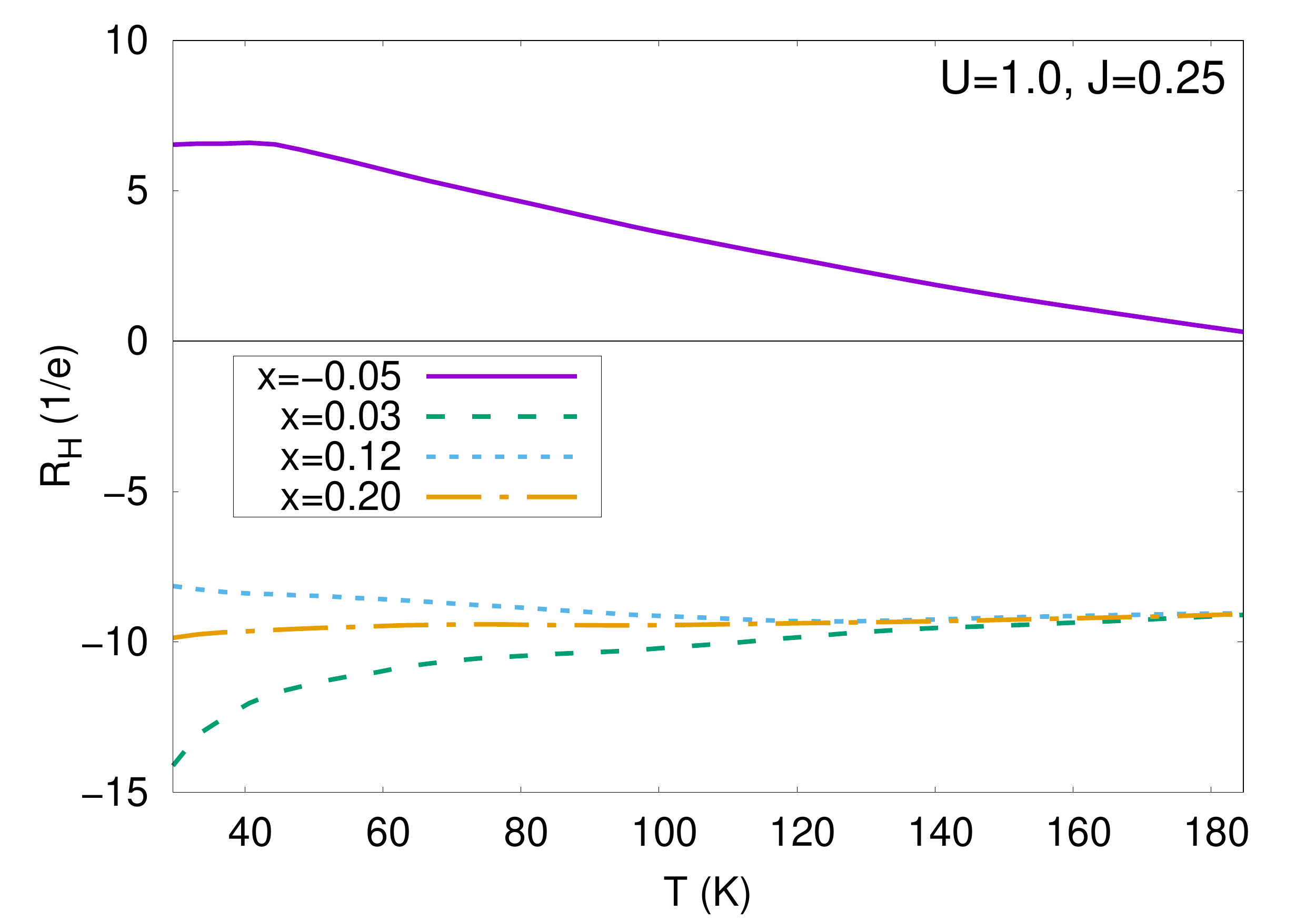}
 \caption{Calculated dependence of the Hall coefficient on the effective temperature for several dopings $x$ with $U=1.0$eV, $J=0.25$eV.}
 \label{fig:RH_T}
\end{figure}

The temperature dependence of $R_H$ deserves additional discussion. Some phenomenological calculations of the self-energy in a two-band model for the pnictides suggest that to reproduce experimentally observed $R_H(T)$ one needs to assume the non-Fermi liquid behavior of the spin susceptibility \cite{Prelovsek_10}. In particular, for large electron dopings, $R_H(T)$ is almost constant but for small $x$ it become an increasing function of temperature \cite{l_fang_09,Rullier-Albenque_10}. Here we argue that the observed temperature dependence can be qualitatively reproduced within our Fermi liquid approach. The resulting $R_H(T)$ from our calculations is shown in Fig.~\ref{fig:RH_T}. Note that the band forming the $\gamma$ Fermi surface pocket for $x<0$ is slightly below the Fermi level for small positive $x$. Thus at finite energy or temperature the scattering to that band contributes to the self-energy and consequently to the transport properties. That is the main reason why $R_H(T)$ for $x=0.03$ is a rapidly changing function of $T$ in Fig.~\ref{fig:RH_T}.

\subsection{Final remarks}

Here I have shown that the quasiparticle scattering due to spin-fluctuations in a multiorbital model with local interactions can be significantly anisotropic. Two factors producing this effect are the orbital matrix elements, which make interactions effectively momentum-dependent, and the momentum dependence of the dynamical susceptibility. In the particular case of the five-orbital model for LaFeAsO, the \dxy portions of the electron Fermi surface experience little scattering due to the small scattering phase space in undoped and electron-doped cases, since there are no \dxy states on the hole sheets available for scattering. This anisotropy on the electron sheets appears to have profound consequences for transport in at least some Fe-based pnictides. There are several factors which together provide experimentally observed disparity between holes and electrons. The first one comes from the longer lifetime of the \dxy states on the electron Fermi surface sheets. Another one is the fact that the maximum of the Fermi velocity is precisely where the lifetime for electrons is largest.


\section{Superconductivity}
\label{sec:sc}

The original proposal of superconducting pairing arising from magnetic interactions was put forward by Emery~\cite{Emery1987} and by Berk and Schrieffer~\cite{BerkSchrieffer}, who were interested primarily in transition metal elements and nearly ferromagnetic metals. Such systems are considered to be close to a ferromagnetic ordering transition in the Stoner sense, so that their susceptibility may be approximated by $\chi = \chi_0/(1 - U\chi_0)$, where $\chi_0$ is the `bare' susceptibility at zero momentum and $U$ is a Hubbard matrix element assumed to be large since $U \chi_0 \simeq 1$. Physically this means a spin up electron traveling through the medium polarizes the spins around it ferromagnetically lowering the system's energy. The spin triplet pairing interaction for such a correlated electron gas is therefore attractive, while the singlet interaction turns out to be repulsive~\cite{BerkSchrieffer}. The ``exchanged'' excitations in such a picture are not well-defined collective modes such as phonons or magnons, but rather ``paramagnons'' defined by the existence of a peak-like structure in the imaginary part of the small-$\q$ susceptibility~\cite{ScalapinoSF}.

While there are many types of spin fluctuation theories, they share more commonalities than differences. Indeed, in the singlet channel exchange of spin fluctuations always leads to a repulsive interaction, and therefore can only result in a superconducting states with the sign-changing gap. If this interaction is sufficiently strong at some particular momentum it will necessarily lead to the superconductivity. 
In the context of heavy fermion systems it was realized~\cite{ScalapinoHF,VarmaHF} that strong AFM spin fluctuations in either the weak or strong coupling limit lead naturally to the spin-singlet $d$-wave pairing~\cite{Scalapino1995}.

\subsection{Simple pairing theory}
\label{subsec:sconeband}

\begin{figure}[t]
 \centering
 \includegraphics[width=\textwidth]{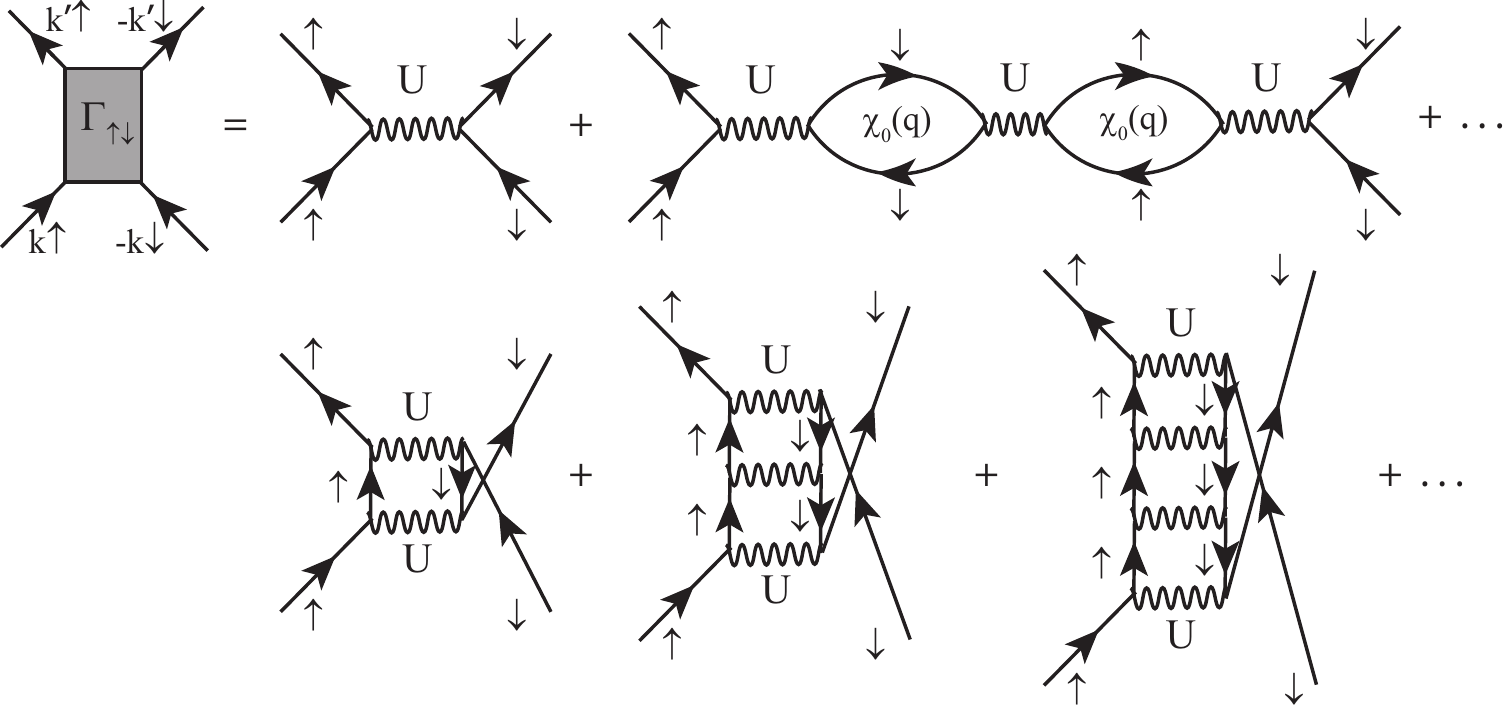}
 \caption{Cooper vertex $\Gamma_{\su\sd}$ for a singlet superconducting state in the RPA. `Bubble' and `ladder'-type diagrams up to a fourth order in the interaction $U$ are shown. Labels for momenta of incoming and outgoing lines are the same as shown on the left side here.}
 \label{fig:sfpairing}
\end{figure}

Before moving to the description of the multiorbital variant of the theory, let us describe how the spin-fluctuation theory of pairing is constructed in the single-band case with a Hubbard interaction of the type
\beq
 H = \sum\limits_f U n_{f\su} n_{f\sd},
\eeq
where $U$ is the on-site Coulomb (Hubbard) repulsion and $n_{f\sigma}$ is the number of particles operator on the site $f$ with a spin $\sigma$. The superconducting interaction in the singlet channel is determined by the Cooper vertex $\Gamma_{\su\sd}$, which in the spirit of the Berk-Schrieffer theory~\cite{BerkSchrieffer,ScalapinoSF,ScalapinoHF}, is given by the RPA series shown in Fig.~\ref{fig:sfpairing}. The basic element is again the `bare' susceptibility~(\ref{eq:chi0}). The sum of bubbles and ladders yields
\bea \label{eq:Gamma1band}
 \Gamma_{\su\sd} &=& U (1 + U^2 \chi_0^2 + ...) + U^2 \chi_0 (1 + U \chi_0 + ... ) = \frac{U}{1 - U^2 \chi_0^2} + \frac{U^2 \chi_0}{1 - U \chi_0}\\
 &=& \frac{3}{2} U^2 \chi_s - \frac{1}{2} U^2 \chi_c + U,
\eea
where $\chi_{s}$ and $\chi_{c}$ are the spin and charge susceptibilities, respectively:
\beq
 \chi_{s} = \frac{\chi_0}{1 - U \chi_0}, \chi_{c} = \frac{\chi_0}{1 + U \chi_0}.
\eeq

A magnetic instability develops in the system if the Stoner criterion is fulfilled, $1 = U \chi_0(\q,\omega=0)$.  The ferromagnetic instability corresponds to $\q = 0$; the AFM instability, which we are interested in, appears at the antiferromagnetic wave
vector $\q = \Q$. If we avoid the development of the instability, e.g., via doping, then the long-range order does not appear, but the product $U \chi_0(\q,\omega=0)$  will be close to unity, thus leading to a large magnitude of the spin susceptibility $\chi_{s}$ and, correspondingly, to its sizeable contribution to the Cooper vertex $\Gamma_{\su\sd}$. However, unlike the electron-phonon attractive interaction in the Bardeen-Cooper-Schrieffer (BCS) theory, $\Gamma_{\su\sd}$ results in the effective \textit{repulsive} interaction $V(\k,\k')$. Writing down the Hamiltonian of the system in terms of the mean-field theory and explicitly separating the superconducting interaction, we have
\beq
 H = \sum_{\k,\sigma} \eps_\k a_{\k\sigma}^\dag a_{\k\sigma} + \frac{1}{2} \sum_{\k,\k',\sigma} V(\k-\k') a_{-\k\sigma}^\dag a_{\k -\sigma}^\dag a_{\k'\sigma} a_{\k' -\sigma},
\eeq
where $a_{\k\sigma}^\dag$ is the creation operator of an electron with a momentum $\k$ and spin $\sigma$. Then gap equation takes the form
\beq
 \Delta_\k(T) = - \sum\limits_{\k'} \frac{V(\k-\k')}{2 E_{\k'}} \Delta_{\k'}(T) \tanh{\frac{E_{\k'}}{2T}},
 \label{eq:delta}
\eeq
where $E_{\k} = \sqrt{\eps_{\k}^2 + \Delta_{\k}^2}$. In the case of electron-phonon
interaction with a coupling constant $g_\mathrm{e-ph}$ in the BCS theory, we have $V(\k-\k') = - g_\mathrm{e-ph}^2$ and equation~(\ref{eq:delta}) has the solution $\Delta_\k = \Delta_0(T)$. This corresponds to the uniform $s$-wave gap. In FeBS, the orbital fluctuations enhanced by electron-phonon interaction can lead to a sign-constant solution, which in the multiband case is called the $s_{++}$ state~\cite{Onari,Kontani}. On the other hand, for the spin-fluctuation interaction we have $V(\k-\k') > 0$ and the uniform $s$-wave solution does not satisfy equation~(\ref{eq:delta}). In the case of spin fluctuations, $V(\k-\k')$ has a maximum at the wave vector $\Q$, and if we use a rough approximation, $V(\k-\k') = |\Lambda| \delta(\k-\k'+\Q)$, then equation~(\ref{eq:delta}) takes the form
\beq
 \Delta_\k(T) = - |\Lambda| \frac{\Delta_{\k+\Q}(T)}{2 E_{\k+\Q}} \tanh{\frac{E_{\k+\Q}}{2T}}.
\eeq
It is obvious that the last equation has a solution if $\Delta_\k$ and $\Delta_{\k+\Q}$ have different signs. In the simplest case of
\beq
 \Delta_\k = - \Delta_{\k+\Q}
 \label{eq:gapchangesign}
\eeq
the equation acquires the form
\beq
 1 = |\Lambda| \frac{1}{2 E_{\k+\Q}} \tanh{\frac{E_{\k+\Q}}{2T}}.
\eeq
The solution defines a gap that reverses sign at the wave vector $\Q$. If this vector connects different bands of the quasiparticles (Fermi surfaces belonging to different bands) that is realized in FeBS, then the solution of this type with an $A_{1g}$ symmetry is called the $s_\pm$ state~\cite{Mazin2008}. The competing states will be those with a $B_{1g}$ and a $B_{2g}$ symmetries, namely, having the $d_{xy}$ and $d_{x^2-y^2}$ types of the order parameter.

Overdoped cuprates can serve us here as a prototypical example of an unconventional pairing due to the spin fluctuations in the case of a large Fermi surface. For the cuprates, susceptibility $\chi$ is peaked at $\Q \simeq (\pi,\pi)$, and the two possible states of this type, which involve pairing on nearest neighbor bonds only, are
\beq
 \Delta_\k^{d,s} = \Delta_0 (\cos k_x \mp \cos k_y).
\eeq
Which state will be stabilized then depends on the Fermi surface in question. So we need to use the fact that the states close to the Fermi surface are the most important in Eq.~(\ref{eq:delta}), and examine the pairing kernel for these momenta. For example, for a $(\pi,0) \to (0,\pi)$ scattering, $\Delta_\k^s$ satisfies Eq.~(\ref{eq:gapchangesign}) by being zero, whereas $\Delta_\k^d$ is nonzero and changes sign, contributing to the condensation energy. It should therefore not be surprising that the end result of a complete numerical evaluation of Eq.~(\ref{eq:delta}) over a large cuprate Fermi surface gives $d$-wave pairing.

Eq.~(\ref{eq:Gamma1band}) for the pairing vertex in the singlet channel can be solved together with equations for the renormalization of the electronic band structure due to the scattering on the spin fluctuations. If this is done self-consistently and spin fluctuations are treated in the RPA, it is called fluctuation-exchange approximation (FLEX). Single-band FLEX~\cite{n_bickers_89,Bickers1989} employs sum of all ladder graphs for the generating functional self-consistently valid for intermediate strength of the correlations. The FLEX equations for the single-particle Green's function $G$, the self-energy $\Sigma$, the effective interaction $V$, the bare ($\chi^0$) and renormalized spin ($\chi^s$) and charge ($\chi^c$) susceptibilities are
\begin{eqnarray}
 G_\k(\omega_n) &=& \left[ \omega_n - \eps_\k + \mu - \Sigma_\k(\omega_n) \right]^{-1}, \\
 \Sigma_\k(\omega_n) &=& \frac{T}{N} \sum\limits_{\p, m} V_{\k-\p}(\omega_n - \omega_m) G_\p(\omega_m), \\
 V_\q(\nu_m) &=& U^2 \left[ \frac{3}{2} \chi^s_\q(\nu_m) + \frac{1}{2} \chi^c_\q(\nu_m) - \chi^0_\q(\nu_m) \right], \\
 \chi^0_\q(\nu_m) &=& - \frac{T}{N} \sum\limits_{\k, n} G_{\k+\q}(\omega_n + \nu_m) G_\k(\omega_n), \\
 \chi^{s,c}_\q(\nu_m) &=& \frac{\chi^0_\q(\nu_m)}{1 \mp U \chi^0_\q(\nu_m)},
\end{eqnarray}
where $\omega_n = \ii \pi T (2n+1)$ and $\nu_m = \ii \pi T (2m)$ are fermionic and bosonic Matsubara frequencies, respectively, and $\eps_\k$ is the bare band dispersion. In the last equation the '$-$' sign in the denominator corresponds to the $\chi^{s}_\q(\nu_m)$, while the '$+$' sign corresponds to the $\chi^{c}_\q(\nu_m)$. FLEX equations are solved numerically for a particular lattice partition and number of $\omega$-points in the wide energy range.

FLEX was applied to the case of the one-band Hubbard model for cuprates~\cite{Lenck1994,Monthoux1995,Dahm1995,Langer1995,Grabowski1996,Altmann2000,Manske2003}, as well as generalized for the multiband case, see, e.g., Refs.~\cite{Esirgen1998,Takimoto2004}. The theory even can explain the shape of the superconducting dome~\cite{Grabowski1996}. Other advantages include description of the low-energy kink feature seen in ARPES as having pure electronic origin and demonstration of the spin resonance asymmetry for electron and hole doped cuprates~\cite{Manske2003}. The main disadvantage, if one discuss the case of cuprates, is the lack of strong electronic correlations in the whole scheme of FLEX. As is known for the cuprates, the Mott-Hubbard physics significantly affects normal state properties. Considering this, while FLEX results looks sometimes very convenient, they have to be regarded as some approximation to the theory of superconductivity in cuprates.

\subsection{Multiorbital systems}
\label{subsec:scmultiband}

Now I am going to discuss the superconducting instability in the multiorbital system. To simplify the task, we consider the system without the spin-orbit coupling. The Cooper vertex $\Gamma_{\su\sd}$ has to be calculated in the normal phase, where there are no anomalous Green's functions. The expression for the $\ppm$ component of the spin susceptibility~(\ref{eq.chipmmu}) in the absence of the SO interaction takes the form
\beq \label{eq.chipmmunorm}
 \chi^{ll',mm'}_{0,+-}(\q,\ii\Omega) = -T \sum\limits_{\omega_n, \p, \mu,\nu} \varphi^{\mu}_{\p m} {\varphi^*}^{\mu}_{\p l} G_{\mu \su}(\p,\ii\omega_n) G_{\nu \sd}(\p+\q,\ii\Omega+\ii\omega_n) \varphi^{\nu}_{\p+\q l'} {\varphi^*}^{\nu s'}_{\p+\q m'}.
\eeq

The Cooper vertex in the multiorbital case is similar to that in the single-band case~(\ref{eq:Gamma1band}),
\beq \label{eq:GammaOrb}
 \Gamma_{\su\sd}^{l_1 l_2 l_3 l_4}(\k,\k',\omega) = \left[ \frac{3}{2} \hat{U_s} \hat\chi_s(\k-\k',\omega) \hat{U_s} - \frac{1}{2} \hat{U_c} \hat\chi_c(\k-\k',\omega) \hat{U_c} + \frac{1}{2} \hat{U_s} + \frac{1}{2} \hat{U_c} \right]_{l_1 l_2 l_3 l_4},
\eeq
where $\hat\chi_{s,c} = \left( \hat{1} \mp \hat\chi_{0} \hat{U}_{s,c} \right)^{-1} \hat\chi_{0}$ is the spin ($s$) and charge ($c$) RPA susceptibilities, $\hat{U}_{s,c}$ are the interaction matrices in the spin and charge channels~\cite{Graser2009}, and $l_1$ to $l_4$ are the orbital indices.

Hubbard interaction~(\ref{eq:Hint}) is local in the orbital basis that makes it easier to calculate susceptibility and Cooper vertex in this basis. However, gap equations are easier to solve in the band basis (especially near the Fermi surface), therefore, we transform the Cooper vertex into a band basis via matrix elements $\varphi^{\mu}_{\k l}$ defined by Eq.~(\ref{eq:varphi}),
\beq \label{eq:GammaBand}
 \Gamma^{\mu\nu}(\k,\k',\omega) = \sum\limits_{l_1,l_2,l_3,l_4} \varphi^{\mu*}_{\k l_2} \varphi^{\mu*}_{-\k l_3} \Gamma_{\su\sd}^{l_1 l_2 l_3 l_4}(\k,\k',\omega) \varphi^{\nu}_{\k' l_1} \varphi^{\nu}_{-\k' l_4}.
\eeq
Calculations show that $\Gamma^{\mu\nu}$ rapidly decreases with increasing $\omega$ in the range of frequencies that are much lower than the bandwidth. Although the equation for the superconducting gap depends on $\mathrm{Im}\Gamma^{\mu\nu}$, the momenta $\k$ and $\k'$ making the main contribution to the pairing should correspond to the small frequencies at which these momenta appears to be close to the Fermi surface. Similarly to the case where the coupling constant for the electron-phonon interaction is determined by the frequency integral of the Eliashberg function $\alpha^2 F(\omega)$, using the Kramers-Kronig relationship, we obtain
\beq
 \int\limits_0^\infty d\omega \frac{\mathrm{Im}\Gamma^{\mu\nu}(\k,\k',\omega)}{\omega} = \mathrm{Re}\Gamma^{\mu\nu}(\k,\k',\omega = 0) \equiv \tilde\Gamma^{\mu\nu}(\k,\k').
\eeq
Thus, the problem of the effective pairing interaction calculation reduces to finding the real part of $\Gamma^{\mu\nu}$ at the zero frequency which substantially simplifies further calculations.

If we represent the order parameter $\Delta_\k$ as a product of the amplitude $\Delta_0$ and the angular part $g_\k$, we can determine the dimensionless coupling $\Lambda$ as a result of the eigenvalue problem solution with the eigenvalues $\Lambda$ and eigenvectors $g_\k$, see Ref.~\cite{Graser2009}:
\beq \label{eq:lambda}
 \Lambda g_\k = -\sum\limits_\nu \oint\limits_\nu \frac{d\k'_{||}}{2\pi} \frac{1}{2\pi v_{F\k'}} \tilde\Gamma^{\mu\nu}(\k,\k') g_{\k'},
\eeq
where $v_{F\k}$ is the Fermi velocity, the contour integral is taken over the parallel to the $\nu$-th Fermi surface component of momenta $\k'_{||}$, and the band $\mu$ is unambiguously determined by which of the Fermi surfaces the momentum $\k$ belongs to. Positive $\Lambda$'s correspond to attraction and the maximal one of them represents the state with the highest $T_c$, i.e., the most favorable pairing symmetry with the corresponding gap function determined by $g_\k$. By arranging $\Lambda$'s in descending order, we can determine which symmetries and gap structures are most favorable and which will be competing with each other.

\begin{figure}[t]
 \centering
 \includegraphics[width=\textwidth]{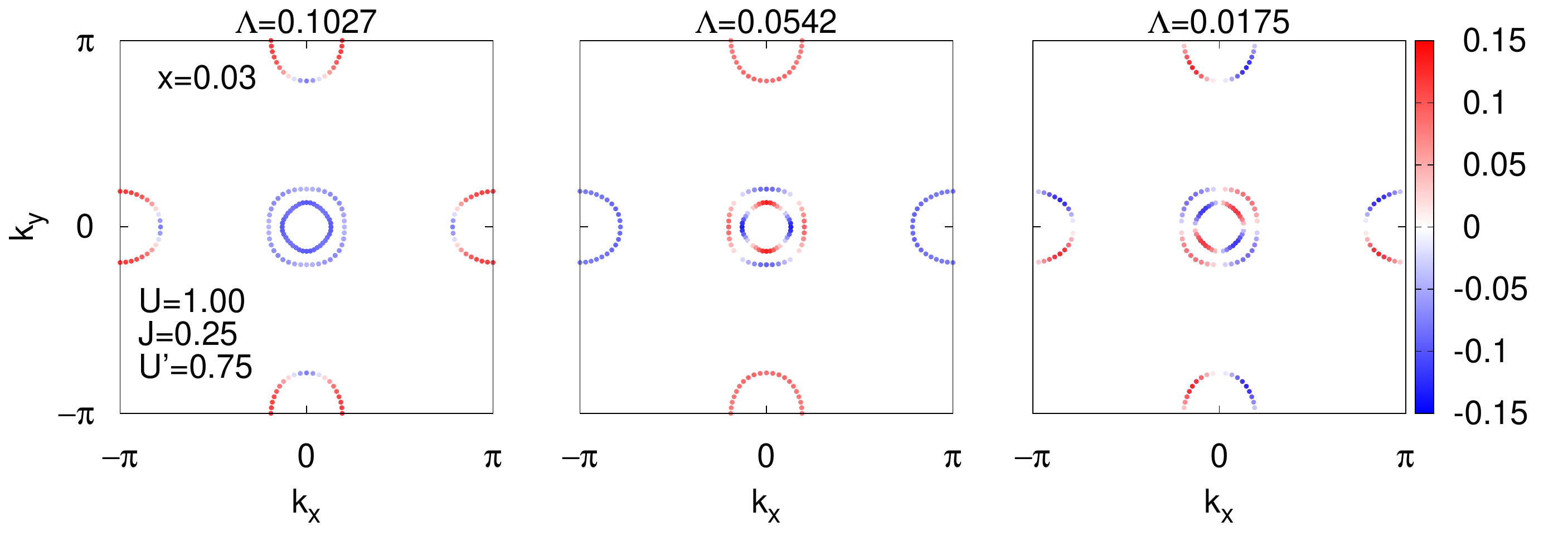}
 \includegraphics[width=\textwidth]{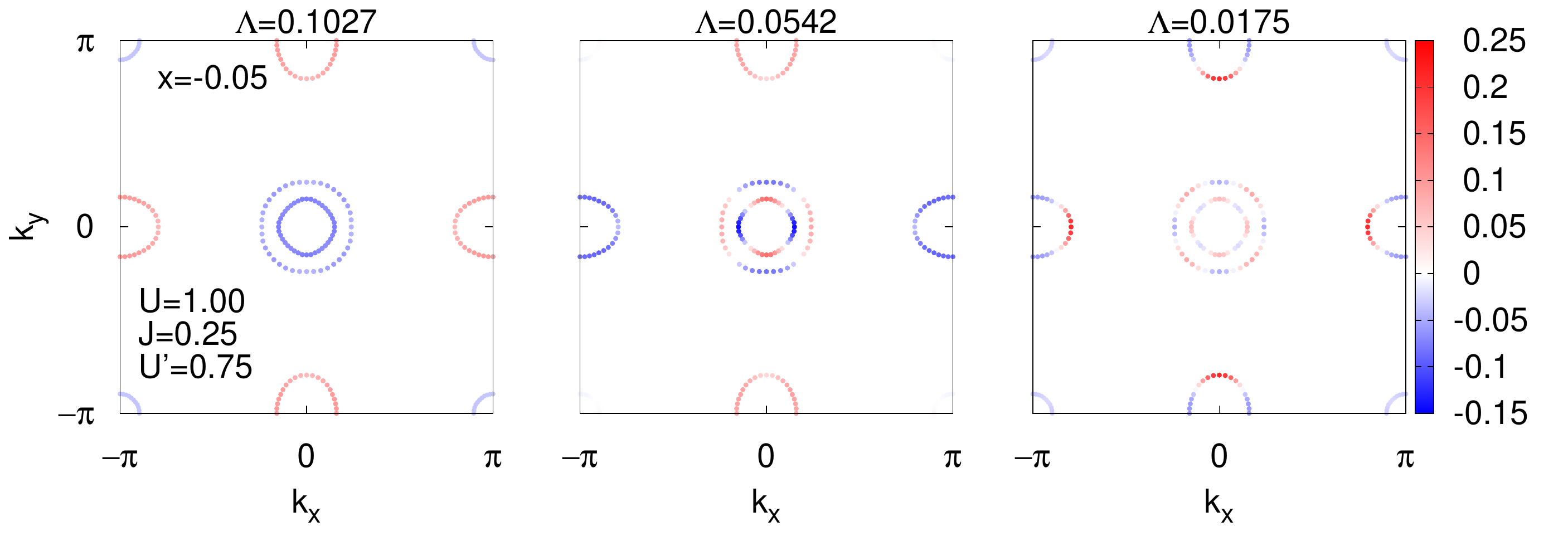}
 \caption{Angular part $g_\k$ of the gap $\Delta_\k$ for three largest values of the dimensionless coupling $\Lambda$ for two different dopings. Top (bottom) row is for $x=0.03$ ($x=-0.05$). Hubbard parameters are $U=1$, $U'=0.75$, and $J=J'=0.25$ (values are given in eV).}
 \label{fig:sfpairing_gk}
\end{figure}

In Fig.~\ref{fig:sfpairing_gk} we show results of solving Eq.~(\ref{eq:lambda}) for one particular set of interaction parameters and for two doping concentrations. Electron doping here shows exemplary situation: largest $\Lambda$ corresponds to nodal $s_\pm$ gap, next one represents the typical $d_{x^2-y^2}$ state, and the third one is the example of $d_{xy}$ gap. For the hole doping, we see the `isotropization' of the $s_\pm$ gap on electron sheets~\cite{Kemper2010}. For the smaller $\Lambda$, the order parameter has a $d_{x^2-y^2}$ symmetry and its nodal lines cross the $(\pi,\pi)$ pocket rendering the gap magnitude on it vanishingly small.

Both spin-fluctuation theories~\cite{Graser2009,Kemper2010,Kuroki2008} with their self-consistent generalizations in the FLEX approximation~\cite{Ikeda2008,Ikeda2010,ZhangPRB2009} and the renormalization group (RG) analysis~\cite{Chubukov2008,Thomale2011} are quite demanding numerical methods. But since it is the amplitude of scattering in the particle-particle channel on the Fermi surface that is important for the pairing, the angular dependence of this amplitude can be expanded in terms of the same harmonics as the expansion of $\Delta_\k$. Such a method, which is called LAHA (lowest angular harmonics approximation), makes it possible to describe pairing in FeBS for a wide range of dopings using a limited set of parameters and without doing complicated calculations~\cite{MaitiKorshunovPRL2011,MaitiKorshunovPRB2011,MaitiKorshunov2012}. The main assumption of the LAHA is the fact that the Cooper vertex $\tilde\Gamma^{\mu\nu}(\k,\k')$ can be factorized in momenta $\k$ and $\k'$,
\beq
 \tilde\Gamma_\eta(\k,\k') = \sum\limits_{m,n} C_{mn}^\eta \Psi_m^\eta(\k) \Psi_n^\eta(\k'),
\eeq
where index $\eta$ corresponds to the symmetry group of the order parameter, $C_{mn}^\eta$ are coefficients, and the function $\Psi$ makes up the expansion in terms of angular harmonics. The expansions, depending on $\eta$, have different functional forms. For example, $\Psi_m^{A_{1g}}(\k) = a_m + b_m \cos{4 \phi_\k} + c_m \cos{8 \phi_\k} + \ldots$ for the $A_{1g}$ representation, and $\Psi_m^{B_{1g}}(\k) = a_m^* \cos{2 \phi_\k} + b_m^* \cos{6 \phi_\k} + c_m^* \cos{10 \phi_\k} + \ldots$ for the $B_{1g}$ representation.

Now the problem is reduced to finding a function $\tilde\Gamma^{\alpha \alpha'}_\eta$, where $\alpha$ and $\alpha'$ correspond to Fermi surface sheets. These indices are unambiguously defined by positions of $\k$ and $\k'$. In Fig.~\ref{fig:FS}, for the electron dopong there are two hole and two electron pockets, which we denote as $\alpha_{1,2}$ and $\beta_{1,2}$, respectively. For $d_{x^2-y^2}$ and the extended $s$ wave components, one can write down the following expressions:
\bea
 \tilde\Gamma^{\alpha_i \alpha_j} &=& U_{\alpha_i \alpha_j} + \tilde{U}_{\alpha_i \alpha_j} \cos 2\phi_i \cos 2\phi_j, \nn\\
 \tilde\Gamma^{\alpha_i \beta_1} &=& U_{\alpha_i \beta} (1 + 2 \gamma_{\alpha_i \beta} ~\cos 2\theta_1) + \tilde{U}_{\alpha_i \beta} (1 + 2{\tilde\gamma}_{\alpha_i \beta} \cos 2\theta_1) \cos 2\phi_i, \nn\\
 \tilde\Gamma^{\beta_1 \beta_1} &=& U_{\beta\beta} \left[1 + 2 \gamma_{\beta\beta}(\cos 2\theta_1 + \cos 2\theta_2) + 4\gamma'_{\beta\beta} \cos 2\theta_1 \cos 2\theta_2 \right] \nn\\
 &+& {\tilde U}_{\beta\beta} \left[1 + 2 {\tilde \gamma}_{\beta\beta} (\cos 2\theta_1 + \cos 2\theta_2) + 4 {\tilde \gamma}'_{\beta\beta} \cos 2\theta_1 \cos 2\theta_2 \right],
\eea
where $U_{ij}$ and ${\tilde U}_{ij}$ are the interactions in the $s$ and $d$ channels, respectively, $\gamma_{\alpha_i \beta}$, ${\tilde\gamma}_{\alpha_i \beta}$, $\gamma_{\beta \beta}$, $\gamma'_{\beta\beta}$, ${\tilde \gamma}_{\beta\beta}$, ${\tilde \gamma}'_{\beta\beta}$ determine the degree of interaction anisotropy, $\phi_i$ and $\theta_i$ are the angles on the hole and electron Fermi surfaces counted off from the $k_x$-axis. The equation for the gap is now reduces to the $4 \times 4$ matrix equation that can be easily solved. Coefficients $C_{mn}^\eta$ and all $a$, $b$, etc. entering the expansion of $\Psi$, can be obtained from a comparison with the calculation of the total $\tilde\Gamma^{\mu\nu}(\k,\k')$ using Eqs.~(\ref{eq:GammaOrb}) and~(\ref{eq:GammaBand}). A comparison of the results for the gap has shown that the LAHA reproduces the RPA results quite well, see Ref.~\cite{MaitiKorshunovPRB2011} for details.

\begin{figure}[t]
 \centering
 \includegraphics[width=\textwidth]{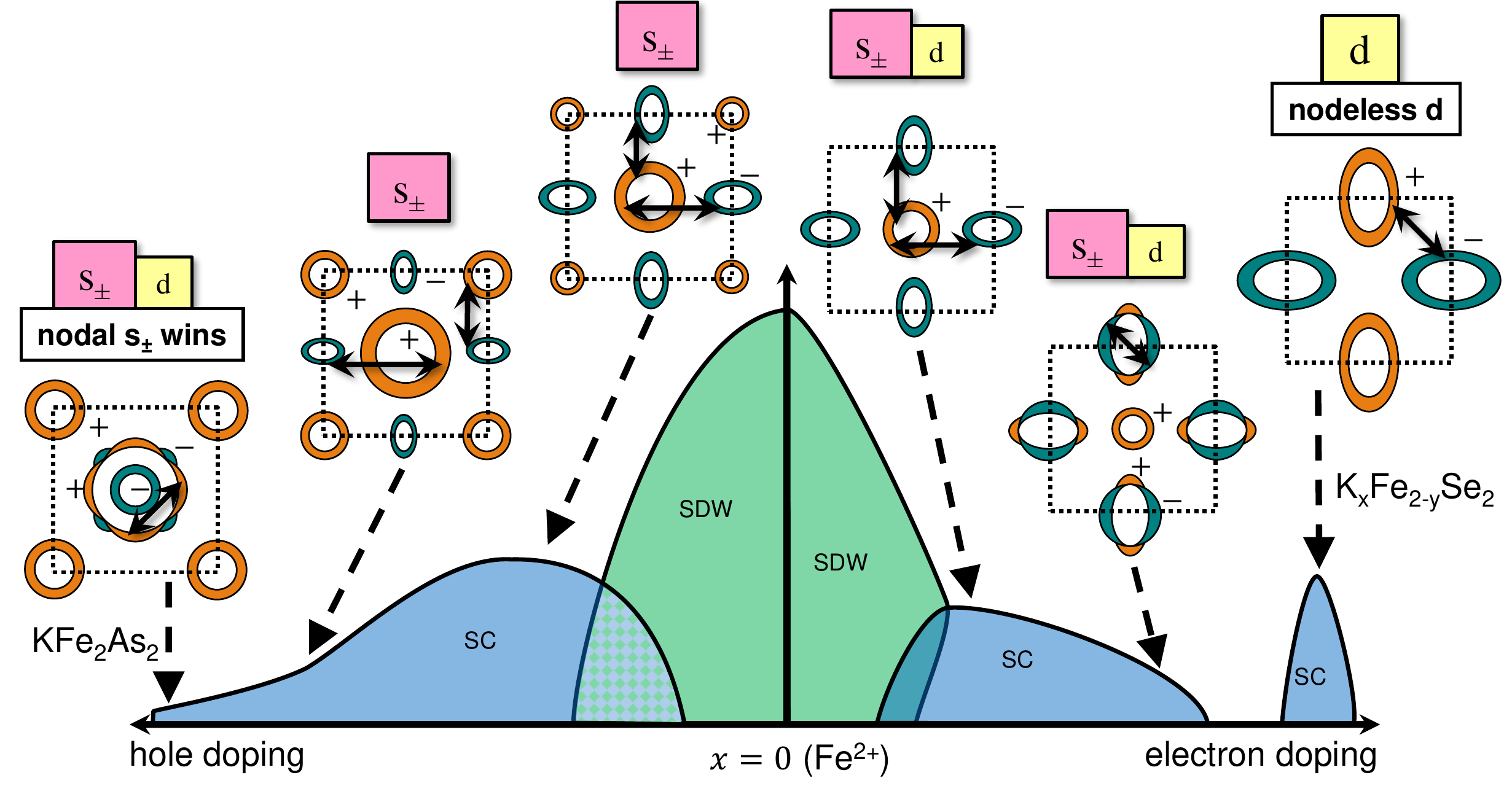}
 \caption{Schematic phase diagram of iron compounds for both hole and electron dopings. The coexistence of AFM (SDW) and superconducting (SC) phases appears on a microscopic level for the case of electron doping, and on the macroscopic level (division into SDW and SC domains) upon hole doping. The qualitative picture of the symmetries of the superconducting parameter that follows from the spin-fluctuation theory~\cite{Graser2009,Kemper2010,HirschfeldKorshunov2011,Korshunov2014eng} and from the LAHA~\cite{MaitiKorshunovPRB2011,MaitiKorshunov2012} for the two-dimensional system is shown on schematical Fermi surfaces in the insets above the phase diagram. Captions ($s_\pm$, $d$) mark the dominant and subdominant symmetries of pairing. Solid lines with an arrow at both ends ($\leftrightarrow$) indicate the dominant interaction at the Fermi surface.}
 \label{fig:phasediag}
\end{figure}

One of the LAHA advantages is the possibility of varying the effective interaction parameters $U_{ij}$ and ${\tilde U}_{ij}$, thereby determining to which extent each solution for the gap is stable. In this fermiological picture, one can determine exactly which effective interaction leads to the pairing.

Fig.~\ref{fig:phasediag} shows a schematic phase diagram and Fermi surfaces for various dopings. Depending on the topology and relative volumes of the hole and electron pockets, a competition between the $s_\pm$ and $d$ states appears. However, it is the $s_\pm$ state that always wins in the presence of both electron and hole pockets. The dominant interactions $U_{ij}$ and ${\tilde U}_{ij}$ that were obtained from the analysis of the LAHA results are shown by arrows connecting the particles on Fermi surfaces. It is easy to see that in the case of low doping the strongest interaction $U_{\alpha_i\beta}$ is between the electron and hole pockets, and $s_\pm$ state is the dominating one. Upon electron doping, the repulsion $U_{\beta\beta}$ inside the electron pocket is large, and it is best for the system to form a sign-changing gap on the electron pockets in order to reduce this contribution. In this case, the $s_\pm$ state has nodal lines at electron Fermi surfaces. If the electron doping is very high (as in K$_x$Fe$_{2-y}$Se$_2$), once the hole pockets disappear, the system forms $d$-type superconductivity because of the strong interaction between the electron pockets. One question remains open: whether such a state would be favorable as compared to the bonding-antibonding $s_\pm$ state~\cite{HirschfeldKorshunov2011,Mazin2011} upon the transformation to the Brillouin zone corresponding to two Fe atoms per unit cell. It seems that, because the spin-orbit interaction is present in this case~\cite{KorshunovTogushovaSO2013}, and because of the hybridization along the symmetry directions following from it, the bonding-antibonding $s_\pm$ state should be most favorable~\cite{Khodas2012}. However, as follows from the calculations in the 10-orbital model for K$_{0.8}$Fe$_{1.7}$Se$_2$ and K$_{0.85}$Fe$_{1.8}$Se$_2$, it is the pairing of the $d_{x^2-y^2}$-type that always
dominates~\cite{Kreisel2013}.

For the hole doping, on the contrary, the emergence of a new hole pocket $\gamma$ near the $(\pi,\pi)$ point leads to the stabilization of the nodeless $s_\pm$ state. This picture is greatly affected by the orbital character of the bands. Since the pocket $\gamma$ is mainly formed by the $d_{xy}$ orbital, as are the small regions on the electron pockets (see Fig.~\ref{fig:FS}), the new scattering channel between this pocket and the electron pockets leads to the `isotropization' of the gap on electron pockets. With a further hole doping, when the electron pockets disappear as in KFe$_2$As$_2$, the strong interaction inside the hole pocket $\alpha_2$ forces the system to form a sign-changing gap with nodes on this pocket. The symmetry of the gap refers, as before, to the $A_{1g}$ representation and corresponds to the $s_\pm$ state with added higher angular harmonics~\cite{MaitiKorshunov2012}. There is, however, an alternative scenario for the pairing that involves spin-orbit coupling~\cite{Vafek2017}.

\subsection{Final remarks}

We conclude that in spite of the variety of the materials the multiorbital spin fluctuation theory of pairing can explain many observed features of iron-based superconductors, in particular, the different variants of the experimentally examined behaviors of the superconducting gap. The anisotropic $s_\pm$ state and its nodal structure on Fermi surfaces are quite sensitive to some details of the electronic
structure, such as the orbital character of the bands, spin-orbit interaction, and changes in the band structure due to the doping.

\section{Spin resonance peak}
\label{sec:spinres}

Since different mechanisms of Cooper pairs formation result in different superconducting gap symmetries and structures in FeBS~\cite{HirschfeldKorshunov2011}, one way to elucidate the mechanism of pairing is to determine the details of the order parameter. For example, as discussed above, spin fluctuation approach gives the $s_{\pm}$ state as the main instability for the wide range of doping concentrations~\cite{Mazin2008,Korshunov2014eng,Graser2009,Kuroki2008,MaitiKorshunovPRB2011}, while orbital fluctuations promote the $s_{++}$ state~\cite{Kontani2011}.

\begin{figure}[t]
\centering
 \includegraphics[width=0.8\textwidth]{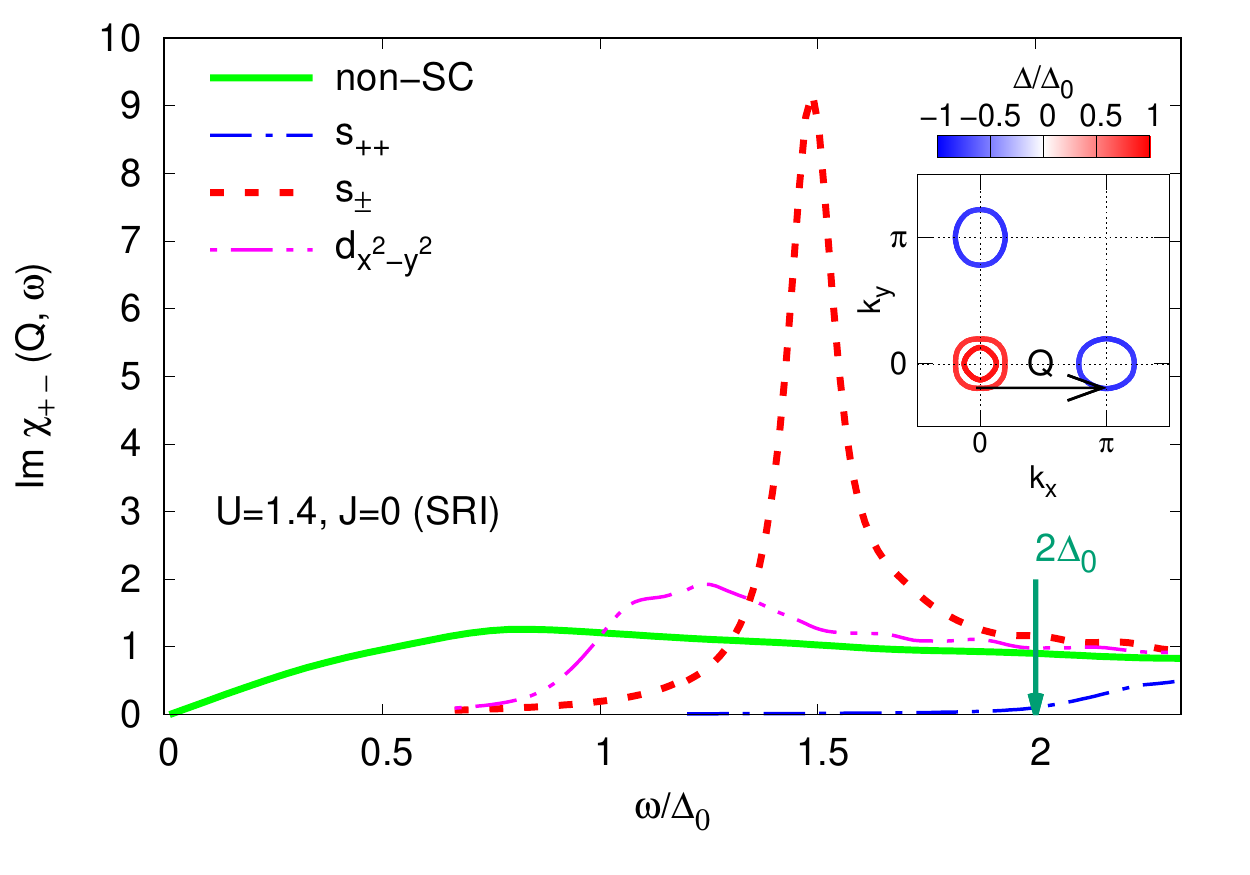}
 \caption{Frequency dependence of the spin susceptibility $\mathrm{Im}\chi(\q=\Q,\omega)$ at $x=0.05$ in the normal state (non-SC) and in the superconducting state with $s_{++}$, $d_{x^2-y^2}$, and $s_\pm$ gaps. The calculation was done for $U=1.4$eV and $J=0$ in the presence of the SRI. For the $s_\pm$ state, a resonance peak appears for $\omega < 2 \Delta_0$. Inset: $s_\pm$ gap at the Fermi surface together with the wave vector $\Q$.}
 \label{fig:spinres}
\end{figure}

Inelastic neutron scattering (INS) is a useful tool here since the measured dynamical spin susceptibility $\chi(\q,\omega)$ in the superconducting state carries information about the gap structure. There are been many reports of a well-defined peak in neutron spectra in 1111, 122, and 11 systems appearing only for $T < T_c$ at or around $\q = \Q$~\cite{LumsdenReview,ChristiansonBKFA,Inosov2010,ArgyriouKorshunov2010,Dai2015}. The common explanation is that the peak is the so-called spin resonance appearing due to the $s_\pm$ state. The reasoning goes as follows. Since $\chi_0(\q,\omega)$ describes the particle-hole excitations and since all excitations at frequencies less than about $2\Delta_0$ (at $T=0$) are absent in the superconducting state, the imaginary part $\mathrm{Im}\chi_0(\q,\omega)$ becomes finite only above $2\Delta_0$. The anomalous Green's functions entering Eq.~(\ref{eq.chipm}) give rise to terms proportional to $\left[1 - \frac{\Delta_\k \Delta_{\k+\q}}{E_{\k} E_{\k+\q}}\right]$. These are the anomalous coherence factors. At the Fermi level, one has $E_{\k} \equiv \sqrt{\eps_\k^2 + \Delta_\k^2} = |\Delta_\k|$. If $\Delta_\k$ and $\Delta_{\k+\q}$ have the same sign, the coherence factors will be equal to zero that leads to a gradual increase in the spin susceptibility with increasing frequency in the range $\omega > \Omega_c$ with $\Omega_c = \min \left(|\Delta_\k| + |\Delta_{\k+\q}| \right)$,  whereas at frequencies lower than $\Omega_c$ we have $\mathrm{Im}\chi_0(\q,\omega) = 0$.
For the $s_{++}$ state, this can be seen from Fig.~\ref{fig:spinres}, where we present results for susceptibilities at the wave vector $\q=\Q$ as functions of frequency $\omega$ obtained via analytical continuation from Matsubara frequencies ($\ii\Omega \to \omega + \ii\delta$ with $\delta \to 0+$). If, however, as in the case of $s_\pm$ and $d$ states, vector $\q = \Q$ connects the Fermi surfaces with different signs of the gap, $\sgn \Delta_\k \neq \sgn \Delta_{\k+\q}$, then the coherence factors are nonzero and a jump appears in the imaginary part of $\chi_0$ at $\omega = \Omega_c$. According to the Kramers-Kronig relations, a logarithmic singularity appears in the real part of the susceptibility. For a certain set of parameters $U$, $U'$, $J$, $J'$ entering the interaction matrix $\hat{U}$, nonzero value of $\mathrm{Re}\chi_0$ and $\mathrm{Im}\chi_0 = 0$ lead to a divergence of the imaginary part of the RPA susceptibility~(\ref{eq.rpa.chippmsol}). The corresponding peak in $\mathrm{Im}\chi(\Q,\omega)$ is called the spin resonance and appears for the frequencies $\Omega_R \leq \Omega_c$. This peak is quite pronounced for the $s_\pm$ state, see Fig.~\ref{fig:spinres}. For the $d_{x^2-y^2}$ gap symmetry (although, in principle, the resonance could arise because of the sign-changing character of the gap), the vector $\Q$ connects the states on the hole Fermi surface near the nodes of the gap $\Delta_\k$. Therefore, the total gap in $\mathrm{Im}\chi_0$ determined by $\Omega_c$, is tiny. Since, $\Omega_c \ll \Delta_0$, the jump in $\mathrm{Im}\chi_0$ is negligibly small, and the susceptibility in the RPA shows a slight increase in comparison with that for the normal state (Fig.~\ref{fig:spinres}). The same is true for $d_{xy}$ and $d_{x^2-y^2} + \ii d_{xy}$ gaps~\cite{KorshunovEreminResonance2008} and for the triplet $p$-wave pairing~\cite{Maier2008}.

Thus, the existence of a spin resonance refers to an exclusive property of the $s_\pm$ state. For iron compounds, the spin resonance was predicted theoretically~\cite{KorshunovEreminResonance2008,Maier2008}, and then revealed experimentally in the 1111, 122, and 11 families of pnictides and chalcogenides~\cite{ChristiansonBKFA,Inosov2010,ArgyriouKorshunov2010,Lumsden2009,ChristiansonBFCA,Park,Osborn,QiuFeSeTe,Babkevich}.

\subsection{Unequal gaps}

Such a simple explanation was indirectly questioned by the angle-resolved photoemission spectroscopy (ARPES) results and recent measurements of gaps via Andreev spectroscopy. Latter clearly shows that there are at least two distinct gaps present in 11, 122, and 1111 systems~\cite{Daghero2009,Tortello2010,Ponomarev2013,Abdel-Hafiez2014,Kuzmicheva2014,Kuzmichev2016,Kuzmicheva2017} and even three gaps in LiFeAs~\cite{Kuzmichev2012,Kuzmichev2013}. Larger gap ($\Delta_L$) is about 9meV and the smaller gap ($\Delta_S$) is about 4meV in BaCo122 materials. From ARPES we know that electron Fermi surface sheets and the inner hole sheet are subject to opening the lager gap while the smaller gap is located at the outer hole Fermi surface~\cite{Ding2008,Evtushinsky2009}. The very existence of the smaller gap rise the question -- what would be the spin resonance frequency in the system with two distinct gaps? Naive expectation is that the frequency shifts to the lower gap scale and $\omega_R < 2 \Delta_S$. Then the observed peak in INS in BaCo122 system at frequency $\omega_{INS} \sim 9.5$meV~\cite{Inosov2010} can not be the spin resonance since it is greater than $2\Delta_S \sim 8$meV~\cite{Tortello2010}. Thus the peak could be coming from the $s_{++}$ state~\cite{Onari2010,Onari2011}, where it forms at frequencies \textit{above} $2\Delta$ due to the redistribution of the spectral weight upon entering the superconducting state and a special form of scattering in the normal state. Below we study this question in details and show that the naive expectation is wrong and that the true minimal energy scale is $\omega_R \leq \Delta_L+\Delta_S$. 

To proceed further, we choose the following values for the interaction parameters: $U=1.4$eV, $J=0$, and make use of the spin-rotational invariance constraint $U'=U-2J$ and $J'=J$. For the $s_{++}$ state we assume a uniform gap $\Delta_{\k\mu} = \Delta_{\mu}$ and for the $s_\pm$ state the gap will be $\Delta_{\k\mu} = \Delta_{\mu} \cos k_x \cos k_y$, where $\mu$ is the band index.

\begin{figure}[ht]
 \centering
 \includegraphics[width=0.8\textwidth]{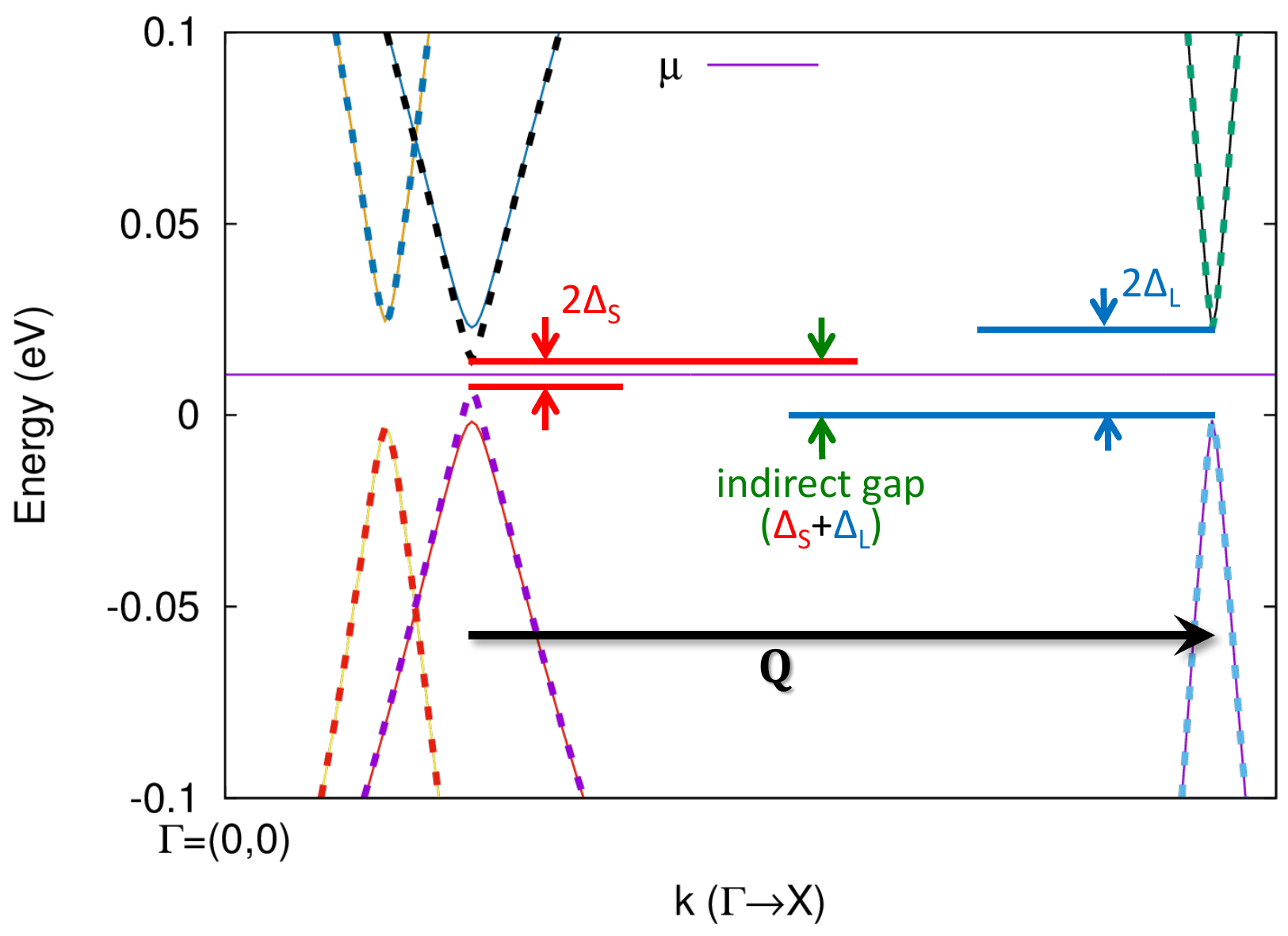}
 \caption{Energy spectrum of the five-orbital model near the Fermi level $\eps_F = \mu$ in the superconducting state, $E_{\k \nu} = \pm\sqrt{\eps_{\k \nu}^2 + \Delta_{\k \nu}^2}$, as a function of momentum $\k$ along the $\Gamma-X$ direction, i.e. $(0,0)-(\pi,0)$. Scattering wave vector $\Q$ entering the spin susceptibility is also shown.}
 \label{fig:5orbDelta}
\end{figure}

\begin{figure}[ht]
 \centering
 \includegraphics[width=0.8\textwidth]{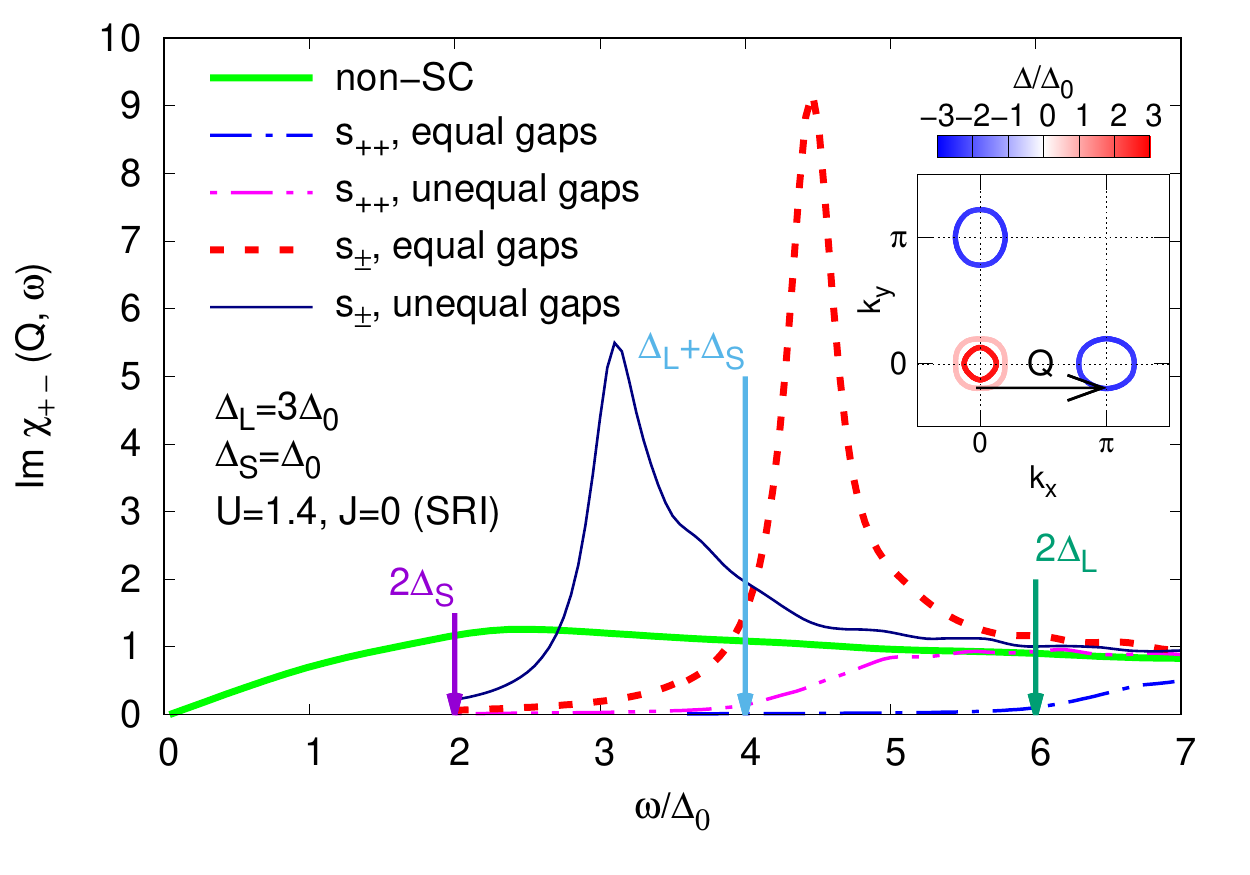}
 \caption{Calculated $\mathrm{Im}\chi_{+-}(\Q,\omega)$ with $\Q=(\pi,0)$ in the 1-Fe BZ for the five-orbital model in the normal, $s_{++}$ and $s_\pm$ superconducting states. Two cases of superconducting states are shown: equal gaps with $\Delta_{\alpha_{1,2}} = \Delta_{\beta_{1,2}} = \Delta_L$, and unequal gaps with $\Delta_{\alpha_{1,2}} = \Delta_{\beta_{1}} = \Delta_L$ and $\Delta_{\beta_2} = \Delta_S$, where $\Delta_S = \Delta_L / 3$. Latter case is shown in the inset, where gaps at the Fermi surface are plotted together with the wave vector $\Q$.}
 \label{fig:5orbImChi}
\end{figure}

\begin{figure}[ht]
 \centering
 \includegraphics[width=0.49\textwidth]{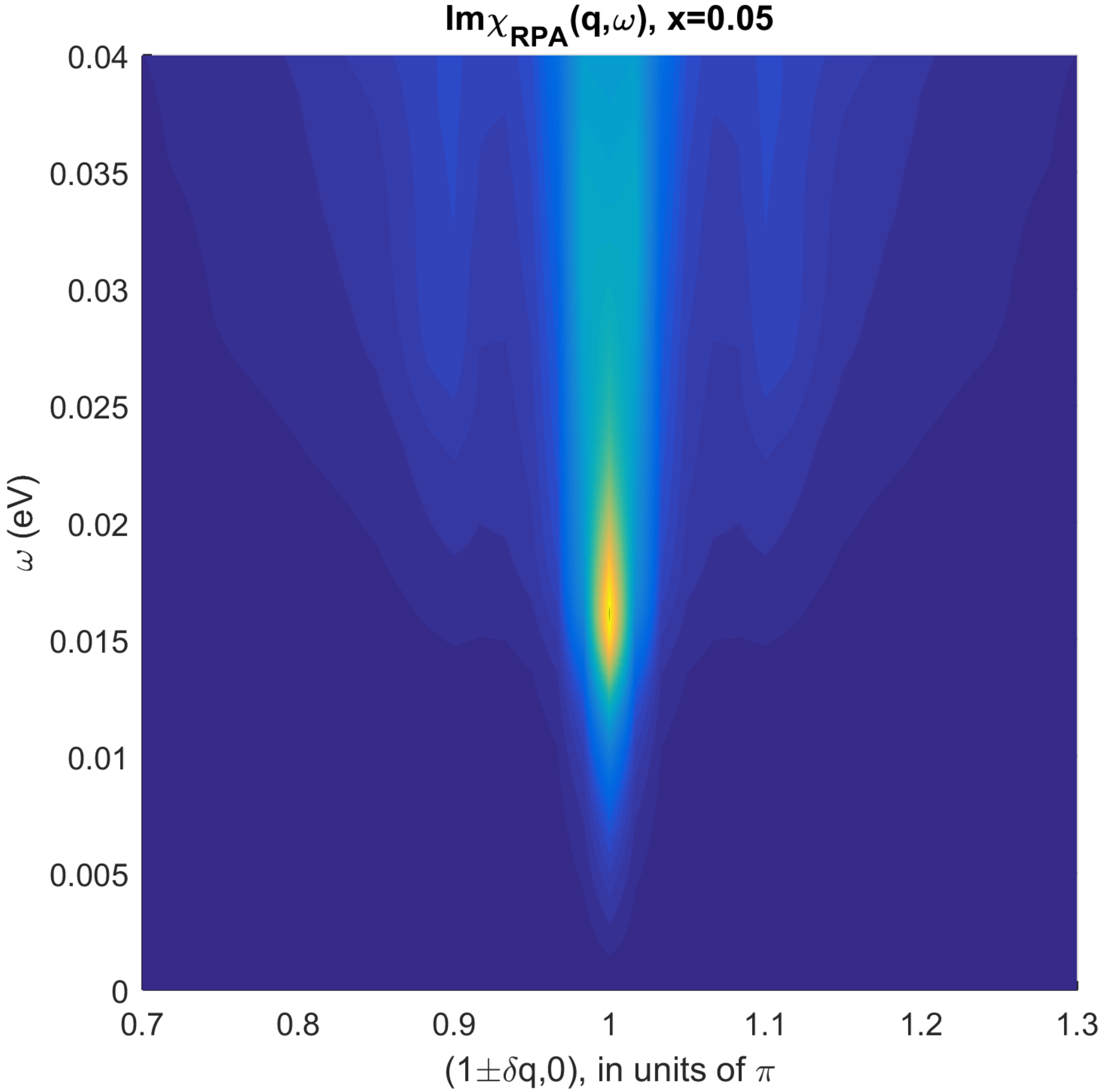}
 \includegraphics[width=0.49\textwidth]{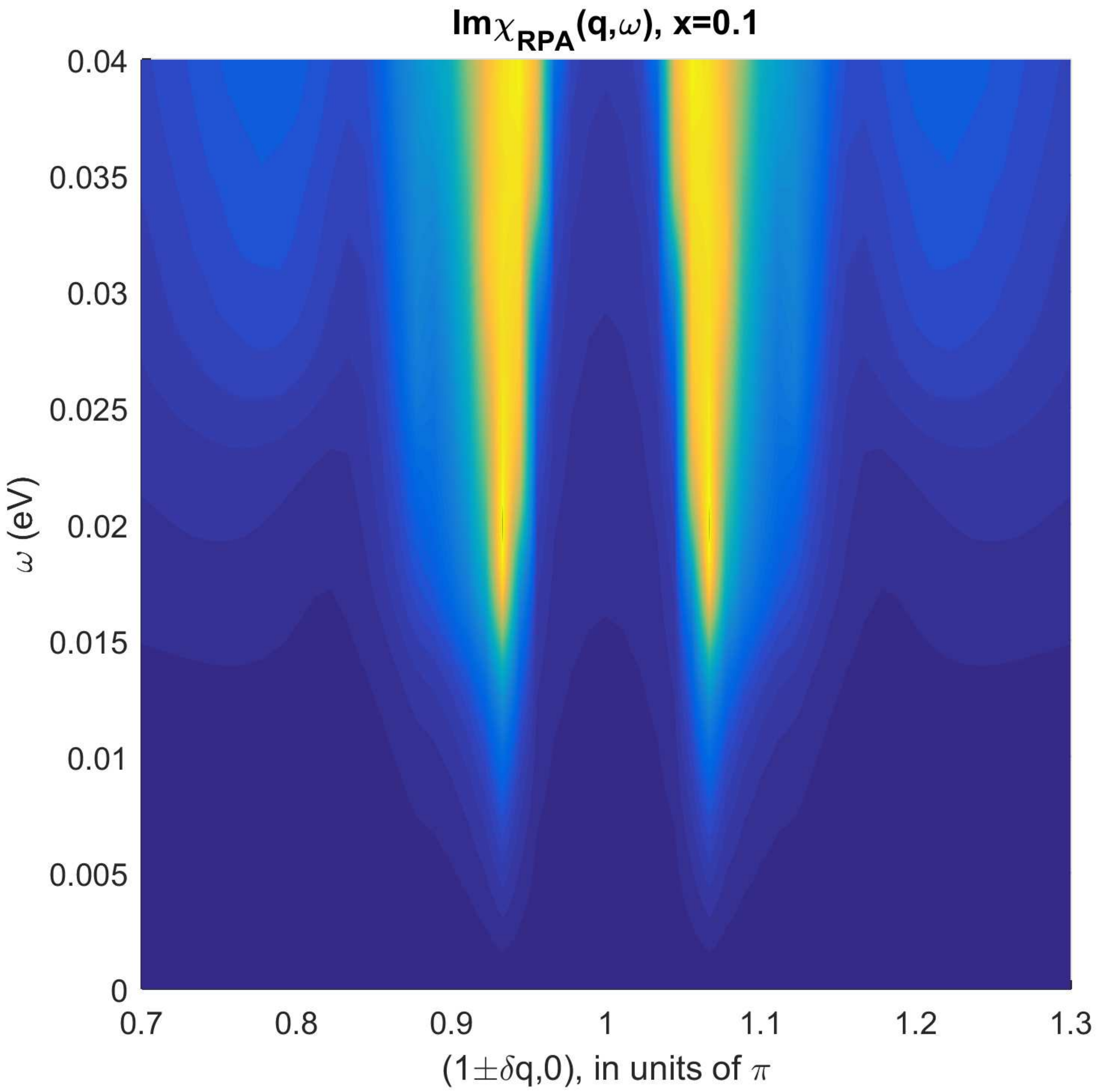}
 \caption{Frequency and momentum dependence of $\mathrm{Im}\chi_{+-}(\Q,\omega)$ around $\Q=(\pi,0)$ in the 1-Fe BZ for the five-orbital model in the $s_\pm$ states with unequal gaps, $\Delta_{\alpha_{1,2}} = \Delta_{\beta_{1}} = 3 \Delta_{\beta_2}$. Left panel is for doping $x=0.05$, and right panel is for $x=0.1$.}
 \label{fig:5orbImChiDoping}
\end{figure}

Energy spectrum of the five-orbital model~(\ref{eq:H0}) near the Fermi level in the superconducting state, $E_{\k \nu} = \pm\sqrt{\eps_{\k \nu}^2 + \Delta_{\k \nu}^2}$, is shown in Fig.~\ref{fig:5orbDelta}. We consider here the case of unequal gaps with the smaller gap $\Delta_{\beta_2} = \Delta_S$ on the outer hole Fermi surface and larger gaps $\Delta_{\alpha_{1,2}} = \Delta_{\beta_{1}} = \Delta_L$ on inner hole and electron Fermi surfaces. To be consistent with the experimental data, we choose $\Delta_S = \Delta_0 = \Delta_L / 3$, see the inset in Fig.~\ref{fig:5orbImChi}. Naturally, the two energy scales, $2\Delta_S$ and $2\Delta_L$, appear in the energy spectrum $E_{\k \nu}$ and they are connected with hole $\alpha_2$ and electron $\beta_{1,2}$ bands, respectively. On the other hand, the susceptibility $\chi_{0+-}(\Q,\omega)$ contains scattering \textit{between} hole and electron bands with the wave vector $\Q$. The energy gap that have to be overcome to excite electron-hole pair is the indirect gap with the scale $\tilde\Delta = \Delta_L + \Delta_S$. That is why spin excitations in the $s_{++}$ state start with the frequency proportional to the indirect gap $\tilde\Delta = 4\Delta_0$, see Fig.~\ref{fig:5orbImChi}. The same is true for the discontinuous jump in $\mathrm{Im}\chi_{0}$ for the $s_\pm$ state -- it shifts to frequency $\approx \tilde\Delta$. This, together with the corresponding $\log$ singularity in $\mathrm{Re}\chi_{0}$, produce the spin resonance peak in RPA at frequency $\omega_R \leq \tilde\Delta$. Such shift of resonance peak to lower frequencies compared to the equal gaps situation is seen in Fig.~\ref{fig:5orbImChi}, where the spin response $\mathrm{Im}\chi_{+-}(\Q,\omega)$ for the cases of equal and distinct gaps is shown.

The changes in the band structure and/or doping level can result in the change of the indirect gap. In particular, since for the hole doping hole Fermi surfaces become larger the wave vector $\Q$ may connect states on the electron Fermi surface and on the \textit{inner} hole Fermi surface. Gaps on both these Fermi surfaces are determined by $\Delta_L$ and thus the indirect gap will be equal to $\tilde\Delta = 2\Delta_L$. This sets up a maximal energy scale for the spin resonance, i.e. $\omega_R \leq 2\Delta_L$.

Thus we conclude that depending on the relation between the wave vector $\Q$ and the exact Fermi surface geometry, the indirect gap in most FeBS can be either $\tilde\Delta = \Delta_L + \Delta_S$ or $\tilde\Delta = 2\Delta_L$. The peak in the dynamical spin susceptibility at the wave vector $\Q$ will be the true spin resonance if it appears below the indirect gap scale, $\omega_R \leq \tilde\Delta$.

Note that with doping, the spin resonance peak becomes incommensurate because the wave vector at which the susceptibility is maximal becomes incommensurate due to the changes in the relative shapes and sizes of hole and electron sheets. It is clearly seen in Fig.~\ref{fig:5orbImChiDoping}, where we show results for two dopings.

\subsection{Analysis of available experimental data}

Since for the Fermi surface geometry characteristic to the most of FeBS materials the indirect gap is either $\tilde\Delta = \Delta_L + \Delta_S$ or $\tilde\Delta = 2\Delta_L$, then we can derive a simple criterion to determine whether the experimentally observed peak in inelastic neutron scattering is the true spin resonance -- if the peak frequency $\omega_R$ is less than the indirect gap $\tilde\Delta$, then it is the spin resonance and, consequently, the superconducting state has the $s_\pm$ gap structure.

Sometimes it is not always clear experimentally which gaps are connected by the wave vector $\Q$. Even without knowing this exactly, one can draw some conclusions. For example, if one of the gaps is $\Delta_L$, then there are three cases possible: (i) $\omega_R \leq \Delta_L + \Delta_S$ and the peak at $\omega_R$ is the spin resonance, (ii) $\omega_R \leq 2\Delta_L$ and the peak is most likely the spin resonance but the definitive conclusion can be drawn only from the calculation of the dynamical spin susceptibility for the particular experimental band structure, and (iii) $\omega_R > 2\Delta_L$ and the peak is definitely not a spin resonance. In the latter case, the peak could be coming from the $s_{++}$ state~\cite{Onari2010,Onari2011}.

Now we can compare energy scales extracted from ARPES, Andreev spectroscopy, and inelastic neutron scattering. Latter gives peak frequency $\omega_{INS} \approx 9.5$meV in BaFe$_{1.85}$Co$_{0.15}$As$_2$ with $T_c = 25$K~\cite{Inosov2010}. For the same system, gap sizes extracted from ARPES are $\Delta_L \approx 6.7$meV and $\Delta_S \approx 4.5$meV~\cite{Terashima2009}, and for a similar system with $T_c = 25.5$K, $\Delta_L \approx 6.6$meV and $\Delta_S \approx 5$meV~\cite{Kawahara2010}. Gap sizes extracted from Andreev spectroscopy are $\Delta_L \approx 9$meV and $\Delta_S \approx 4$meV in BaFe$_{1.8}$Co$_{0.2}$As$_2$ with $T_c = 24.5$K~\cite{Tortello2010}. Evidently, $\omega_{INS} < \Delta_L + \Delta_S$ and we can safely state that the peak in INS is the spin resonance.

For the hole doped systems, peak frequency in INS is about $14$meV in Ba$_{0.6}$K$_{0.4}$Fe$_2$As$_2$ with $T_c = 38$K~\cite{ChristiansonBKFA}. There is a slight discrepancy between gap sizes extracted from ARPES and Andreev spectra. Former gives $\Delta_L \approx 12$meV and $\Delta_S \approx 6$meV in the same material with $T_c = 37$K~\cite{Ding2008}, thus $\omega_{INS} < \Delta_L + \Delta_S$. Gap sizes from Andereev spectroscopy are $\Delta_L \approx 8$meV and $\Delta_S \approx 2$meV in Ba$_{0.65}$K$_{0.35}$Fe$_2$As$_2$ with lower $T_c = 34$K~\cite{Abdel-Hafiez2014}. In this case, $\omega_{INS} > \Delta_L + \Delta_S$ but $\omega_{INS} < 2\Delta_L$ and we still can assume that the peak in INS is the spin resonance. However, in such a case definitive conclusion can be given only by the calculation of spin response for the particular experimental band structure.
%
%

\begin{figure}[ht]
 \centering
 \includegraphics[width=0.8\linewidth]{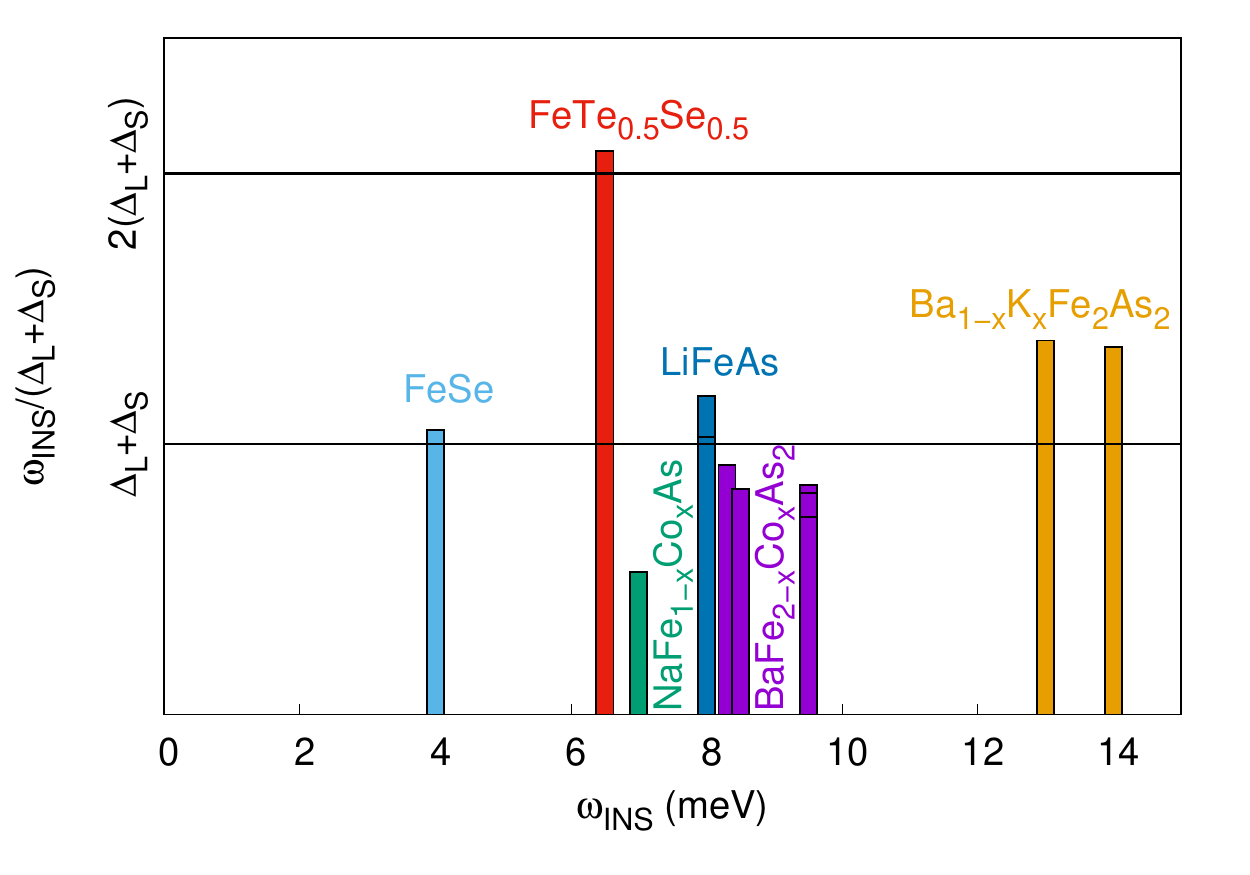}
 \caption{Data from Table~\ref{tab} grouped by materials. Each bar height is determined by $\omega_{INS}/(\Delta_L+\Delta_S)$. If it's below $\Delta_L + \Delta_S$ boundary, then case (i) is realized; case (ii) occurs once it's below $2\Delta_L$ line, and situation (iii) corresponds to the intersection of the $2(\Delta_L+\Delta_S) > 2\Delta_L$ limit.}
 \label{fig:omega_Delta_exp}
\end{figure}

To proceed further, we combine data on the peak frequency $\omega_R$ and maximal and minimal gap sizes $\Delta_L$ and $\Delta_S$ available in the literature. Results are presented in Table~\ref{tab} and illustrated in Fig.~\ref{fig:omega_Delta_exp}. Unfortunately, for many materials either the INS data or gaps estimations are absent that gives a whole set of tasks for future experiments. Here are some conclusions which we can make based on the available data:
\begin{enumerate}
  \item In electron-doped BaFe$_{1-x}$Co$_{x}$As$_2$ system, NaFe$_{1-x}$Co$_{x}$As system, and FeSe, $\omega_R < \Delta_L + \Delta_S$ and, thus the peak in INS is the true spin resonance evidencing sign-changing gap.

  \item Some hole doped Ba$_{1-x}$K$_{x}$Fe$_2$As$_2$ materials satisfy $\omega_R \leq \Delta_L + \Delta_S$ condition, and some satisfy $\omega_R < 2\Delta_L$ condition. Latter comes especially from newer tunneling~\cite{Shan2012,Shimojima2011} and Andreev reflection~\cite{Abdel-Hafiez2014} data revealing smaller gap values. The fact that $\omega_R < 2\Delta_L$ is still consistent with the sign-changing gap, but as we mentioned before, the calculation of the spin response for the particular experimental band structure is required to make a final conclusion.

  \item The only case when $\omega_{INS}>2\Delta_L$ is FeTe$_{0.5}$Se$_{0.5}$. According to our analysis, there should be no sign-changing gap structure. But before concluding this since this is the single case only, gap data coming from $\mu$SR~\cite{Biswas.PhysRevB.81.092510,Bendele.PhysRevB.81.224520} should be double checked by independent techniques.

  \item Interesting to note that ARPES in all cases gives gaps values larger than extracted from other techniques. Natural question arise -- whether the ARPES overestimates or all other methods underestimates superconducting gaps?
\end{enumerate}

\begin{table}[ht]
\caption{\label{tab} Comparison of peak energies in INS and larger and smaller gaps in various FeBS. Data on gap sizes $\Delta_L$ and $\Delta_S$ extracted from the Andreev experiments data, tunneling spectra, muon spin rotation ($\mu$SR), the BCS fit of $H_{c1}(T)$, and angle-resolved photoemission spectroscopy (ARPES) are presented. If the peak frequency and gaps satisfy condition $\omega_{INS}<\Delta_L+\Delta_S$, gaps are written in \DLS{bold face}, and if they satisfy condition $\omega_{INS}<2\Delta_L$, gaps are written in \DLL{italic}. \GDLL{Red} color is used in the case of $\omega_{INS}>2\Delta_L$.}
\begin{tabular}{|c|c|c|c|c|c|}
\hline
\centering{Material} & $T_c$ (K) & $\omega_{INS}$ (meV) & $\Delta_L$, $\Delta_S$ (meV)\\
\hline
BaFe$_{1.9}$Co$_{0.1}$As$_2$ & 19 & 8.3~\cite{Wang2016} & \DLS{5.0, 4.0} (ARPES)~\cite{Wang2016} \\
BaFe$_{1.866}$Co$_{0.134}$As$_2$ & 25 & 8.0~\cite{Wang2016} & \DLS{6.5, 4.6} (ARPES)~\cite{Wang2016} \\
BaFe$_{1.81}$Co$_{0.19}$As$_2$ & 19 & 8.5~\cite{Wang2016} & \DLS{5.6, 4.6} (ARPES)~\cite{Wang2016} \\
BaFe$_{1.85}$Co$_{0.15}$As$_2$ & 25 & 9.5~\cite{Inosov2010,Park} & \DLS{6.7, 4.5} (ARPES)~\cite{Terashima2009} \\
BaFe$_{1.85}$Co$_{0.15}$As$_2$ & 25.5 & $\sim 9.5$? & \DLS{6.6, 5} (ARPES)~\cite{Kawahara2010} \\
BaFe$_{1.8}$Co$_{0.2}$As$_2$ & 24.5 & $\sim 9.5$? & \DLS{9, 4} (Andreev refl.)~\cite{Tortello2010} \\
%
\hline
Ba$_{0.6}$K$_{0.4}$Fe$_2$As$_2$ & 38 & 14~\cite{ChristiansonBKFA,Castellan2011,Shan2012} & \DLS{12.5, 5.5} (ARPES)~\cite{Ding2008,Wray2008} \\
Ba$_{0.6}$K$_{0.4}$Fe$_2$As$_2$ & 38 & 14~\cite{ChristiansonBKFA,Castellan2011,Shan2012} & \DLS{7-11.5, 4-7} (ARPES)~\cite{Zhang2010} \\
Ba$_{0.6}$K$_{0.4}$Fe$_2$As$_2$ & 38 & 14~\cite{ChristiansonBKFA,Castellan2011,Shan2012} & \DLL{8.4, 3.2} (Tunneling)~\cite{Shan2012,Shimojima2011} \\
Ba$_{0.6}$K$_{0.4}$Fe$_2$As$_2$ & 35 & 14~\cite{Castellan2011} & \DLS{10-12, 7-8} (ARPES)~\cite{Zhao2008} \\
Ba$_{0.6}$K$_{0.4}$Fe$_2$As$_2$ & 37.5 & 14~\cite{Castellan2011} & \DLL{8.5-9.3, 1.7-2.3} ($H_{c1}$)~\cite{Ren2008} \\
Ba$_{0.65}$K$_{0.35}$Fe$_2$As$_2$ & 34 & 13~\cite{Castellan2011} & \DLL{7.4-8, 1.4-2} (Andreev spec.)~\cite{Abdel-Hafiez2014} \\
Ba$_{1-x}$K$_{x}$Fe$_2$As$_2$ & 32 & 14~\cite{Castellan2011} & \DLL{9.2, 1.1} (ARPES)~\cite{Evtushinsky2009,Evtushinsky2009NJP} \\
%
\hline
FeSe & 8 & 4~\cite{Wang.nmat4492} & \DLS{2.5, 3.5} (Tunneling)~\cite{Kasahara.PNAS.111.16309} \\
FeSe & 8 & 4~\cite{Wang.nmat4492} & \DLS{0.6-1, 2.4-3.2} (Andreev refl.)~\cite{Ponomarev2013} \\
FeTe$_{0.5}$Se$_{0.5}$ & 14 & 6-7~\cite{Mook.PhysRevLett.104.187002,LeePRB2010,Wen2010} & \GDLL{2.61, 0.51-0.87} ($\mu$SR)~\cite{Biswas.PhysRevB.81.092510,Bendele.PhysRevB.81.224520} \\
%
%
\hline
%
%
LiFeAs & 18 & 8~\cite{Taylor.PhysRevB.83.220514} & \DLS{5-6, 2.8-3.5} (ARPES)~\cite{Borisenko.PhysRevLett.105.067002,Borisenko.symmetry-04-00251,Umezawa.PhysRevLett.108.037002} \\
LiFeAs & 18 & 8~\cite{Taylor.PhysRevB.83.220514} & \DLL{5.4, 1.4} (Andreev refl.)~\cite{Kuzmichev2012,Kuzmichev2013} \\
LiFeAs & 18 & 8~\cite{Taylor.PhysRevB.83.220514} & \DLL{5.3, 2.5} (Tunneling)~\cite{Chi.PhysRevLett.109.087002,Hanaguri.PhysRevB.85.214505,Nag.srep27926} \\
\hline
%
NaFe$_{0.935}$Co$_{0.045}$As & 18 & 7~\cite{Zhang.PhysRevB.88.064504} & \DLS{6.8, 6.5} (ARPES)~\cite{Liu.PhysRevB.84.064519} \\
NaFe$_{0.95}$Co$_{0.05}$As & 18 & $\sim 7$? & \DLS{6.8, 6.5} (ARPES)~\cite{Liu.PhysRevB.84.064519} \\
%
\hline
\end{tabular}
\end{table}

On the separate note, I would like to mention that the appearance of a hump structure in the superconducting state at frequencies larger than the main peak frequency (the so-called double resonance feature) may be related to the $2\Delta_L$ energy scale, see Fig.~\ref{fig:5orbImChi}. Such hump structure was observed in NaFe$_{0.985}$Co$_{0.015}$As~\cite{Zhang.PhysRevLett.111.207002,Zhang.PhysRevB.90.140502} and FeTe$_{0.5}$Se$_{0.5}$~\cite{Mook.PhysRevLett.104.187002}. Somehow similar structure was found in polarized inelastic neutron studies of BaFe$_{1.9}$Ni$_{0.1}$As$_2$~\cite{Lipscombe.PhysRevB.82.064515} and Ba(Fe$_{0.94}$Co$_{0.06}$)$_2$As$_2$~\cite{Steffens.PhysRevLett.110.137001}, but its origin may be related to the spin-orbit coupling~\cite{KorshunovTogushovaSO2013} rather than the simple $2\Delta_L$ energy scale. However, more plausible explanation of the double resonance feature is related to the pre-existing magnon mode, i.e. the dispersive low-energy peak in underdoped materials is associated with the spin excitations of the magnetic order with the intensity enhanced below $T_c$ due to the suppression of the damping~\cite{Wang2016}.

\subsection{Final remarks}

The main results here is that the true spin resonance in the $s_\pm$ state appears below the indirect gap scale $\tilde\Delta$ that is determined by the sum of gaps on two different Fermi surface sheets connected by the scattering wave vector $\Q$. In the $s_{++}$ state, spin excitations are absent below $\tilde\Delta$ unless additional scattering mechanisms are assumed. For the Fermi surface geometry characteristic to the most of FeBS materials, the indirect gap is either $\tilde\Delta = \Delta_L + \Delta_S$ or $\tilde\Delta = 2\Delta_L$. Whether the minimal or maximal energy scale will be realized depends on the relation between the exact band structure of a particular material and the wave vector of the spin resonance $\Q$. Comparison of energy scales extracted from INS, Andreev spectroscopy, ARPES and other techniques allowing to determine superconducting gaps, for most materials gives confidence that the observed feature in INS is the spin resonance peak. 

\section{Conclusion}
\label{sec:conclusion}

Spin fluctuations stem from the basic interaction of electrons and thus they are a natural companion of the many-electron system. The variety of physics arising due to the spin fluctuations are truly fascinating. Here I concentrated on metallic multiband systems exemplarily represented by the iron-based superconductors. To be specific, the five-orbital model for pnictides that is composed of the kinetic part and the Hubbard-type interactions was considered. Also, how the spin-orbit interaction enters the model was discussed. Spin excitations were described via the dynamical spin susceptibility. In the multiorbital case, it is a complicated though straightforwardly calculated quantity. Addition of the spin-orbit interaction makes the calculations more cumbersome and can result in the disparity between its $\xx$, $\yy$, and $\zz$ components. For the finite Hubbard interaction, here we used RPA.

To consider the scattering of electrons on spin excitations, we calculated electron's self-energy and kinetic coefficients. Second-order diagram for the self-energy depends on the particle-hole `bubble' that was replaced with the dynamical spin susceptibility in RPA, thereby, advancing beyond the second-order. The quasiparticle scattering due to spin-fluctuations can be significantly anisotropic. The anisotropy on the electron sheets is larger than on the hole sheets and this disparity significantly affects transport properties of pnictides.

Exchange of spin fluctuations produces effective electron-electron interaction that can lead to the Cooper pairing. In a multiband system, this results in the unconventional superconducting state with the gap having a complicated structure. In most cases, it is either $s_\pm$ or $d$-wave gap. Multiorbital spin fluctuation theory can explain many features of Fe-based superconductors. Making the expansion of the Cooper vertex in the leading angular harmonics, we were able (by varying the effective interaction parameters) to determine the stability of each solution for the gap and say which effective interaction leads to the pairing. Depending on the topology and relative volumes of the hole and electron sheets, a competition between the $s_\pm$ and $d$ states appears, however, the $s_\pm$ state always wins in the presence of both electron and hole pockets.

As for the experimental observation of the gap structure, the $s_\pm$ state is characterized by the true spin resonance that appears below the indirect gap scale. Later  is determined by the sum of gaps on two different Fermi surface sheets connected by the scattering wave vector. This gives the simple criterion to determine whether the experimentally observed peak in inelastic neutron scattering is the true spin resonance -- if the peak frequency is less than the indirect gap energy, then it is the spin resonance and, consequently, the superconducting state has the $s_\pm$ gap structure. Comparison of energy scales from the available experimental data on the peak in inelastic neutron scattering and on superconducting gaps extracted from various experimental techniques gives the confidence that the observed peak for most iron-based materials is the true spin resonance. This indirectly points to the spin fluctuation mechanism of pairing.

\subsection{Acknowledgements}

I'm grateful to A.V. Chubukov, O.V. Dolgov, I.M. Eremin, P.J. Hirschfield, H. Kontani, A. Kordyuk, A. Kreisel, S.A. Kuzmichev, T.E. Kuzmicheva, I.I. Mazin, S.G. Ovchinnikov, V.M. Pudalov, M.V. Sadovskii, V.A. Shestakov, and Yu.N. Togushova for useful discussions. This work was supported in part by the Russian Foundation for Basic Research (grant 16-02-00098), Government Support of the Leading Scientific Schools of the Russian Federation (NSh-7559.2016.2), and ``BASIS'' Foundation for Development of Theoretical Physics.

\bibliography{mmkbibl2}
\bibliographystyle{aipnum4-1}

\end{document}